%% file: cpv-rpp.tex
\documentclass[12pt]{iopart}

\input{preamble}

\usepackage{amsfonts}
\newcommand{\tickYes}{\checkmark}
\newcommand{\tickYesYes}{$^{\small \checkmark}$\kern-1.3ex\checkmark}
\usepackage{pifont}
\newcommand{\tickNo}{\hspace{1pt}\ding{55}}
\usepackage{rotating}
\usepackage{multirow}
\usepackage{alphalph}
\usepackage{enumitem}
\usepackage{wasysym} 

\begin{document}

\review{\CP violation in the $B$ system}

\author{T Gershon$^1$ and V V Gligorov$^{2}$}

\address{$^1$ Department of Physics, University of Warwick, Coventry, United Kingdom}
\address{$^2$ LPNHE, Universit\'{e} Pierre et Marie Curie, Universit\'{e} Paris Diderot, CNRS/IN2P3, Paris, France}
\eads{\mailto{T.J.Gershon@warwick.ac.uk}, \mailto{Vladimir.Gligorov@cern.ch}}

\begin{abstract}
The phenomenon of \CP violation is crucial to understand the asymmetry between matter and antimatter that exists in the Universe.
Dramatic experimental progress has been made, in particular in measurements of the behaviour of particles containing the $b$ quark, where \CP violation effects are predicted by the Kobayashi-Maskawa mechanism that is embedded in the Standard Model.
The status of these measurements and future prospects for an understanding of \CP violation beyond the Standard Model are reviewed.
\end{abstract}


\submitto{\RPP}

\maketitle


\renewcommand{\thefootnote}{\AlphAlph{\arabic{footnote}}}
\setcounter{footnote}{0}

\input{introduction}
\clearpage
\input{theory}
\clearpage
\input{experiment}

\clearpage
\input{tree-dominated}

\clearpage
\input{penguin-dominated}

\clearpage
\input{tree-penguin}

\clearpage
\input{baryons}

\clearpage
\input{fits}

\clearpage
\input{summary}

\ack
This work is supported by the 
Science and Technology Facilities Council (United Kingdom) and by the 
European Research Council under FP7. 
The authors acknowledge support from CERN while the review was in preparation,
and thank A.~Bondar and M.~Jung for useful comments.

\clearpage
\section*{Bibliography}
\bibliographystyle{unsrt} 
\bibliography{main,LHCb-PAPER,LHCb-CONF,LHCb-DP}

\end{document}

%% file: preamble.tex
\usepackage{microtype}
\usepackage{lineno}  
\usepackage{xspace} 
\usepackage{caption} 

\usepackage{graphicx}  
\usepackage{color}
\usepackage{colortbl}
\graphicspath{{./figs/}} 

\usepackage{upgreek} 

\usepackage{ifthen} 
\newboolean{pdflatex}
\setboolean{pdflatex}{true} 
\newboolean{articletitles}
\setboolean{articletitles}{true} 
\newboolean{uprightparticles}
\setboolean{uprightparticles}{false} 

\usepackage{hyperref}    
\usepackage[all]{hypcap} 

\input{lhcb-symbols-def} 

\usepackage{cite} 
\usepackage{mciteplus}

%% file: lhcb-symbols-def.tex
\def\lhcb {\mbox{LHCb}\xspace}
\def\atlas  {\mbox{ATLAS}\xspace}
\def\cms    {\mbox{CMS}\xspace}

\def\babar  {\mbox{BaBar}\xspace}
\def\belle  {\mbox{Belle}\xspace}
\def\cleo   {\mbox{CLEO}\xspace}
\def\cdf    {\mbox{CDF}\xspace}
\def\dzero  {\mbox{D0}\xspace}

\def\lhc    {\mbox{LHC}\xspace}
\def\lep    {\mbox{LEP}\xspace}
\def\tevatron {Tevatron\xspace}

\ifthenelse{\boolean{uprightparticles}}%
{

 \def\Peta        {\ensuremath{\upeta}\xspace}

 \def\Pmu         {\ensuremath{\upmu}\xspace}                 
 \def\Pnu         {\ensuremath{\upnu}\xspace}                 
                  
 \def\Ppi         {\ensuremath{\uppi}\xspace}                 
                  
 \def\Prho        {\ensuremath{\uprho}\xspace}

 \def\Ppsi        {\ensuremath{\uppsi}\xspace}

 \def\PDelta      {\ensuremath{\Delta}\xspace}                 
 \def\PXi      {\ensuremath{\Xi}\xspace}                 
 \def\PLambda      {\ensuremath{\Lambda}\xspace}                 
 \def\PSigma      {\ensuremath{\Sigma}\xspace}                 
 \def\POmega      {\ensuremath{\Omega}\xspace}                 
 \def\PUpsilon      {\ensuremath{\Upsilon}\xspace}                 
 
 \def\PB      {\ensuremath{\mathrm{B}}\xspace}                 
                  
 \def\PD      {\ensuremath{\mathrm{D}}\xspace}

 \def\PJ      {\ensuremath{\mathrm{J}}\xspace}                 
 \def\PK      {\ensuremath{\mathrm{K}}\xspace}

 \def\Pb      {\ensuremath{\mathrm{b}}\xspace}                 
 \def\Pc      {\ensuremath{\mathrm{c}}\xspace}                 
 \def\Pd      {\ensuremath{\mathrm{d}}\xspace}                 
 \def\Pe      {\ensuremath{\mathrm{e}}\xspace}

 \def\Pi      {\ensuremath{\mathrm{i}}\xspace}

 \def\Pp      {\ensuremath{\mathrm{p}}\xspace}

 \def\Ps      {\ensuremath{\mathrm{s}}\xspace}

}
{

 \def\Peta        {\ensuremath{\eta}\xspace}

 \def\Pmu         {\ensuremath{\mu}\xspace}                 
 \def\Pnu         {\ensuremath{\nu}\xspace}                 
                  
 \def\Ppi         {\ensuremath{\pi}\xspace}                 
                  
 \def\Prho        {\ensuremath{\rho}\xspace}

 \def\Ppsi        {\ensuremath{\psi}\xspace}                 
                  
 \mathchardef\PDelta="7101
 \mathchardef\PXi="7104
 \mathchardef\PLambda="7103
 \mathchardef\PSigma="7106
 \mathchardef\POmega="710A
 \mathchardef\PUpsilon="7107
                  
 \def\PB      {\ensuremath{B}\xspace}                 
                  
 \def\PD      {\ensuremath{D}\xspace}

 \def\PJ      {\ensuremath{J}\xspace}                 
 \def\PK      {\ensuremath{K}\xspace}

 \def\Pb      {\ensuremath{b}\xspace}                 
 \def\Pc      {\ensuremath{c}\xspace}                 
 \def\Pd      {\ensuremath{d}\xspace}                 
 \def\Pe      {\ensuremath{e}\xspace}

 \def\Pi      {\ensuremath{i}\xspace}

 \def\Pp      {\ensuremath{p}\xspace}

 \def\Ps      {\ensuremath{s}\xspace}

}

\def\epem       {\ensuremath{\Pe^+\Pe^-}\xspace}

\def\mup        {\ensuremath{\Pmu^+}\xspace}
\def\mun        {\ensuremath{\Pmu^-}\xspace} 
\def\mumu       {\ensuremath{\Pmu^+\Pmu^-}\xspace}

\def\lepton     {\ensuremath{\ell}\xspace}

\def\neu        {\ensuremath{\Pnu}\xspace}
\def\neub       {\ensuremath{\overline{\Pnu}}\xspace}
\def\neum       {\ensuremath{\neu_\mu}\xspace}

\def\neul       {\ensuremath{\neu_\ell}\xspace}

\def\dquark    {\ensuremath{\Pd}\xspace}

\def\squark    {\ensuremath{\Ps}\xspace}

\def\cquark    {\ensuremath{\Pc}\xspace}

\def\bquark    {\ensuremath{\Pb}\xspace}

\def\pion  {\ensuremath{\Ppi}\xspace}
\def\piz   {\ensuremath{\pion^0}\xspace}

\def\pip   {\ensuremath{\pion^+}\xspace}
\def\pim   {\ensuremath{\pion^-}\xspace}
\def\pipm  {\ensuremath{\pion^\pm}\xspace}
\def\pimp  {\ensuremath{\pion^\mp}\xspace}

\def\rhomeson  {\ensuremath{\Prho}\xspace}
\def\rhoz   {\ensuremath{\rhomeson^0}\xspace}
\def\rhop   {\ensuremath{\rhomeson^+}\xspace}
\def\rhom   {\ensuremath{\rhomeson^-}\xspace}
\def\rhopm  {\ensuremath{\rhomeson^\pm}\xspace}

\def\kaon  {\ensuremath{\PK}\xspace}
  \def\Kbar  {\kern 0.2em\overline{\kern -0.2em \PK}{}\xspace}

\def\Kz    {\ensuremath{\kaon^0}\xspace}
\def\Kzb   {\ensuremath{\Kbar^0}\xspace}
\def\Kp    {\ensuremath{\kaon^+}\xspace}
\def\Km    {\ensuremath{\kaon^-}\xspace}
\def\Kpm   {\ensuremath{\kaon^\pm}\xspace}
\def\Kmp   {\ensuremath{\kaon^\mp}\xspace}
\def\KS    {\ensuremath{\kaon^0_{\rm\scriptscriptstyle S}}\xspace} 
\def\KL    {\ensuremath{\kaon^0_{\rm\scriptscriptstyle L}}\xspace} 

\def\Kstarz  {\ensuremath{\kaon^{*0}}\xspace}
\def\Kstarzb {\ensuremath{\Kbar^{*0}}\xspace}
\def\Kstar   {\ensuremath{\kaon^*}\xspace}

\def\Kstarp  {\ensuremath{\kaon^{*+}}\xspace}
\def\Kstarm  {\ensuremath{\kaon^{*-}}\xspace}
\def\Kstarpm {\ensuremath{\kaon^{*\pm}}\xspace}

\def\KorKstarz{\ensuremath{\kaon^{(*)0}}\xspace}
\def\KorKstarzb{\ensuremath{\Kbar^{(*)0}}\xspace}

\newcommand{\etapr}{\ensuremath{\Peta^{\prime}}\xspace}

  \def\Dbar    {\kern 0.2em\overline{\kern -0.2em \PD}{}\xspace}
\def\D       {\ensuremath{\PD}\xspace}
\def\Db      {\ensuremath{\Dbar}\xspace}
\def\Dz      {\ensuremath{\D^0}\xspace}
\def\Dzb     {\ensuremath{\Dbar^0}\xspace}
\def\Dp      {\ensuremath{\D^+}\xspace}
\def\Dm      {\ensuremath{\D^-}\xspace}

\def\Dmp     {\ensuremath{\D^\mp}\xspace}
\def\Dstar   {\ensuremath{\D^*}\xspace}

\def\Dstarp  {\ensuremath{\D^{*+}}\xspace}
\def\Dstarm  {\ensuremath{\D^{*-}}\xspace}

\def\Dstarmp {\ensuremath{\D^{*\mp}}\xspace}
\def\DorDstar   {\ensuremath{\D^{(*)}}\xspace}

\def\DorDstarp  {\ensuremath{\D^{(*)+}}\xspace}
\def\DorDstarm  {\ensuremath{\D^{(*)-}}\xspace}

\def\DorDstarmp {\ensuremath{\D^{(*)\mp}}\xspace}

\def\Dsp     {\ensuremath{\D^+_\squark}\xspace}
\def\Dsm     {\ensuremath{\D^-_\squark}\xspace}

\def\Dsmp    {\ensuremath{\D^{\mp}_\squark}\xspace}

\def\Dssmp   {\ensuremath{\D^{*\mp}_\squark}\xspace}

\def\B       {\ensuremath{\PB}\xspace}
\def\Bbar    {\ensuremath{\kern 0.18em\overline{\kern -0.18em \PB}{}}\xspace}

\def\Bz      {\ensuremath{\B^0}\xspace}
\def\Bzb     {\ensuremath{\Bbar^0}\xspace}
\def\Bu      {\ensuremath{\B^+}\xspace}
\def\Bub     {\ensuremath{\B^-}\xspace}
\def\Bp      {\ensuremath{\Bu}\xspace}
\def\Bm      {\ensuremath{\Bub}\xspace}
\def\Bpm     {\ensuremath{\B^\pm}\xspace}
\def\Bmp     {\ensuremath{\B^\mp}\xspace}
\def\Bd      {\ensuremath{\B^0}\xspace}
\def\Bs      {\ensuremath{\B^0_\squark}\xspace}
\def\Bds     {\ensuremath{\B^0_{(\squark)}}\xspace}
\def\Bdb     {\ensuremath{\Bbar^0}\xspace}
\def\Bsb     {\ensuremath{\Bbar^0_\squark}\xspace}
\def\Bdsb    {\ensuremath{\Bbar^0_{(\squark)}}\xspace}

\def\jpsi     {\ensuremath{{\PJ\mskip -3mu/\mskip -2mu\Ppsi\mskip 2mu}}\xspace}
\def\psitwos  {\ensuremath{\Ppsi{(2S)}}\xspace}

  \def\Y#1S{\ensuremath{\PUpsilon{(#1S)}}\xspace}

\def\proton      {\ensuremath{\Pp}\xspace}

\def\Lz {\ensuremath{\PLambda}\xspace}
\def\Lbar {\ensuremath{\kern 0.1em\overline{\kern -0.1em\PLambda}}\xspace}

\def\Lb      {\ensuremath{\Lz^0_\bquark}\xspace}

\def\Lc      {\ensuremath{\Lz^+_\cquark}\xspace}

\def\Xibm    {\ensuremath{\Xi^-_\bquark}\xspace}

\def\Omegabm {\ensuremath{\Omega^-_\bquark}\xspace}

\def\to                 {\ensuremath{\rightarrow}\xspace}

\def\CP                {\ensuremath{C\!P}\xspace}

\def\rhobar {\ensuremath{\overline \rho}\xspace}
\def\etabar {\ensuremath{\overline \eta}\xspace}

\newcommand{\dms}{\ensuremath{\Delta m_{\squark}}\xspace}

\newcommand{\DGs}{\ensuremath{\Delta\Gamma_{\squark}}\xspace}
\newcommand{\DGd}{\ensuremath{\Delta\Gamma_{\dquark}}\xspace}
\newcommand{\Gs}{\ensuremath{\Gamma_{\squark}}\xspace}

\newcommand{\phis}{\ensuremath{\phi_{\squark}}\xspace}
\newcommand{\betas}{\ensuremath{\beta_{\squark}}\xspace}

\newcommand{\asld}{\ensuremath{a^{d}_{sl}}\xspace}
\newcommand{\asls}{\ensuremath{a^{s}_{sl}}\xspace}
\newcommand{\aslq}{\ensuremath{a^{q}_{sl}}\xspace}
\newcommand{\absl}{\ensuremath{A^{b}_{sl}}\xspace}

\newcommand{\tev}{\ensuremath{\mathrm{\,Te\kern -0.1em V}}\xspace}
\newcommand{\gev}{\ensuremath{\mathrm{\,Ge\kern -0.1em V}}\xspace}
\newcommand{\mev}{\ensuremath{\mathrm{\,Me\kern -0.1em V}}\xspace}
\newcommand{\kev}{\ensuremath{\mathrm{\,ke\kern -0.1em V}}\xspace}
\newcommand{\ev}{\ensuremath{\mathrm{\,e\kern -0.1em V}}\xspace}
\newcommand{\gevc}{\ensuremath{{\mathrm{\,Ge\kern -0.1em V\!/}c}}\xspace}
\newcommand{\mevc}{\ensuremath{{\mathrm{\,Me\kern -0.1em V\!/}c}}\xspace}
\newcommand{\gevcc}{\ensuremath{{\mathrm{\,Ge\kern -0.1em V\!/}c^2}}\xspace}
\newcommand{\gevgevcccc}{\ensuremath{{\mathrm{\,Ge\kern -0.1em V^2\!/}c^4}}\xspace}
\newcommand{\mevcc}{\ensuremath{{\mathrm{\,Me\kern -0.1em V\!/}c^2}}\xspace}

\def\mub{\ensuremath{{\rm \,\upmu b}}\xspace}
\def\nb {\ensuremath{\rm \,nb}\xspace}

\def\invfb   {\ensuremath{\mbox{\,fb}^{-1}}\xspace}

\def\ms   {\ensuremath{{\rm \,ms}}\xspace}

\def\mhz  {\ensuremath{{\rm \,MHz}}\xspace}
\def\khz  {\ensuremath{{\rm \,kHz}}\xspace}
\def\hz   {\ensuremath{{\rm \,Hz}}\xspace}

\newcommand{\stat}{\ensuremath{\mathrm{\,(stat)}}\xspace}
\newcommand{\syst}{\ensuremath{\mathrm{\,(syst)}}\xspace}

\def\gsim{{~\raise.15em\hbox{$>$}\kern-.85em
          \lower.35em\hbox{$\sim$}~}\xspace}
\def\lsim{{~\raise.15em\hbox{$<$}\kern-.85em
          \lower.35em\hbox{$\sim$}~}\xspace}

\def\pt         {\mbox{$p_{\rm T}$}\xspace}

\newcommand{\dedx}{\ensuremath{\mathrm{d}\hspace{-0.1em}E/\mathrm{d}x}\xspace}

\def\degrees{\ensuremath{^{\circ}}\xspace}

\def\rad{\ensuremath{\rm \,rad}\xspace}

\def\tell1  {TELL1\xspace}
\def\ukl1   {UKL1\xspace}

\newcommand{\eg}{\mbox{\itshape e.g.}\xspace}
\newcommand{\ie}{\mbox{\itshape i.e.}\xspace}
\newcommand{\etc}{\mbox{\itshape etc.}\xspace}

%% file: introduction.tex
\section{Introduction}
\label{sec:introduction}

Among the possible discrete symmetries of nature, the combined symmetry under charge conjugation ($C$) and parity ($P$) is particularly interesting.
The $C$ operation conjugates all internal quantum numbers, and therefore transforms particles into antiparticles and vice versa, while under $P$ all spatial co-ordinates are inverted.
Violation of \CP symmetry allows an absolute distinction between matter and antimatter to be made~\cite{Landau:1957tp,Okubo:1958zza}, and is a prerequisite for the evolution of a matter dominated universe~\cite{Sakharov:1967dj}.

The observation of the long-lived neutral kaon decaying to two charged pions provided the first experimental observation of \CP violation~\cite{Christenson:1964fg}.
Among the various ideas put forward to explain these phenomena, only that proposed by Kobayashi and Maskawa~\cite{Kobayashi:1973fv} has survived the test of time, and is now an integral part of the Standard Model (SM).
By extending the quark mixing concept of Cabibbo~\cite{Cabibbo:1963yz} to include three quark pairs (or ``families''), a single irreducible phase appears and gives rise to \CP violation.

The phase of the Cabibbo-Kobayashi-Maskawa (CKM) matrix is the sole source of \CP violation in the Standard Model.
Therefore, all possible \CP violating observables in the quark sector are related to this single quantity, which allows for a wide range of tests of the SM predictions to be made.
Remarkably, the theory has survived all such tests to date.
In particular, there are a wide range of observables in the \B sector, \ie\ with particles involving \bquark quarks, that are useful to test the SM predictions for \CP violation.
It is the purpose of this review to describe these observables, giving both the current experimental status as well as future prospects.

It is a striking feature of \CP violation in the \B sector that many of the experimental measurements can be interpreted without the need for detailed calculation of hadronic effects in the initial and final states.
Consequently, the review is focussed on experimental aspects, with only brief discussions of the relevant theoretical methods given at appropriate places.
Progress in theory that proceeds in parallel with improved experimental measurements is, of course, nonetheless essential.
Reviews of theoretical methods appropriate for \B physics can be found in Refs.~\cite{Hocker:2006xb,Antonelli:2009ws,Artuso:2015swg}.

It must be stressed that the purpose of ongoing investigations into \CP violation in the \B sector is not simply to measure the SM parameters ever more precisely, but to uncover evidence of physics beyond the SM.  
In addition to the general arguments that lead to an expectation for discoveries of non-SM physics in the near future (see, for example, Ref.~\cite{Feng:2013pwa}), it is known that the SM cannot explain the baryon asymmetry of the Universe~\cite{Morrissey:2012db} and therefore new sources of \CP violation must exist.
There is, however, no guarantee that the new \CP violation will be observable in \B physics, and it is important to search also in other areas.
Nonetheless, since the precision of current \B physics measurements is still far from the limiting theoretical uncertainties for many important observables, there is a window of possible discovery that must be thoroughly investigated.
Moreover, the rapidly increasing size of the available data samples provides good prospects for discoveries in the next 5--10 years. 

While this review is intended to be self-contained, it is impossible to include details of all of the huge range of \B system \CP violation measurements that have been proposed and performed.
Detailed reviews of quark flavour physics can be found in Refs.~\cite{Antonelli:2009ws,HFAG,Blake:2015tda,NirGershon,PDG2016}, and extensive discussions of the physics programmes of various \B physics experiments in Refs.~\cite{Brodzicka:2012jm,Bevan:2014iga,Browder:2008em,Aushev:2010bq,LHCb-PAPER-2012-031}.
The reader may also be interested in recent reviews on kaon~\cite{Bryman:2011zz,Cirigliano:2011ny} or charm~\cite{Artuso:2008vf,Gersabeck:2012rp} physics, on the strong \CP problem~\cite{Peccei:2006as,Kim:2008hd} and searches for electric dipole moments~\cite{Engel:2013lsa}, or on prospects for \CP violation measurements in the lepton sector~\cite{Nunokawa:2007qh,Branco:2011zb}.

The remainder of the review is organised as follows.
In Sec.~\ref{sec:theory} the notations and conventions for discussion of \CP violation effects are described.
Sec.~\ref{sec:experiment} gives an overview of the experimental facilities and techniques that are used for measurements of \CP violation parameters.
Sections~\ref{sec:tree-dominated}--\ref{sec:tree-penguin} contain descriptions of the current status of measurements, organised by quark-level process.
A dedicated, though brief, discussion of \CP violation effects in $b$ baryon decays in given in Sec.~\ref{sec:baryons}.
The combination of experimental results in global fits is reviewed in Sec.~\ref{sec:fits}, where a discussion of future prospects is also given.
Finally, a summary can be found in Sec.~\ref{sec:summary}.

%% file: theory.tex
\section{Notations and conventions}
\label{sec:theory}

\subsection{Types of \CP violation in the quark sector}
\label{sec:CPV-types}

Since quarks are charged under the strong interaction, they are never observed directly but are always bound into hadrons. 
This has important implications for the phenomenology of quark mixing; in particular, a $\Bds$ meson can oscillate into its antiparticle $\Bdsb$.
Consequently, the physical states (\ie\ the states that have well-defined masses and lifetimes) are admixtures of the flavour eigenstates
\begin{equation}
  \label{eq:physicalStates}
  B^0_{(\squark)\,\rm H} = p \Bds + q \Bdsb ~~ {\rm and} ~~
  B^0_{(\squark)\,\rm L} = p \Bds - q \Bdsb \, ,
\end{equation}
where $p$ and $q$ are complex coefficients that satisfy $p^2 + q^2 = 1$.
The physical states are labelled $B^0_{(\squark)\,\rm H}$ and $B^0_{(\squark)\,\rm L}$ to distinguish the heavier (H) from the lighter (L), and have mass difference $\Delta m_{(\squark)} = m_{B^0_{(\squark)\,\rm H}} - m_{B^0_{(\squark)\,\rm L}}$ and width difference $\Delta \Gamma_{(\squark)} = \Gamma_{B^0_{(\squark)\,\rm L}} - \Gamma_{B^0_{(\squark)\,\rm H}}$. 
Note that $\Delta m_{(\squark)}$ is positive by definition while with the given sign convention $\Delta \Gamma_{(\squark)}$ is predicted to be positive, for both \Bd and \Bs mesons, in the SM.
The values of $\Delta m_{(\squark)}$ and $\Delta \Gamma_{(\squark)}$ have important consequences for the decay-time-dependent decay rates as will be discussed in Sec.~\ref{sec:rates}.

The amplitude for the decay of a \B hadron to a final state $f$ can be expressed as $A_f$, and that for the decay of a \Bbar hadron to the same final state as $\bar{A}_f$.
The corresponding amplitudes for the conjugate processes ($\Bbar \to \bar{f}$ and $\B \to \bar{f}$ decays) are $\bar{A}_{\bar{f}}$ and $A_{\bar{f}}$.
For hadrons carrying any conserved quantum numbers (\ie\ charged mesons and all baryons), it is not possible for both \B and \Bbar to decay to the same final state, and therefore the only possible manifestation of \CP violation is when $\bar{A}_{\bar{f}}$ and $A_f$ are not equal in magnitude.
For neutral \B mesons, however, additional \CP violation effects related to particle-antiparticle mixing can occur.
Considering the concrete case where $f$ is a \CP-eigenstate, \ie\ $f = \bar{f} = f_{\CP}$, the possible \CP violation effects depend on the quantity
\begin{equation}
  \label{eq:lambda}
  \lambda_{f_{\CP}} = \frac{q}{p}\frac{\bar{A}_{f_{\CP}}}{A_{f_{\CP}}}\,.
\end{equation}
Consequently there are two additional possible ways for asymmetries to arise.
The different types of \CP violation can be summarised as
\begin{itemize}
\item \CP violation in mixing: $\left| q/p \right| \neq 1$;
\item \CP violation in mixing/decay interference: ${\rm arg}(\lambda_{f_{\CP}})\neq 0$;
\item \CP violation in decay: $\left|\bar{A}_{f_{\CP}}/A_{f_{\CP}}\right| \neq 1$.
\end{itemize}
A summary of which of these types of \CP violation have been observed, and in which different hadronic systems, is shown in Table~\ref{tab:scoreboard}.

\begin{table}[!htb]
  \centering
  \caption{
    \label{tab:scoreboard}
    Summary of the systems where \CP violation effects have been observed.
    A five standard deviation ($\sigma$) significance threshold is required
    for a \tickYes; several such observations in different channels are
    required for a \tickYesYes.
    Note that \CP violation in decay is the only possible category for particles that do not undergo oscillations.
  }
  \vspace{0.3pc}
  \begin{tabular}{cccccccccccc}
    \hline \\ [-2.8ex]
    & $K^0$ & $K^+$ & $\Lz$ & $D^0$ & $D^+$ & $D_s^+$ & $\Lc$ & $B^0$ & $B^+$ & $B_s^0$ & $\Lb$ \\
    \hline \\ [-2.5ex]
    \CP violation in mixing & \tickYesYes & -- & -- & \tickNo & -- & -- & -- &
    \tickNo & -- & \tickNo & -- \\ [0.5ex]
    \CP violation in & \multirow{2}{*}{\tickYes} & \multirow{2}{*}{--} & \multirow{2}{*}{--} & \multirow{2}{*}{\tickNo} & \multirow{2}{*}{--} & \multirow{2}{*}{--} & \multirow{2}{*}{--} & \multirow{2}{*}{\tickYesYes} & \multirow{2}{*}{--} & \multirow{2}{*}{\tickNo} & \multirow{2}{*}{--} \\
    mixing/decay interference\\ [0.5ex]
    \CP violation in decay & \tickYes & \tickNo & \tickNo & \tickNo & \tickNo & \tickNo & \tickNo & \tickYesYes & \tickYesYes & \tickYes & \tickNo \\
    \hline
  \end{tabular}
\end{table}

Another categorisation of \CP violation effects, as either indirect or direct, can be found in the literature.
This is of mainly historical interest, related to the 1964 proposal of Wolfenstein~\cite{Wolfenstein:1964ks} that the \CP violation effects observed in the kaon system could be due to a new ``superweak'' force that contributes only to meson-antimeson mixing.
In this scenario, all observed \CP violation effects must be consistent with a universal phase in the mixing amplitude.  
The phrase ``indirect \CP violation'' refers to effects that are consistent with this hypothesis, while ``direct \CP violation'' refers to effects that are not; this categorisation is thus useful to test the superweak hypothesis.
Since direct \CP violation has been observed in all of the kaon~\cite{Fanti:1999nm,AlaviHarati:1999xp}, \Bd~\cite{Aubert:2004qm,Chao:2004mn} and \Bs~\cite{LHCb-PAPER-2013-018} systems, there is little reason for further consideration of the superweak hypothesis, and this terminology is in principle obsolete.

\subsection{\boldmath Strong and weak phases}
\label{sec:phases}

As discussed in Sec.~\ref{sec:CPV-types}, the condition for \CP violation in decay is $\left|\bar{A}_{\bar{f}}/A_f \right| \neq 1$.
In order for this to be realised, the amplitudes $\bar{A}_{\bar{f}}$ and $A_f$ must have at least two components with different strong and weak phases.
The labelling of phases as ``strong'' or ``weak'' simply reflects their behaviour under the \CP operation: ``strong'' phases (denoted by $\delta$ in the following) do not change sign under \CP, while weak phases ($\phi$) do.
In the SM, the strong and weak phases arise from hadronic interactions and from the CKM matrix, respectively.

\begin{figure}[!htb]
\centering
\includegraphics[width=0.45\textwidth]{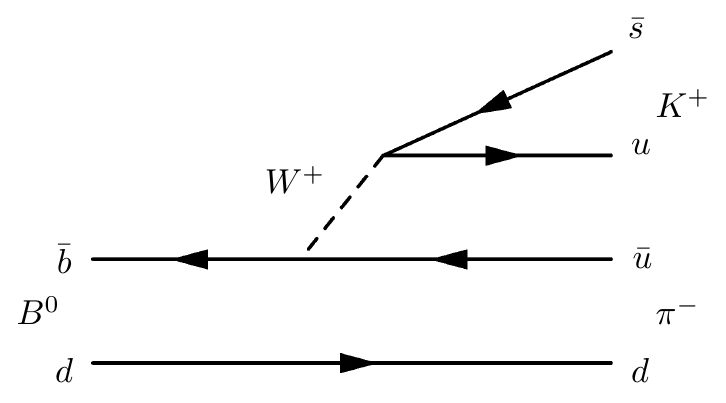}
\includegraphics[width=0.45\textwidth]{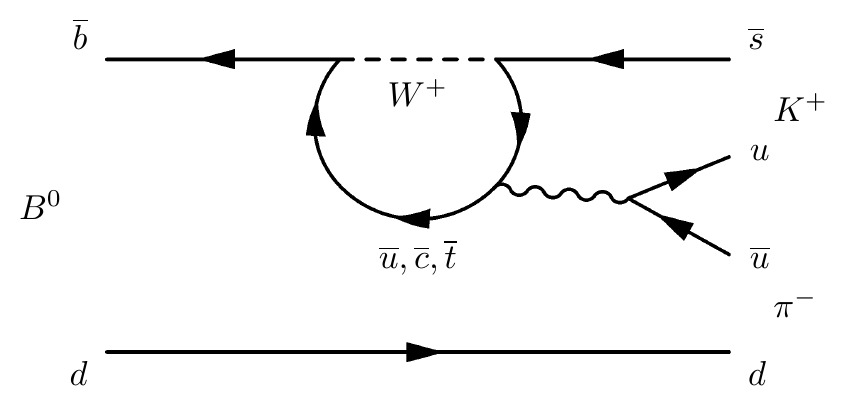}
\caption{
  SM (left) tree and (right) penguin diagrams for the decay $\Bz\to \Kp\pim$.
}
\label{fig:TandP}
\end{figure}

The separation of the amplitudes into components can be done in many ways.
It is particularly useful to separate the components according to the CKM elements involved, whose relative weak phase is therefore known in the SM.
Another approach, of heuristic benefit, is to distinguish ``tree'' ($T$) from ``penguin'' ($P$) amplitudes (example diagrams are shown in Fig.~\ref{fig:TandP}), so that
\begin{equation}
  A_f = \left| T \right| e^{i(\delta_T - \phi_T)} + \left| P \right| e^{i(\delta_P -\phi_P)} \, ,~~
  \bar{A}_{\bar{f}} = \left| T \right| e^{i(\delta_T + \phi_T)} + \left| P \right| e^{i(\delta_P + \phi_P)} \, .
\end{equation}
The \CP asymmetry is defined from the rate difference between the particle involving the quark ($\bar{B}$) and that containing the antiquark ($B$),
\begin{eqnarray}
  {\cal A}_{\CP} & = & \frac{\left| \bar{A}_{\bar{f}} \right|^2 - \left| A_{f} \right|^2}{\left| \bar{A}_{\bar{f}} \right|^2 + \left| A_{f} \right|^2} \, , \nonumber \\
  & = & 
  \frac{2 \left| T \right| \left| P \right| \sin(\delta_T - \delta_P) \sin(\phi_T - \phi_P)} {\left| T \right|^2 + \left| P \right|^2 + 2 \left| T \right| \left| P \right| \cos(\delta_T - \delta_P) \cos(\phi_T - \phi_P)} \, ,   \label{eq:Acp} \\
  & = & 
  \frac{2 r \sin(\delta_T - \delta_P) \sin(\phi_T - \phi_P)} {1 + r^2 + 2 r \cos(\delta_T - \delta_P) \cos(\phi_T - \phi_P)} \, , \label{eq:acp-decay}
\end{eqnarray}
where $r = \frac{\left| P \right|}{\left| T \right|}$ or $\frac{\left| T \right|}{\left| P \right|}$ (it is conventional to choose $r < 1$).

For the \CP asymmetry to be non-zero, it is therefore necessary that all of $\left| T \right|$, $\left| P \right|$, $\sin(\delta_T - \delta_P)$ and $\sin(\phi_T - \phi_P)$ are also non-zero.
For any given \B decay to a final state $f$ there are only two observables (the branching fraction and $A_{\CP}$) and so it is impossible to determine all four underlying parameters without additional input.
Such external information could ideally be taken from first-principles calculations, and a great deal of effort has been invested in developing theoretical methods to enable such calculations (see, for example, Refs.~\cite{Li:1994iu,Li:1995jr,Keum:2000wi,Beneke:1999br,Beneke:2000ry,Beneke:2001ev,Beneke:2002jn,Beneke:2003zv,Bauer:2000yr,Bauer:2001cu,Bauer:2001yt,Beneke:2002ph,Bell:2007tv,Bell:2009nk}).
However, the associated uncertainties are often large.
Fortunately, there are many cases where the parameters of different decay modes can be related, for example by flavour symmetries, and so data-driven methods can be used to determine the weak phase difference.
In such cases, theoretically clean interpretation of \CP violation effects is still possible in processes involving more than one weak phase.
Processes where only a single weak phase contributes can also, in general, be interpreted with low theoretical uncertainty.

An important subtlety in Eq.~(\ref{eq:Acp}) is that the final state $f$ should be a unique point in phase space.
This equation can therefore be trivially applied to two-body decays, but care is needed when discussing multibody final states.
In particular, it is often useful to consider multibody final states in terms of their resonant components, \eg\ in a Dalitz plot analysis of a three-body decay, or an angular analysis of a ($B \to VV$) decay mediated by two vector particles.
In such cases, the resonant lineshapes of contributing intermediate states and the different angular distributions of contributing partial waves can guarantee the required strong phase difference even if none is present in the expressions for the intermediate amplitudes.
Consequently, analysis of multibody \B decays is a powerful approach to investigate \CP violation.

\subsection{\boldmath The Unitarity Triangle}
\label{sec:UT}

\begin{figure}[!htb]
  \centering
  \includegraphics[width=0.5\textwidth]{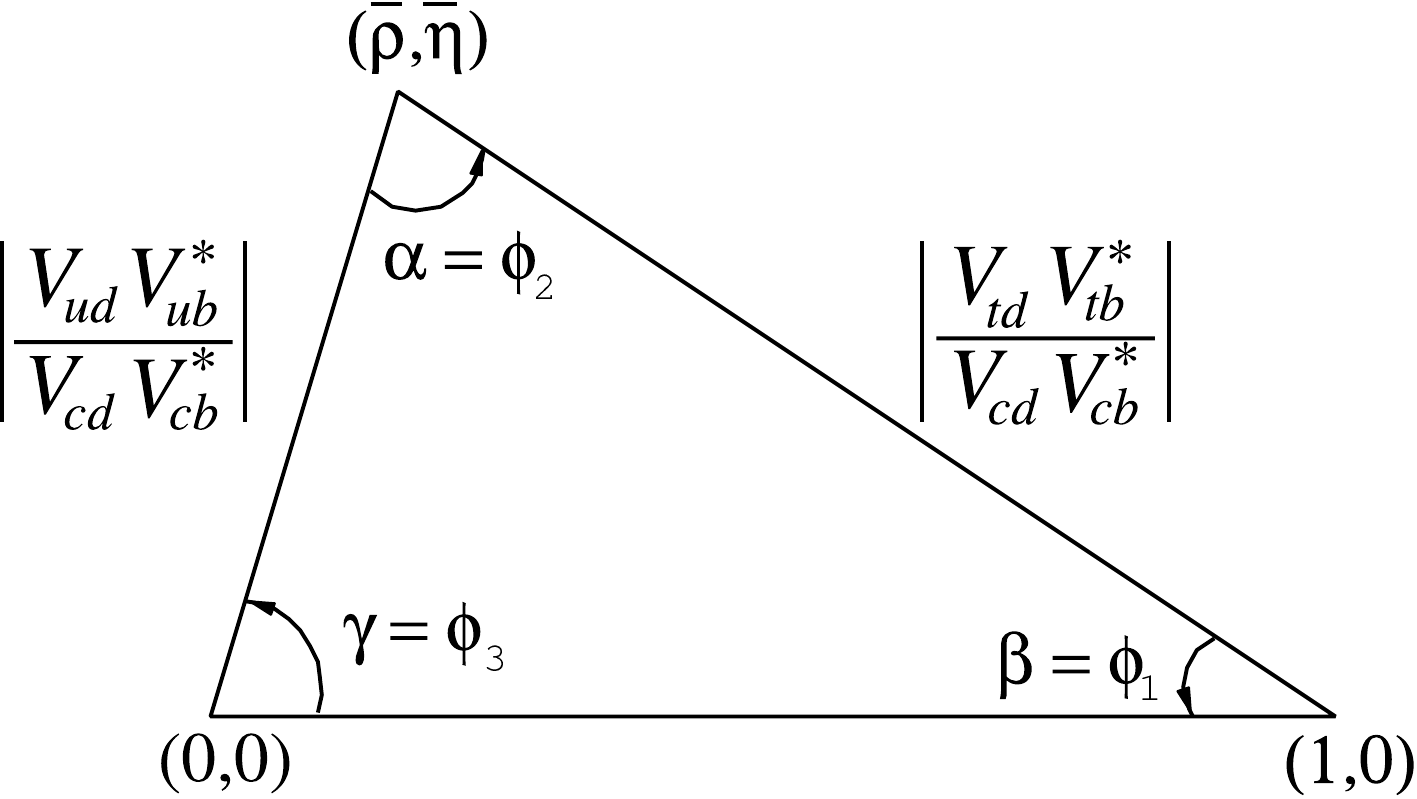}
  \caption{\small 
    The Unitarity Triangle, with both the $\alpha,\beta,\gamma$ and $\phi_1,\phi_2,\phi_3$ notation for the angles. 
    Reproduced from Ref.~\cite{Gligorov:1106345}.
  }
  \label{fig:UT}
\end{figure}

The usefulness of the \B sector for \CP violation tests can be conveniently visualised by considering the unitarity relation between the first and third columns of the CKM matrix
\begin{equation}
  V_{\rm CKM} =
  \left( \begin{array}{ccc}
    V_{ud} & V_{us} & V_{ub} \\
    V_{cd} & V_{cs} & V_{cb} \\
    V_{td} & V_{ts} & V_{tb} \\
  \end{array} \right) \, ,
  \label{eq:CKM}
\end{equation}
that is
\begin{equation}
  V_{ud}^{}V_{ub}^* + V_{cd}^{}V_{cb}^* + V_{td}^{}V_{tb}^* = 0 \, .
  \label{eq:UT}
\end{equation}
The result of Eq.~(\ref{eq:UT}) can, after a trivial rescaling, be expressed as a triangle in the complex plane that is known as the Unitarity Triangle and is shown in Fig.~\ref{fig:UT}.

The apex of the Unitarity Triangle is located at the point $(\rhobar,\etabar)$ given by~\cite{Wolfenstein:1983yz,Buras:1994ec}
\begin{equation}
  \label{eq:rhoetabar}
  \rhobar + i\etabar
  \equiv -\frac{V_{ud}V_{ub}^*}{V_{cd}V_{cb}^*}
  \equiv 1 + \frac{V_{td}V_{tb}^*}{V_{cd}V_{cb}^*} \, .
\end{equation}
The angles of the Unitarity Triangle are defined as
\begin{equation}
  \alpha \equiv
  \arg\left(-\frac{V_{td}^{}V_{tb}^*}{V_{ud}^{}V_{ub}^*}\right) \, , ~~
  \beta \equiv
  \arg\left(-\frac{V_{cd}^{}V_{cb}^*}{V_{td}^{}V_{tb}^*}\right) \, , ~~
  \gamma \equiv
  \arg\left(-\frac{V_{ud}^{}V_{ub}^*}{V_{cd}^{}V_{cb}^*}\right) \, ,
  \label{eq:angles}
\end{equation}
each of which can be measured from various different \B decay processes, as described below.
Among several alternative notations that can be found in the literature, that with 
$\phi_2 \equiv \alpha,~\phi_1 \equiv \beta,~\phi_3 \equiv \gamma$ is particularly widely used.
The $(\alpha, \beta, \gamma)$ notation is, however, prevalent, and is used in this review.
A similar triangle that can be formed from the second and third columns of the CKM matrix has an angle defined by 
\begin{equation}
  \beta_s \equiv
  \arg\left(-\frac{V_{ts}^{}V_{tb}^*}{V_{cs}^{}V_{cb}^*}\right) \, .
  \label{eq:angles-betas}
\end{equation}
This quantity is particularly relevant for \CP violation studies involving oscillations of \Bs mesons.

\subsection{Decay-time-dependent decay rates}
\label{sec:rates}

Following Refs.~\cite{Dunietz:2000cr,HFAG}, the most general decay-time-dependent rates for a \B meson that is initially (at time $t=0$) known to be in the $B^0_{(\squark)}$ or $\overline{B}^0_{(\squark)}$ flavour eigenstate to decay into a final state $f$ are 
\begin{eqnarray}
  \frac{d\Gamma_{B^0_{(\squark)} \to f}(t)}{dt} & \propto & e^{-\Gamma_{(s)}t} 
  \Big[
    \cosh(\frac{\Delta \Gamma_{(s)} t}{2}) + 
    A_{f}^{\Delta \Gamma_{(s)}} \sinh(\frac{\Delta \Gamma_{(s)} t}{2}) + \nonumber \\
    & & \hspace{30mm} C_f \cos(\Delta m_{(s)} t) - S_f \sin(\Delta m_{(s)} t)
    \Big] \, , \label{eg:tdcpv1} \\  
  \frac{d\Gamma_{\overline{B}^0_{(\squark)} \to f}(t)}{dt} & \propto & e^{-\Gamma_{(s)}t}
  \Big[
    \cosh(\frac{\Delta \Gamma_{(s)} t}{2}) +
    A_{f}^{\Delta \Gamma_{(s)}} \sinh(\frac{\Delta \Gamma_{(s)} t}{2}) - \nonumber \\
    & & \hspace{30mm} C_f \cos(\Delta m_{(s)} t) + S_f \sin(\Delta m_{(s)} t)
  \Big] \, , \label{eg:tdcpv2}
\end{eqnarray}
where 
\begin{equation}
  A^{\Delta \Gamma}_f \equiv - \frac{2\, \Re(\lambda_f)}{1 + |\lambda_f|^2} \, , ~~
  C_f \equiv \frac{1 - \left|\lambda_f\right|^2}{1 + \left|\lambda_f\right|^2} \, , ~~
  S_f \equiv \frac{2\, \Im(\lambda_f)}{1 + \left|\lambda_f\right|^2} \, ,
  \label{eq:ACS-defs}
\end{equation}
and $\lambda_f$ is the quantity defined in Eq.~(\ref{eq:lambda}).
The relation $|C_f|^2+|S_f|^2+|A^{\Delta \Gamma}_f|^2 = 1$ holds by definition.
(An alternative notation, $A_f \;\equiv\; -C_f$ is also in use in the literature; moreover, some texts use a different sign convention for $A^{\Delta \Gamma}_f$.)
The decay-time-dependent asymmetry is then
\begin{equation}
  \label{eg:tdcpv-asym}
  \frac{\Gamma_{\overline{B}^0_{(\squark)} \to f}(t) - \Gamma_{B^0_{(\squark)} \to f}(t)}
       {\Gamma_{\overline{B}^0_{(\squark)} \to f}(t) + \Gamma_{B^0_{(\squark)} \to f}(t)} = 
       \frac{S_f \sin(\Delta m_{(s)} t) - C_f \cos(\Delta m_{(s)} t)}
            {\cosh(\frac{\Delta \Gamma_{(s)} t}{2}) +
              A_{f}^{\Delta \Gamma_{(s)}} \sinh(\frac{\Delta \Gamma_{(s)} t}{2})} \, .
\end{equation}
If $f$ is not self-conjugate, similiar decay rates hold for the conjugate final state $\bar{f}$, with parameters $C_{\bar{f}}$, $S_{\bar{f}}$ and $A^{\Delta \Gamma}_{\bar{f}}$. 
For convenience, the $C_f$ and $S_f$ parameters will be referred to as ``sinusoidal'' observables in what follows, while $A_{f}^{\Delta \Gamma}$ terms will be referred to as ``hyperbolic'' observables. 
Since $\DGd=0$ to a good approximation in the SM, the hyperbolic observables are only relevant when considering decays of \Bs mesons.

Depending on the process in question the final state might be flavour specific  (either $\bar{A}_f = 0$ or $A_f = 0$), in which case only \CP violation in mixing or decay are possible, or it might be accessible to both flavour eigenstates (both $\bar{A}_f, A_f \neq 0$), so that \CP violation can also arise in the interference of mixing and decay.
Again, it should be stressed that in Eqs.~(\ref{eg:tdcpv1}) and~(\ref{eg:tdcpv2}), $f$ refers to a particular point in phase space.
In case that the same formalism is used to refer to a state that is an admixture of different components, as in the case of a $B \to VV$ decay such as $\Bs\to\jpsi\phi$ or in a three-body decay such as $\Bd \to\KS\pip\pim$, similar expressions can be formed for each individual component of the final state, but there are additional effects arising from interference between the different components which must
also be accounted for.

The importance of Eqs.~(\ref{eg:tdcpv1}) and~(\ref{eg:tdcpv2}) is that in certain cases they allow straightforward measurement of underlying \CP violation parameters. 
Considering the case that $f$ is a \CP eigenstate, $S_{f_{\CP}}$ probes \CP violation in mixing/decay interference, while $C_{f_{\CP}}$ is sensitive to both \CP violation in mixing and \CP violation in decay -- in practice since \CP violation in mixing is universal and is known to be small, $C_{f_{\CP}}$ can in many cases be interpreted as probing the latter.
If only a single weak phase appears in the decay amplitudes, then $C_{f_{\CP}}$ will be zero and $S_{f_{\CP}}$ can be cleanly interpreted in terms of that phase plus the phase of the mixing amplitude.
Measurement of the hyperbolic parameter $A^{\Delta \Gamma}_{f_{\CP}}$ provides additional sensitivity to the same sum of phases.

In the case that $f$ is not a \CP eigenstate, the situation is slightly more complicated.  
None of the parameters correspond directly to \CP violation, but their combinations do: if $S_f \neq - S_{\bar{f}}$ there is \CP violation in mixing/decay interference, while if $C_f \neq - C_{\bar{f}}$ there is \CP violation in decay. 
An additional probe of \CP violation in decay comes from the normalisation of Eqs.~(\ref{eg:tdcpv1}) and~(\ref{eg:tdcpv2}) and their equivalents for the $\bar{f}$ final state -- any difference in the normalisation factors for $f$ and $\bar{f}$ corresponds to \CP violation. 
Again, it is only in 
the case that a single weak phase contributes
that \CP violation measurements can be cleanly interpreted, but this situation is expected to occur for several decay modes, as discussed below.

It is also possible to obtain weak phase information, through the $A_{f}^{\Delta \Gamma_{(s)}}$ term, in an untagged decay-time-dependent analysis of \Bs decays
\begin{eqnarray}
  \frac{d\left(\Gamma_{\Bs \to f}(t) + \Gamma_{\Bsb \to f}(t)\right)}{dt} 
  & \propto & e^{-\Gs t} 
    \Big[
      \cosh\left(\frac{\DGs t}{2}\right) + A_{f}^{\DGs} \sinh\left(\frac{\DGs t}{2}\right)
    \Big] \, . \nonumber \\
    \label{eg:tdcpv-untagged}
\end{eqnarray}
This distribution can be fitted with a single exponential function to determine the effective lifetime $\tau_{\rm eff}$ given by~\cite{Fleischer:2011cw}
\begin{eqnarray}
  \tau_{\rm eff} \Gs 
  & = & \frac{1}{1-y_s^2} \left( \frac{1+ 2A_{f}^{\DGs}\,y_s + y_s^2}{1+A_{f}^{\DGs}\,y_s} \right) \, ,
\nonumber \\
  & = & 1 + A_{f}^{\DGs}y_s + 
  \left[ 2 - (A_{f}^{\DGs})^2 \right] y_s^2 + ... \, 
  \label{eq:tau-eff}
\end{eqnarray}
where $y_s = \left(\frac{\DGs}{2\Gs}\right)$ and the ellipses denote higher order terms.
The effective lifetimes of several \Bs decays have been measured (for reviews, see Refs.~\cite{HFAG,Stone:2014pra}).
However, since tagged analyses give more precise information on the weak phase and thereby on \CP violation, results on effective lifetimes are not mentioned in this review except in a few specific cases.  

%% file: experiment.tex
\section{Experimental facilities and techniques}
\label{sec:experiment}

The measurements which delimit our current knowledge of \CP violating observables in the \B sector can be split into two kinds: those performed at $e^+e^-$ colliders, and those performed at hadron
colliders. The available experimental techniques also largely divide into those feasible at $e^+e^-$ colliders, at hadron colliders, and at both. 
In this Section, following a brief introduction to the different features relevant for $B$ physics of each type of collider facility, the techniques used in the measurements are discussed.
For reasons of brevity some early experiments which searched for \CP violation in the \B sector, including the \lep detectors and \cleo, are omitted. 

The term ``stable charged particles'' is used throughout this section to mean the electron, muon, pion, kaon and proton, while ``stable charged leptons/hadrons'' refers to the appropriate subset of these. 
The term ``trigger'' is used to refer to real-time preselection of interesting events which are then kept for later analysis. 

\subsection{Features of $B$ physics experiments at $e^+e^-$ colliders}

\begin{figure}[!htb]
\centering
\includegraphics[width=0.46\textwidth]{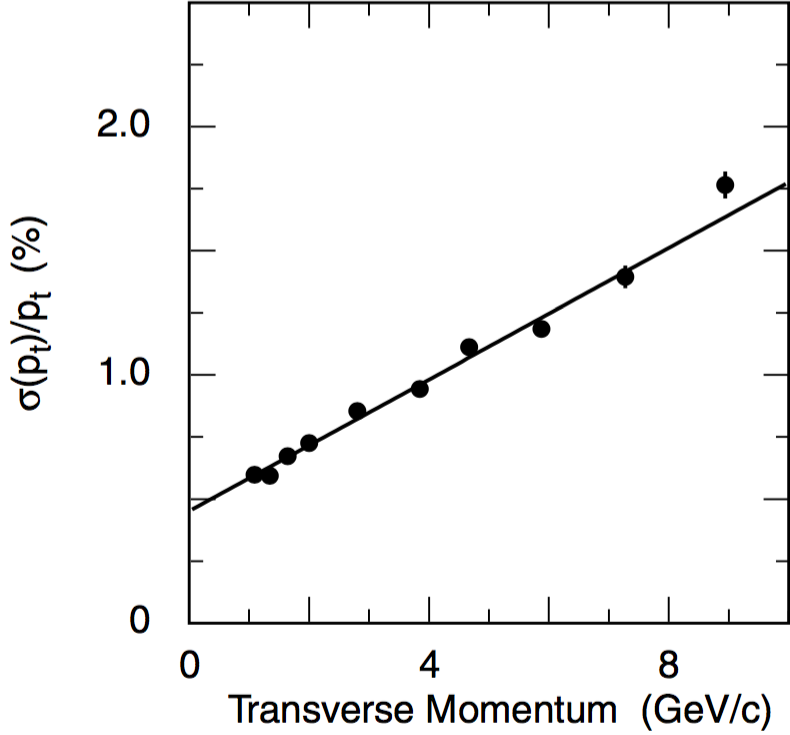}
\includegraphics[width=0.08\textwidth]{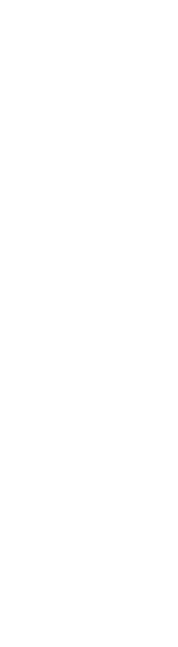}
\includegraphics[width=0.44\textwidth]{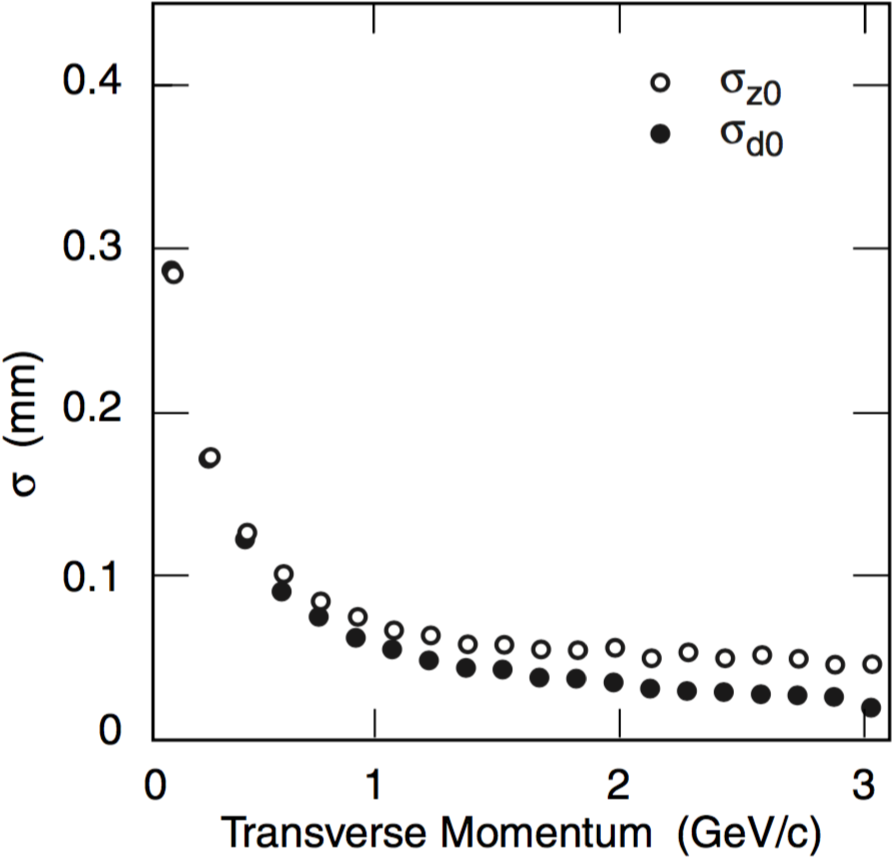}
\caption{
  (Left) transverse momentum ($p_{\rm T}$) and (right) impact parameter resolutions of the \babar detector, reproduced from Ref.~\cite{Aubert:2001tu}. 
}
\label{fig:epemperf}
\end{figure}

A great deal of $B$ physics has been achieved by the asymmetric $e^+e^-$ collider experiments \babar~\cite{Aubert:2001tu,BABAR:2013jta} and \belle~\cite{Abashian:2000cg}, located at the SLAC and KEK collider facilities, respectively. 
They share, and their physics reach is defined by, the following characteristics:
\begin{itemize}
\item 
  The production of \B meson/antimeson pairs via the decay of a $b\bar{b}$ resonance produced in the $e^+e^-$ collision.
  When the collision energy corresponds to the mass of the $\Upsilon(4S)$ resonance, the only species produced are the \Bu and \Bd (anti)mesons.
  The production is coherent, so that the wavefunction of the produced \Bd and \Bdb pair evolves in phase until one or the other decays.
  By operating at the $\Upsilon(5S)$ resonance it is, however, possible to
produce also $\Bs$--$\Bsb$ pairs.
\item 
  Asymmetric $e^+e^-$ beam energies which provide a boost large enough to resolve \Bz, but not \Bs, oscillations. 
  As the centre-of-mass energy of each collision is known, it can be used to distinguish genuine \B hadrons, each carrying half of the centre-of-mass energy, from backgrounds.
\item 
  Detectors with close to $4\pi$ solid angle coverage, providing the ability to efficiently reconstruct all visible decay products of the produced \B particles, both charged and neutral.
  Solenoidal magnetic fields allow track momenta to be determined; the field strength ($1.5 \, {\rm T}$ for both \babar\ and \belle) is optimised to provide good resolution while allowing all but the lowest momentum tracks to escape the inner detector. 
  Particle identification devices~\cite{Iijima:2000uv,Adam:2004fq} provide the capability to distinguish efficiently between all the different species of stable charged particles over the momentum range of interest.
\item 
  Trigger systems that are highly efficient for $\B\Bbar$ events, allowing essentially all such events to be saved to tape without any inefficiencies.
\end{itemize}
Because of the clean production environment and hermetic detectors, $e^+e^-$ collider experiments are particularly useful for studying \CP violating decays involving one or more neutral particles in the final state. 
Fig.~\ref{fig:epemperf} shows the typical transverse momentum and impact parameter resolutions achieved in $e^+e^-$ collider experiments.

\subsection{Features of $B$ physics experiments at hadron colliders}

The most significant hadron collider experiments to date, in terms of $B$ physics and \CP violation, are \dzero and \cdf at the \tevatron proton-antiproton collider, as well as \cms, \atlas and \lhcb at the \lhc proton-proton collider. 
Although there are some important differences in the production environment between these two colliders, the experiments largely share the following characteristics:
\begin{itemize}
\item 
  The production of \B (anti)hadrons via gluon fusion (LHC) or $q\bar{q}$ annihilation (Tevatron), which gives access to all species of \B (anti)hadrons. 
  The production is incoherent, so that any combination of beauty hadron and antihadron can in principle be produced together.
\item
  A geometrical acceptance that is far from hermetic.  
  Although \dzero, \cdf, \cms and \atlas all cover close to $4\pi$ of solid angle (and are therefore equipped with solenoidal magnets), the production of $b\bar{b}$ pairs in high energy hadron collisions is predominantly at large pseudorapidity, \ie\ close to the beam-line, so that many of the decay products pass through the uninstrumented region.
  The \lhcb detector is designed as a forward spectrometer, covering the approximate pseudorapidity region $2 < \eta < 5$, in order to maximise the acceptance for \B physics.
  A dipole magnet, with bending power of about $4~{\rm T\,m}$, deflects charged particles and allows their momentum to be determined.
\item 
  A sufficiently large boost of the produced (anti)hadrons to enable, together with precision vertex detectors, both \Bz and \Bs oscillations to be resolved.
\item 
  Large numbers of additional particles from the underlying proton-(anti)proton interaction which make it infeasible to perform the same kind of full event reconstruction as at $e^+e^-$ colliders. 
\item 
  Very high production cross-sections, that necessitate the separation of $b\bar{b}$ events from other types of events using non-trivial trigger systems.
  The trigger systems introduce substantial inefficiencies for certain \B (anti)hadron decay channels.
\item 
  Multiple inelastic collisions per beam crossing, known as ``pile-up''.
  Since the \cms\ and \atlas\ experiments are primarily focussed on searching for signatures of rare high-\pt processes, significant pile-up helps to achieve high integrated luminosity, although it can introduce significant challenges for the detector operations.
  For \lhcb, limited pile-up is necessary to obtain acceptable detector performance and to associate \B-hadron candidates with the correct primary $pp$ vertex.
  Since the instantaneous luminosity is below the maximum available, it can be tuned to remain constant throughout each LHC fill, providing stable data taking-conditions.
\item 
  An ability to distinguish efficiently the different kinds of charged leptons, but no general ability to distinguish between the different kinds of stable charged hadrons. 
  There are two exceptions: \cdf was able to achieve around $1.5\sigma$ separation between kaons and pions using \dedx information, while \lhcb is unique among hadron collider experiments as its ring imaging Cherenkov detectors~\cite{LHCb-DP-2012-003} provide the capability to distinguish efficiently between all the different species of stable charged particles over the momentum range $2$--$100 \gevc$.
\end{itemize}

Because hadron collider experiments have highly selective triggers, their physics reach greatly depends on what kinds of signatures these triggers can select. 
The \cms, \atlas, and \dzero experiments rely predominantly on the signature of one or more relatively high transverse momentum muons, but this approach restricts the $\B$ physics programme to the class of decays that produce such muons.
On the other hand, since \B hadrons are produced in pairs and decay producing a muon around $11\,\%$ of the time,
signatures of the decay of the other \B in the event can provide a minimum trigger efficiency even for channels that are hard to reconstruct.

\begin{figure}[!htb]
\centering
\includegraphics[width=0.485\textwidth]{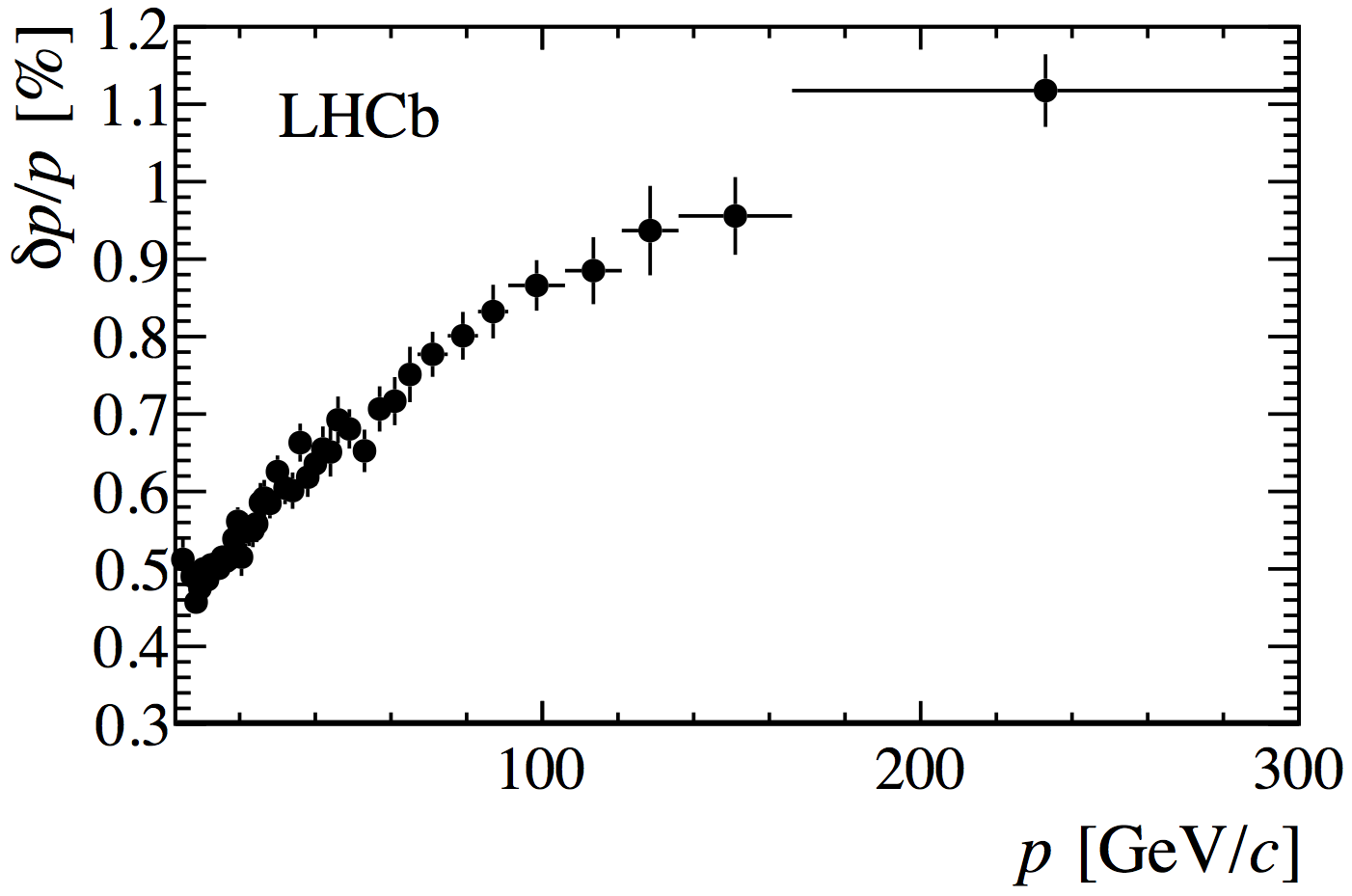}
\includegraphics[width=0.470\textwidth]{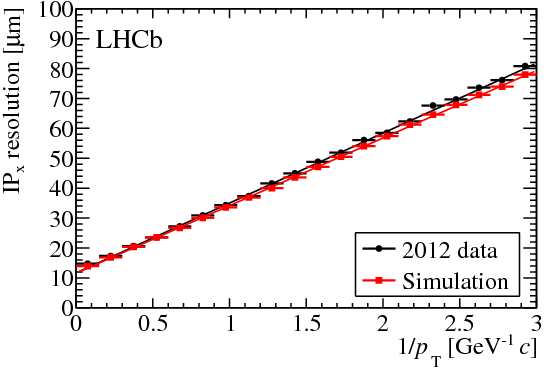}
\caption{
  (Left) momentum and (right) impact parameter resolution of the \lhcb detector, reproduced from Ref.~\cite{LHCb-DP-2014-002}. 
}
\label{fig:hadronperflhcb}
\end{figure}

\begin{figure}[!htb]
\centering
\includegraphics[width=0.46\textwidth]{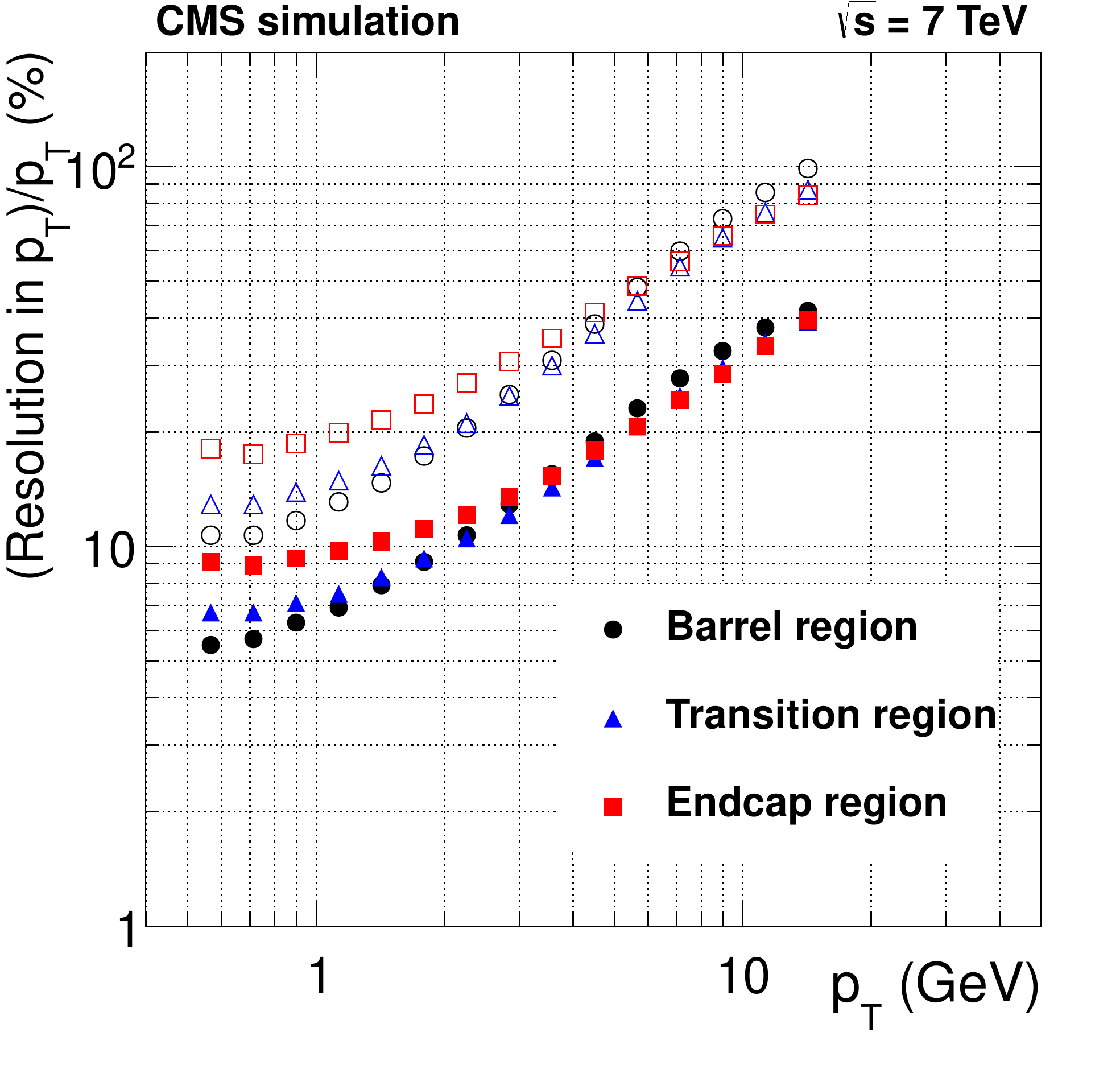}
\includegraphics[width=0.46\textwidth]{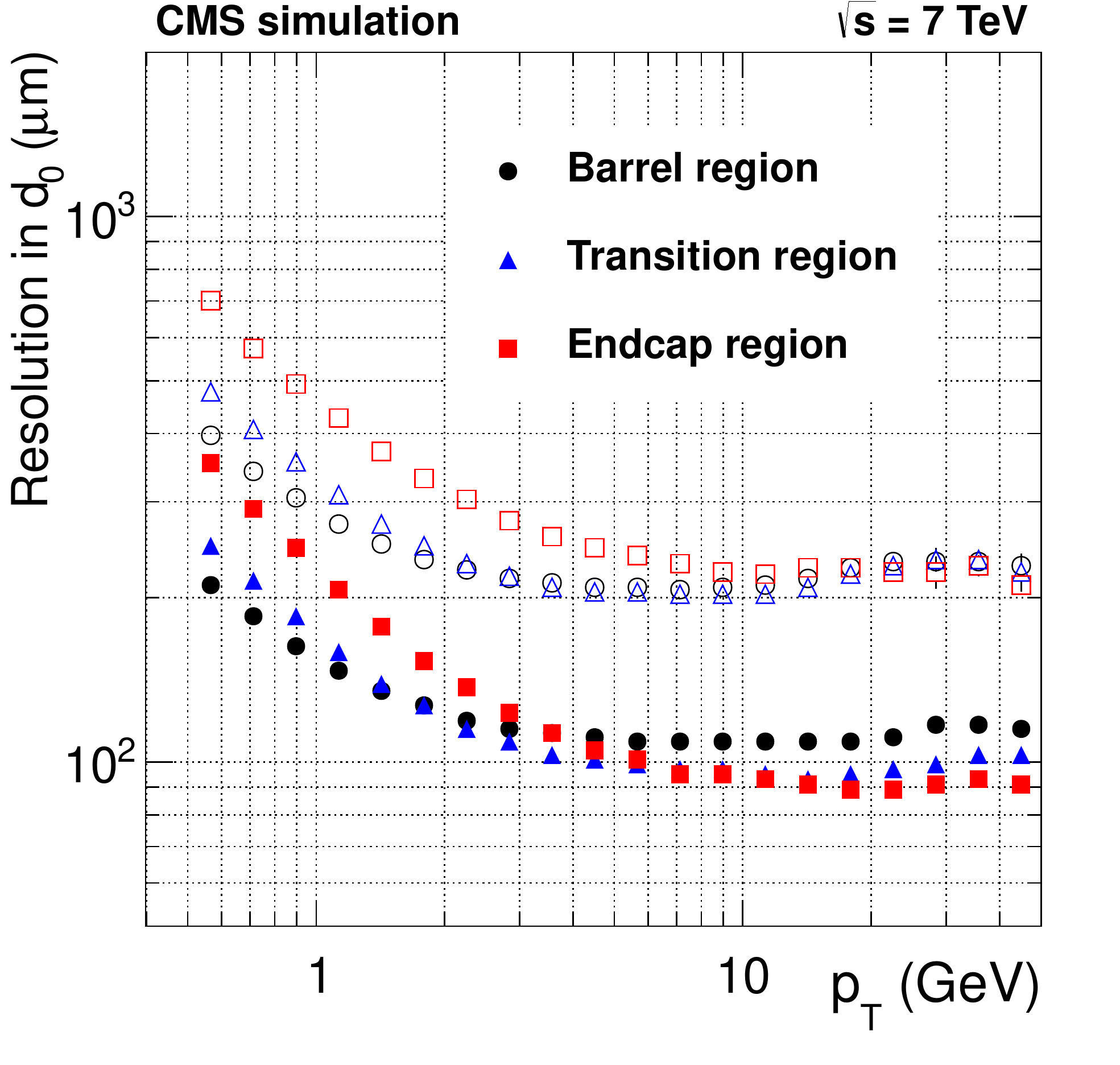}
\caption{
  (Left) transverse momentum and (right) impact parameter resolution of the \cms detector, reproduced from Ref.~\cite{Chatrchyan:2014fea}. 
  The solid (open) symbols correspond to the half-width for $68\,\%$ ($90\,\%$) intervals centred on the mode of the distribution in residuals.
}
\label{fig:hadronperfcms}
\end{figure}

The most distinctive feature of \B decays at a hadron collider experiment that can be exploited in a trigger system is the displaced vertex that is a consequence of the \B lifetime and the significant Lorentz boost.
If the \B decay (secondary) vertex can be efficiently separated from the primary ($pp$ or $p\bar{p}$) interaction vertex, large backgrounds due to the high multiplicity of charged particles (tracks) originating from the primary vertex can be avoided.
\cdf was the first hadron collider experiment whose trigger could use information about the \B decay vertex displacement at the earliest stage of the event selection~\cite{Ristori:2010zz}. 
This gave \cdf the ability to efficiently select hadronic as well as leptonic \B decays, a feature shared with \lhcb~\cite{LHCb-DP-2014-001,LHCb-DP-2012-004}.
While hadron collider experiments are able to study certain final states containing neutral particles, the high occupancies make background discrimination very difficult even in experiments with excellent calorimeter resolution. 
Figs.~\ref{fig:hadronperflhcb}~and~\ref{fig:hadronperfcms} shows the momentum and vertex resolutions achieved by the \lhcb and \cms experiments, respectively.


\subsection{Techniques for \CP violation measurements}

Regardless of the type of \CP violation in question, all measurements of \CP violation in \B decays amount to quantifying the difference in the total or differential decay rates of \B and \Bbar hadrons to a particular final state. 
Depending on the specific measurement in question, systematic uncertainties may arise from asymmetries in the contributing backgrounds, asymmetries in the way the \B hadrons are produced, or asymmetries in the signal detection efficiency for each \B hadron flavour. 

\subsubsection{Signal reconstruction}

The efficiency to reconstruct most final state particles within the detector acceptance is broadly similar across $e^+e^-$ and hadron collider experiments: if
a particle traverses the detector, it will deposit sufficient energy to allow it to be reconstructed. 
Specific techniques can, in certain cases, be used to reconstruct particles such as neutrinos for which this is not the case.
The challenge is therefore to achieve good background rejection while retaining high signal efficiency.
At $e^+e^-$ collider experiments operating near the $\Upsilon(4S)$ resonance this is done by exploiting the event shape, since continuum $e^+e^- \to q\bar{q}$ ($q=u,d,s,c$) events tend to be more jet-like compared to $e^+e^- \to B\bar{B}$ events which are more spherical. 
At hadron collider experiments, the long lifetime of the \B hadron and its large mass (which leads to large transverse momentum for its decay products) results in a distinctive decay topology which can be used to reject background.
Consequently, for many \B hadron decays of interest the selection efficiency is broadly comparable between $e^+e^-$ and hadron collider experiments, especially for final states which involve only charged particles. 
The larger rate of \B hadron production at hadron colliders then leads to larger yields, if the trigger efficiency is not too low, as can be seen in Table~\ref{tab:dacproundup}.

\begin{table}[!tb]
  \caption{
    Summary of yields per unit luminosity and sensitivity of certain representative $A_{\CP}$ measurements at different experiments.
    The first quoted uncertainties are statistical and the second systematic.
  }
  \label{tab:dacproundup}
  \centering
\begin{tabular}{llcr@{$\,\pm\,$}r@{$\,\pm\,$}rc}
  \hline
  Mode & Experiment & Yield/fb$^{-1}$ & \multicolumn{3}{c}{$A_{\CP}$ (\permil)} & Reference \\
  \hline
  $\Bp \to \jpsi \pip$ & \babar & 3 & $123$ & $85$ & $4$ & \cite{Aubert:2004pra} \\
                       & \lhcb  & 5100 & $5$ & $27$ & $11$ & \cite{LHCb-PAPER-2011-024} \\
                       & \dzero & 300 & $-42$ & $44$ & $9$ & \cite{Abazov:2013sqa} \\
  \hline
  $\Bz \to \Kp\pim$    & \babar & 12.5 & $-107$ & \multicolumn{2}{r}{$\!\!16\quad^{+6}_{-4}$} & \cite{Lees:2012mma} \\
                       & \belle & 10.5 & $-69$ & $14$ & $7$ & \cite{Duh:2012ie} \\
                       & \lhcb  & 41400 & $-80$ & $7$ & $3$ & \cite{LHCb-PAPER-2013-018} \\
                       & \cdf   & 1300 & $-83$ & $13$ & $4$ & \cite{Aaltonen:2014vra} \\
  \hline
  $\Bz \to \Kstarz\gamma$ & \babar & 7.5 & $-16$ & $22$ & $7$ & \cite{Aubert:2009ak} \\
                          & \lhcb  & 5300 & $8$ & $17$ & $9$ & \cite{LHCb-PAPER-2012-019} \\
  \hline
\end{tabular}
\end{table}

\subsubsection{Background suppression}

Backgrounds in \B\ hadron \CP violation measurements generally divide into two categories: random combinations of particles produced in the underlying event (hadron collision or \epem\ continuum), and backgrounds due to other \B~hadron decays which are mistaken for the signal. 
Combinatorial background is generally suppressed through topological selection criteria and the kinematic variables used to identify signal \B decays.
In hadron collider experiments, the \B candidate mass is almost always used, as the good momentum resolution leads to a narrow peak for signal while the combinatorial background has a slowly varying shape.
In the \epem\ environment, constraints from the known beam energies can be used, and it is typical to use the two almost uncorrelated variables $\Delta E$ and  $m_{\rm ES}$.
The former is the difference between the energy of the \B\ candidate and $\sqrt{s}/2$ and is strongly correlated with the \B candidate mass, while $m_{\rm ES}=\sqrt{s/4-{\bf p}^2_\B}$, where $\sqrt{s}$ is the total energy, ${\bf p}_\B$ is the momentum of the candidate \B\ meson, and all variables are determined in the centre-of-mass frame.

Backgrounds due to \B\ hadron decays typically occur when another \B\ hadron decay is partially reconstructed and the missing particle has low energy, for example $\Bz\to\Kstarp \pim$ as a background to $\Bz \to \Kp\pim$, or when a final state particle is misidentified, for example $\Bz \to \Kp\pim$ as a background to $\Bs \to \Kp\Km$.
Partially reconstructed backgrounds have a lower reconstructed mass than the true mass of the decaying particle, with a distribution that is smeared depending on the missing energy.
Thus good $\Delta E$ resolution, which is ultimately dependent on good final state particle momentum resolution, is critical to reject such backgrounds in \epem\ experiments.  
Good resolution is also essential for hadron collider experiments, but the situation is more complicated since decays of higher mass \B\ hadrons can lead to partially reconstructed backgrounds that peak near the signal region, \eg\ $\Bs \to \jpsi\pip\pim$ as a background to $\Bp \to \jpsi\pip$ decays.  
In most cases such backgrounds can be separated from the signal due to their different reconstructed mass distributions.  
Further rejection of this type of background can be achieved, to some extent, through isolation criteria.
However, in some cases the residual background level must be estimated from the (absolute or relative) production rates of the relevant \B\ hadrons, their branching fractions to the final states of interest, and the selection efficiencies.
For this reason, it is important for experiments to determine systematically the branching fractions of \B\ hadrons into all accessible final states.

Misidentified backgrounds are also shifted in \B mass with respect to the signal, by an amount that depends on the correct and assumed particle-type hypotheses and on the particle's momentum.
Such backgrounds can be suppressed using particle identification, but typically remain at non-negligible levels after optimisation of selection criteria.
Knowledge of both the level and shape of these backgrounds in the reconstructed mass distribution are critical to reduce associated systematic uncertainties.

\subsubsection{Production and detection asymmetries}

In principle all experiments face issues due to production and detection asymmetries, though in many cases the effects are small enough to be negligible.
Although \B and \Bbar mesons are produced in pairs, and hence in equal numbers, at \epem colliders, there is in principle a forward-backward asymmetry that arises due to electroweak interference effects.
When coupled with an asymmetric detector, this can result in a production asymmetry, as has been seen for $\D/\Dbar$ production~\cite{delAmoSanchez:2011zza,Ko:2012pe}.  
The effect is, however, suppressed by the resonant production of \B mesons in \epem collisions and is negligible for all measurements to date. 
Similarly, the production of \B and \Bbar hadrons in $p\bar{p}$ collisions is symmetric, and therefore asymmetries are negligible.
The $pp$ collision environment, in contrast, does {\it a priori} introduce a production asymmetry, that may vary as a function of collision energy and of the kinematics of the produced \B~hadron.
The asymmetry can be determined from the data, using control modes in which \CP violation can be assumed to be negligible. 
Such determinations have been performed for various \B~hadron species~\cite{LHCb-PAPER-2014-042,LHCb-PAPER-2015-032}.
Since the effects are below the percent level, consistent with expectations~\cite{Norrbin:2000zc}, they concern only the most precise measurements.

The dominant source of detection asymmetries is the different interaction of positively and negatively charged particles with the detector material, which causes their reconstruction efficiencies to differ. 
Because of their nature, detection asymmetries are difficult to simulate accurately and it is essential to measure them directly from the data for each experiment, typically using tag-and-probe techniques.  
A control decay which can be selected with a good signal-to-background ratio 
without reconstructing all final state tracks (tag) is required, and the detection asymmetry is measured from the efficiency to add the missing (probe) track. 
A good example of this technique is the case of $\Dstarp \to \Dz(\Km\pip\pip\pim)\pip$ decays, in which the near-threshold $\Dstarp$--$\Dz$ mass difference remains a powerful signal-to-background discriminant even when one of the four $\Dz$ decay products is not reconstructed. 
Similar methods can be applied for all stable charged particles. 
In general, the most challenging asymmetries are those of protons, both because the difference in material interaction is largest and because tag-and-probe methods rely on decays of charmed baryons, which are less plentiful and less well-known than charmed mesons.
It is also possible to induce detection asymmetries through certain specific selection criteria, for example, because the calorimeter response may be different for positively or negatively charged electrons. 
However, plentiful control samples exist for all particle species and consequently the only challenge is to select a control sample with the same kinematic and geometric distribution as the signal.
Detection asymmetries can also be minimised by reversing the polarity of the magnetic field; this is not possible for all detectors, but is the case in the \lhcb and \dzero experiments.
Measurements of detection asymmetries are discussed for example in Refs.~\cite{LHCb-PAPER-2012-009,LHCb-DP-2013-002,LHCb-PAPER-2014-013,LHCb-PAPER-2014-053,Lees:2015rka,Nakano:2005jb,Abazov:2013uma}.

\subsubsection{Decay-time measurement and flavour tagging}

Several important \CP violation observables can be determined from asymmetries in the decay-time distributions of \B and \Bbar mesons to specific final states.
These measurements require precise determination of vertex positions, in order to determine the decay time, together with ``flavour tagging'', \ie\ information on whether the decaying meson was in a \B or \Bbar state at the time it was tagged.  
Silicon vertex detectors~\cite{Bozzi:2000ic,Natkaniec:2006rv,LHCb-DP-2014-001} are used to determine vertex positions. 
In the case of experiments at $e^+e^-\to\Upsilon(4S)\to\B\Bbar$ colliders, the decay time is determined from the difference between the positions of the vertices of the two \B mesons; the known boost of the $\Upsilon(4S)$ system in the laboratory frame caused by the asymmetric beam energies results in a linear relation between the vertex separation and the decay time difference. 
Hence, $t$ (with range $0 \leq t \leq \infty$) is replaced with $\Delta t$ (with range $-\infty \leq \Delta t \leq \infty$) in Eqs.~(\ref{eg:tdcpv1}),~(\ref{eg:tdcpv2}), and~(\ref{eg:tdcpv-asym}).
For experiments at hadron colliders, the relevant quantity is instead the distance between the \B meson decay vertex and the primary collision vertex; the boost in this case is determined directly from measurement of the \B momentum vector.
Since selection of signal in the trigger often involves requirements on the displacement of the candidate \B vertex, the efficiency can vary significantly as a function of decay time.
In certain analyses, this leads to a potential source of systematic uncertainty that is not relevant for experiments at $e^+e^-$ colliders.

There are distinct differences in the methods used for flavour tagging at $e^+e^-$ and hadron colliders.
In the former case, signatures that can indicate the flavour of the other \B meson in the $\Upsilon(4S)\to\B\Bbar$ event are combined.
These include the charge of a lepton produced in a semileptonic \B decay and the charge of a kaon produced in the $\B \to \D \to \kaon$ decay sequence.  
Multivariate techniques can be used to determine the optimum combination of the information that is available in any given event.
The relative preponderance and clarity of these signatures leads to good performance~\cite{Aubert:2002rg,Kakuno:2004cf}, as quantified in terms of the effective flavour tagging efficiency $\epsilon_{\rm eff} = \epsilon D^2$
where $\epsilon$ is the efficiency to obtain a tag and $D = 1-2w$ is the dilution caused by the probability $w$ to incorrectly determine the flavour.
However, the use of tags from decays which are not fully flavour-specific, such as $\Bz \to \Dstarm \pip$, results in interference effects which lead to systematic uncertainties~\cite{Long:2003wq}.

Hadron collisions produce, in general, a large number of particles in addition to those from \B meson decays, and therefore flavour tagging is more challenging.
The methods used can be characterised as either ``opposite-side'' or ``same-side'' taggers.
Opposite-side algorithms~\cite{LHCb-PAPER-2011-027,LHCb-PAPER-2015-027} are similar in concept to those used in $e^+e^-$ colliders, and search for decay products of the other \B meson produced from the primary collision.
Same-sign algorithms~\cite{LHCb-PAPER-2015-056,LHCb-PAPER-2016-039} exploit the fact that additional particles are produced in the fragmentation processes that produces the \B meson of interest.
These are particularly useful for tagging \Bs mesons, since conservation of strangeness in strong interactions implies that an associated charged kaon can effectively tag the \Bs flavour at production.
The effective tagging efficiency achieved depends on both whether a \Bd\ or \Bs\ decay is considered and on the kinematics (particularly the \pt) of the selected decays, which may be affected for example by trigger requirements.

\subsection{Complementarity of experimental techniques}

Table~\ref{tab:gibson} summarises some key properties of hadron and $e^+e^-$ collider experiments relevant for measurements of \CP violation in the \B sector. 
It can be seen that the two kinds of environments complement each other: $e^+e^-$ collider experiments have much more powerful flavour tagging capabilities and are better able to study final states involving multiple neutral particles or missing energy.
On the other hand, hadron colliders provide access to all flavours of \bquark hadrons and generate much larger yields for final states involving only charged particles, especially leptons. 
Because of this complementarity, a new generation of both hadron and $e^+e^-$ collider experiments is planned. 
The \belle~II~\cite{Aushev:2010bq,Abe:2010gxa} and upgraded \lhcb~\cite{LHCb-TDR-012,LHCb-PAPER-2012-031} experiments aim to begin data-taking within the next five years and will improve on their predecessor's integrated luminosity by around an order of magnitude, while maintaining or improving detector resolutions and particle identification performance. 
In the further future, the Future Circular Collider (FCC) initiative is investigating possibilities for both \epem~\cite{Gomez-Ceballos:2013zzn} and hadron colliders, operating at significantly higher energies than their \lep\ and \lhc\ predecessors; the potential for $b$ physics at each of these is currently under active investigation.

\begin{table}[!htb]
  \caption{
    Summary of some relevant properties for $b$ physics in different experimental environments. Adapted from Ref.~\cite{Gershon:2013aca}.
  }
  \label{tab:gibson}
  \centering
\begin{tabular}{p{3.8cm}p{4.2cm}@{\hspace{3mm}}cc}
  \hline\noalign{\smallskip}
  & $e^+e^- \to \Upsilon(4S) \to B\bar{B}$ & $p\bar{p} \to  b\bar{b}X$ & $pp \to  b\bar{b}X$ \\
  & & ($\sqrt{s} = 2 \tev$) & ($\sqrt{s} = 13 \tev$) \\
  & \multicolumn{1}{c}{PEP-II, KEKB} & Tevatron & LHC \\
\noalign{\smallskip}\hline\noalign{\smallskip}
  Production & \multicolumn{1}{c}{$1 \nb$} & $\sim 100 \mub$ & $\sim 500 \mub$ \\
  \multicolumn{1}{r}{cross-section} \\ [0.2ex]
  Typical $b\bar{b}$ rate & \multicolumn{1}{c}{$10 \hz$} & $\sim 100 \khz$ & $\lsim 1 \mhz$ \\ [0.2ex]
  Pile-up & \multicolumn{1}{c}{0} & 1.7 & 1--40 \\ [0.2ex]
  Trigger efficiency & \multicolumn{1}{c}{$100\,\%$} & \multicolumn{2}{c}{$20$--$80\,\%$} \\ [0.2ex]
  \B\ hadron mixture & $\Bp\Bm\,(\sim 50\,\%)$, & \multicolumn{2}{l}{\Bp (40\,\%), \Bz (40\,\%), \Bs (10\,\%),} \\
  & \multicolumn{1}{r}{$\Bz\Bzb\,(\sim 50\,\%)$} & \multicolumn{2}{r}{$\Lb$ (10\,\%), others ($<1\,\%$)} \\ [0.2ex]
  \B\ hadron boost & \multicolumn{1}{c}{small ($\beta\gamma \sim 0.5$)} & \multicolumn{2}{c}{large ($\beta\gamma \sim 100$)} \\  [0.2ex]
  Underlying event & \multicolumn{1}{c}{$B\Bbar$ pair alone} & \multicolumn{2}{c}{Many additional particles} \\ [0.2ex]
  Production vertex & \multicolumn{1}{c}{Not reconstructed} & \multicolumn{2}{c}{Reconstructed from many tracks} \\ [0.2ex]
  $\B\Bbar$ pair production & Coherent & \multicolumn{2}{c}{Incoherent} \\
  & \multicolumn{1}{r}{(from $\Upsilon(4S)$ decay)} \\ [0.2ex]
  Effective flavour & \multicolumn{1}{c}{$\sim 30\,\%$} & \multicolumn{2}{c}{$\lsim 6\,\%$} \\
  \multicolumn{1}{r}{tagging efficiency} \\
\noalign{\smallskip}\hline\noalign{\smallskip}
\end{tabular}
\end{table}

%% file: tree-dominated.tex
\section{\boldmath \CP violation in tree-dominated transitions}
\label{sec:tree-dominated}

\input{semileptonic}

\input{ccq}

\input{cud}

\input{gamma}

%% file: semileptonic.tex
\subsection{Studies of \CP violation using semileptonic decays}
\label{sec:SL}

Semileptonic decays of \B mesons such as $\Bz \to \Dm \mup \nu_\mu$ can be considered the archetypal tree-dominated transition.  
In the SM there is only a single amplitude;
the decay is therefore expected to be both \CP-conserving and flavour-specific.\footnote{
  Neither of these features has been subjected to rigorous experimental tests~\cite{Brod:2014bfa}.}
Although extensions to the SM, in particular models with charged partners of the Higgs boson, could introduce additional diagrams, these features are not expected to change.
Semileptonic decays are therefore ideal to provide theoretically clean measurements of \CP violation in \Bd or \Bs mixing, parametrised by $\asld$ and $\asls$ respectively.
These quantities can be predicted with good precision in the SM from~\cite{Lenz:2011ti,Artuso:2015swg}
\begin{equation}
  \aslq = {\rm Im}\left(\frac{\Gamma_{12}^q}{M_{12}^q}\right) \equiv \left| \frac{\Gamma_{12}^q}{M_{12}^q} \right| \sin \phi_{12}^q = \frac{\Delta\Gamma_{(\squark)}}{\Delta m_{(\squark)}} \tan \phi_{12}^q \, ,
\end{equation}
where $\Gamma_{12}^q$ and $M_{12}^q$ are the off-diagonal elements of the effective weak Hamiltonian that describes $\Bds$--$\Bdsb$ mixing (and that was diagonalised in Eq.~(\ref{eq:physicalStates}) to obtain the physical states), and $\phi_{12}^q$ is their relative phase.
From these expressions and the measured (or predicted) values of $\Delta\Gamma_{(\squark)}$ and $\Delta m_{(\squark)}$, it can be seen that the SM values of both $\asls$ and $\asld$ are smaller than a permille. 

The experimental signature of \CP violation in mixing is an asymmetry in the yield of mixed decays (\eg\ $\Bz\to\Bzb\to\Dp\mun\neub_\mu$ {\it vs.} $\Bzb\to\Bz\to\Dm\mup\neu_\mu$).
To measure such an asymmetry requires knowledge of detection and production asymmetries, as well as tagging of the initial flavour of the \B meson.
However, since there should be no asymmetry in the unmixed yields, the effect of \CP violation in mixing can also be probed from the asymmetry in the total yields.
This approach is attractive for experiments at hadron colliders, where the comparatively small effective tagging efficiency would lead to an important reduction in sensitivity.
The untagged asymmetry is diluted compared to the mixed asymmetry by a factor given by the inverse of the mixed fraction.
Due to the large value of $\Delta m_s/\Gamma_s$, this factor is effectively 2 in the \Bs system; for the \Bd system it is somewhat larger and depends on the decay time acceptance.
Although the asymmetry is independent of the \B decay time, analysis of the decay time distribution can help to separate signal from background, and to isolate the asymmetry in mixing from production or detection related effects.

A powerful method to obtain high yields with reasonable tagging efficiency is to reconstruct same-sign lepton pairs.  
If both leptons originate from semileptonic \B decays, this signature can only arise when a neutral \B meson has oscillated, and therefore the asymmetry between positive and negative same-sign lepton pairs depends on \CP violation in mixing.  
In the $e^+e^- \to \Upsilon(4S) \to \B\Bbar$ environment, only $\asld$ can contribute, while for higher energy $\epem$, $p\bar{p}$ or $pp$ collisions the inclusive dilepton asymmetry is given by 
\begin{equation}
  \label{eq:absl}
  \absl = C_{d}\;\asld + C_{s}\;\asls \, ,
\end{equation}
where $C_{d}$ and $C_{s}$ depend on the relative production rates of \Bd and \Bs mesons, as well as their respective probabilities to have mixed (which may depend on the selection requirements).
A possible additional term in Eq.~(\ref{eq:absl}) is discussed below.

Another approach to determine these asymmetries inclusively is to tag \B particles produced in top quark decays~\cite{Gedalia:2012sx}.
This method, recently implemented by ATLAS~\cite{Aaboud:2016bmk}, however results in low yields so that the measurements do not currently have competitive precision.

\begin{table}[!htb]
  \caption{
    Summary of the latest results for the \Bd mixing (\asld) and \Bs mixing (\asls) \CP asymmetries, as well as the inclusive dimuon asymmetry \absl measured at \dzero. 
    In all cases the statistical uncertainty is quoted first and the systematic second. 
    All values are percentages. 
    The world averages~\cite{HFAG} are from a fit to all $\asld$, $\asls$ and $\absl$ results, except for the latest LHCb \asls\ result~\cite{LHCb-PAPER-2016-013}; an earlier result~\cite{LHCb-PAPER-2013-033} is included instead.
    The latest SM predictions~\cite{Lenz:2011ti,Artuso:2015swg} are given for comparison.
  }
  \label{tab:semilepcpv}
  \centering
\resizebox{\textwidth}{!}{
\begin{tabular}{lccc}
  \hline
  & \asld~(\%) & \asls~(\%) & \absl~(\%) \\
  \hline
  \babar $K$-tag~\cite{Lees:2013sua,Lees:2015rka} & $0.06 \pm 0.17\,^{+0.38}_{-0.32}$ & -- & --  \\
  \babar $\lepton\lepton$~\cite{Lees:2014qma} & $-0.39 \pm 0.35 \pm 0.19$ & -- & -- \\
  \belle $\lepton\lepton$~\cite{Nakano:2005jb} & $-0.11 \pm 0.79 \pm 0.70$& -- & --  \\
  \lhcb~\cite{LHCb-PAPER-2014-053,LHCb-PAPER-2016-013}  & $-0.02 \pm 0.19 \pm 0.30$ & $\phantom{-}0.39 \pm 0.26 \pm 0.20$ & --  \\
  \dzero~\cite{Abazov:2012hha,Abazov:2012zz,Abazov:2013uma} & $\phantom{-}0.68 \pm 0.45 \pm 0.14$ &$-1.12 \pm 0.74 \pm 0.17$ & $-0.496 \pm 0.153 \pm 0.072$  \\
  \hline
  World average~\cite{HFAG} & $-0.15 \pm 0.17$ & $-0.75 \pm 0.41$ \\
  \hline
  SM & $-0.00047 \pm 0.00006$ & $\phantom{-}0.0000222 \pm 0.0000027$ & $-0.023 \pm 0.004$ \\
  \hline
\end{tabular}
}
\end{table}

\begin{figure}[!htb]
  \centering
  \includegraphics[width=0.60\textwidth]{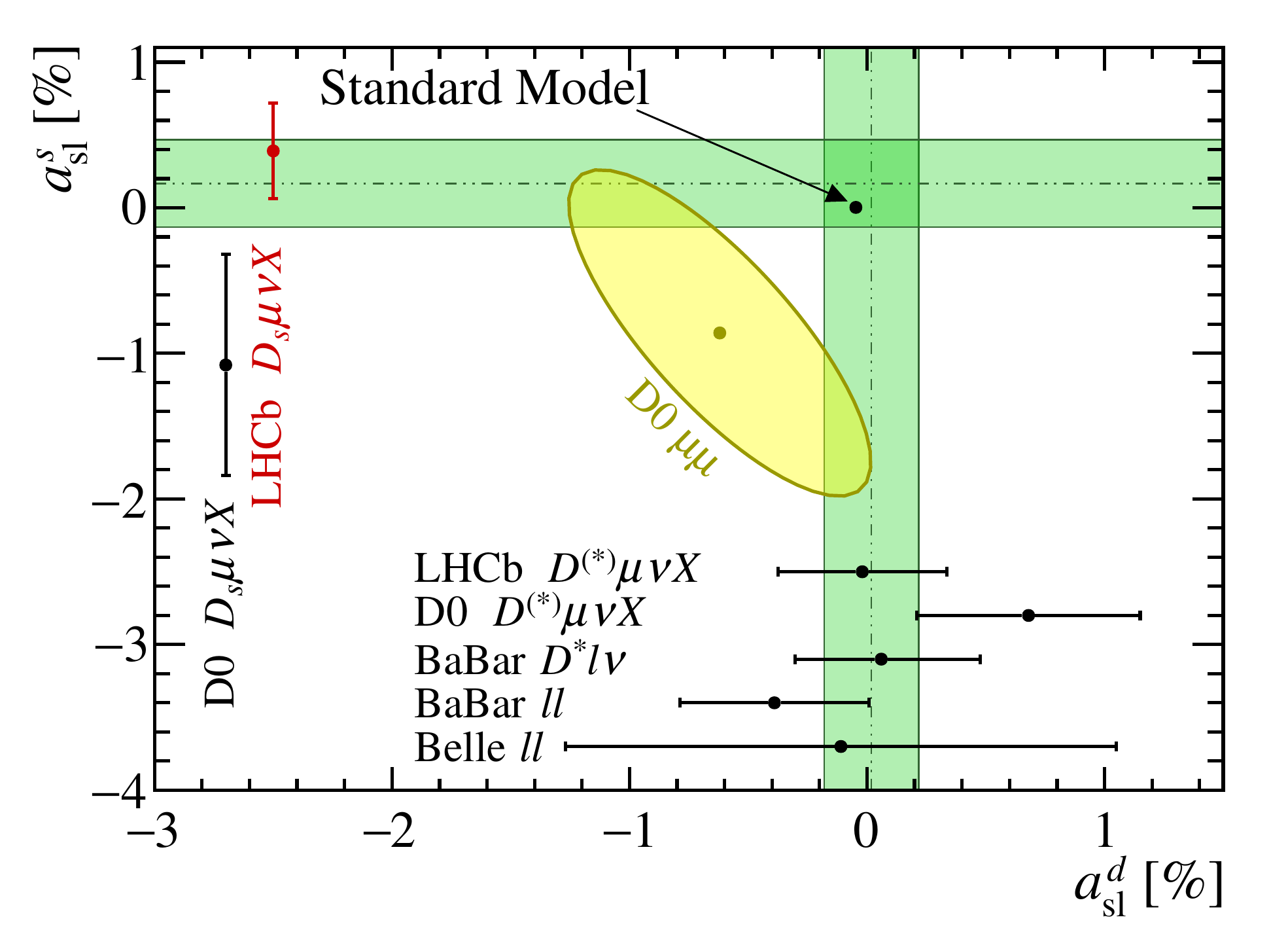}
  \caption{
    Measurements of \asls\ and \asld, with simple one-dimensional averages (that differ from the values shown in Table~\ref{tab:semilepcpv}) shown as horizontal and vertical bands, respectively~\cite{LHCb-PAPER-2016-013}.
    The yellow ellipse represents the D0 inclusive dimuon measurement~\cite{Abazov:2013uma} with $\Delta \Gamma_d$ set to its SM expectation value. 
  } 
\label{fig:asls_asld}
\end{figure} 

Measurements of $\asld$, $\asls$ and $\absl$ have been performed by the \babar, \belle, \lhcb, and \dzero collaborations.
The latest results are collected in Table~\ref{tab:semilepcpv} together with the world averages and SM predictions, which are also shown in Fig.~\ref{fig:asls_asld}.
\belle have measured $\asld$ using the inclusive dilepton approach~\cite{Nakano:2005jb}.
The data analysed correspond to approximately one-tenth of the final \belle $\Upsilon(4S)$ sample, and therefore an update of this analysis would be well motivated.
\babar also have a measurement with an inclusive dilepton sample~\cite{Lees:2014qma} and in addition have a measurement in which the signal $\Bd \to \Dstarm X\lepton^+\neul$ decay is partially reconstructed and a charged kaon reconstructed
on the opposite side of the event ($K$-tag) determines whether the \B meson has mixed or not~\cite{Lees:2013sua,Lees:2015rka}, allowing the determination of detection asymmetries as well as $\asld$.
The measurements of $\asld$ from \lhcb~\cite{LHCb-PAPER-2014-053} and \dzero~\cite{Abazov:2012hha} are obtained from untagged samples of $\Bd \to \DorDstarm X\mu^+\neum$ decays.
Measurements of $\asls$ from \lhcb~\cite{LHCb-PAPER-2016-013} and \dzero~\cite{Abazov:2012zz} are obtained from untagged samples of $\Bs \to \Dsm X\mu^+\neum$ decays.
Due mainly to the common method to determine the muon detection asymmetry, there is a correlation of $+0.13$ between the LHCb \asls\ and \asld\ results.

The determination of the inclusive dimuon asymmetry $\absl$ by \dzero~\cite{Abazov:2013uma} has attracted a great deal of attention due to the deviation, estimated to be $2.8\sigma$, from the SM prediction.
Since the probabilities for \Bd and \Bs mesons to have mixed depend on the selection criteria, \dzero have performed the measurement in bins of the distances of closest approach of the muons to the primary $p\bar{p}$ interaction.
This provides separation between regions where $\absl$ is dominated by \asld and those where it is dominated by \asls. 
There is no significant difference between the values of \absl measured in the different regions, however the consistency of the binned data with the SM prediction gives a more significant deviation of $3.6\sigma$.
It has been noted~\cite{Artuso:2015swg,Borissov:2013wwa} that there is another possible contribution to \absl, in addition to those given in Eq.~(\ref{eq:absl}).
This contribution arises from \CP violation in interference between mixing and decay and is expected to be negligible in the SM due to the small value of $\Delta \Gamma_d$.
Current constraints, both direct and indirect, on $\Delta \Gamma_d$ do not, however, exclude a significant enhancement~\cite{Gershon:2010wx,Bobeth:2014rda}, and therefore improved determinations are needed.
A similar possible contribution due to interference between mixing and decay in \Bs\ oscillations is negligible due to the small value of $2\beta_s$.
Thus, all measurements of the possible contributions to the inclusive dimuon asymmetry are currently consistent with zero, and so the inclusion of the D0 result causes some tension in the global fit.

It should be noted that many of the results given in Table~\ref{tab:semilepcpv} are dominated by systematic uncertainties. 
Although some of these, for example due to uncertainty in lepton misidentication probabilities, can be expected to scale with increased data samples, it may prove hard to reduce the uncertainties below the permille level.  
The systematic uncertainties in results from \dzero tend to be smaller than those from other experiments, as symmetries in the initial state and in the detector configuration can be exploited to reduce the sizes of possible biases in the measurement.
There is no immediate prospect of larger data samples becoming available in such a symmetric environment, although the possibility of this type of measurement at a future high luminosity $Z$ factory should be considered.

%% file: ccq.tex
\subsection{Measurement of $\beta$ and $\beta_s$ using $b \to c\bar{c}q$ transitions}
\label{sec:ccq}

The $b \to c\bar{c}q$ transitions are characterised by a dominant $b\to c$ tree diagram, as well as subleading $b\to q$ penguin diagrams in which the $c\bar{c}$ quark pair is emitted from the loop. 
In the limit of negligible penguin contribution in a decay to a \CP eigenstate $f$,
the parameter $\lambda_f$ has unit magnitude and phase determined by the relevant CKM matrix elements involved in the mixing and decay amplitudes.
Therefore effects due to \CP violation in the interference of mixing and decay can be measured from the $S_f$ parameter of Eq.~(\ref{eq:ACS-defs}) giving theoretically clean determinations of $\beta$ and $\beta_s$ in the $\Bd$ and $\Bs$ systems, respectively:
\begin{eqnarray}
  S_f(\Bd\to\jpsi\KS) & = & -\eta_{\CP} \sin(2\beta) \, , \label{eq:sin2b} \\
  S_f(\Bs\to\jpsi\phi) & = & +\eta_{\CP} \sin(2\beta_s) \, . \label{eq:sin2bs}
\end{eqnarray}
In the above, $\eta_{\CP}$ is the \CP eigenvalue of the final state, which is $-1$ for $\jpsi\KS$ and depends on the transversity amplitude in the $\Bs\to\jpsi\phi$ decay (these and other decay channels are discussed in more detail in Secs.~\ref{sec:sin2b} and~\ref{sec:psiphiandbuddies} below).\footnote{
  Here, and throughout the review, the symbols $\rho$, $\omega$, $K^*$ and $\phi$ refer to the lightest vector meson of the corresponding family: $\rho(770)$, $\omega(782)$, $K^*(892)$ and $\phi(1020)$.
  Only a few measurements have been performed with final states involving higher excitations; these are not discussed for reasons of brevity.  
}
Note that the change of sign between Eq.~(\ref{eq:sin2b}) and Eq.~(\ref{eq:sin2bs}) arises due to the conventional definition of Eq.~(\ref{eq:angles-betas}) that makes the SM expectation for $\beta_s$ positive.

The expected level of suppression of the penguin/tree ratio for $b \to c\bar{c}s$ transitions is ${\cal O}(\lambda^2) \times f_{\rm loop}$, where $\lambda = \sin \theta_C \approx 0.23$ is the Wolfenstein parameter~\cite{Wolfenstein:1983yz} ($\theta_C$ is the Cabibbo angle~\cite{Cabibbo:1963yz}) and $f_{\rm loop}$ is the loop suppression factor.
Since there is no reliable first principles calculation of the size of this factor, it is of great interest to study also $b \to c\bar{c}d$ transitions, where the penguin/tree ratio is not CKM-suppressed.
Several methods have been proposed that use flavour symmetries to relate decay modes mediated by the two sets of quark-level transitions, and thereby to constrain the possible amount of ``penguin pollution'' in the determination of $\beta$ and $\beta_s$, as discussed in Sec.~\ref{sec:ccd}.
Explicit calculations suggest that the effects of ``penguin pollution'' on the determinations of $2\beta$ and $2\beta_s$ are $\lsim 1\degrees$~\cite{Jung:2012mp,DeBruyn:2014oga,Frings:2015eva}.

\subsubsection{Measurements of $\beta$} 
\label{sec:sin2b}

The determination of $\sin(2\beta)$ from $\Bd \to \jpsi \KS$~\cite{Carter:1980tk,Bigi:1981qs} has long been considered a ``golden mode'' of \CP violation in the \B system.
The experimental challenge is to measure the coefficient of the sinusoidal oscillation of the decay-time asymmetry of Eq.~(\ref{eg:tdcpv-asym}).
This motivated the design of the asymmetric $\epem$ \B factory experiments, \babar and \belle, in which the boost of the produced \B mesons in the laboratory frame results in a separation of their decay vertices.
Due to the quantum correlations of the $B$ mesons produced in $\Upsilon(4S)$ decay, the decay of one into a final state that tags its flavour ($\Bz$ or $\Bzb$) can be used to specify the flavour of the other at that instant.

\begin{figure}[!htb]
\centering
\includegraphics[width=0.45\textwidth]{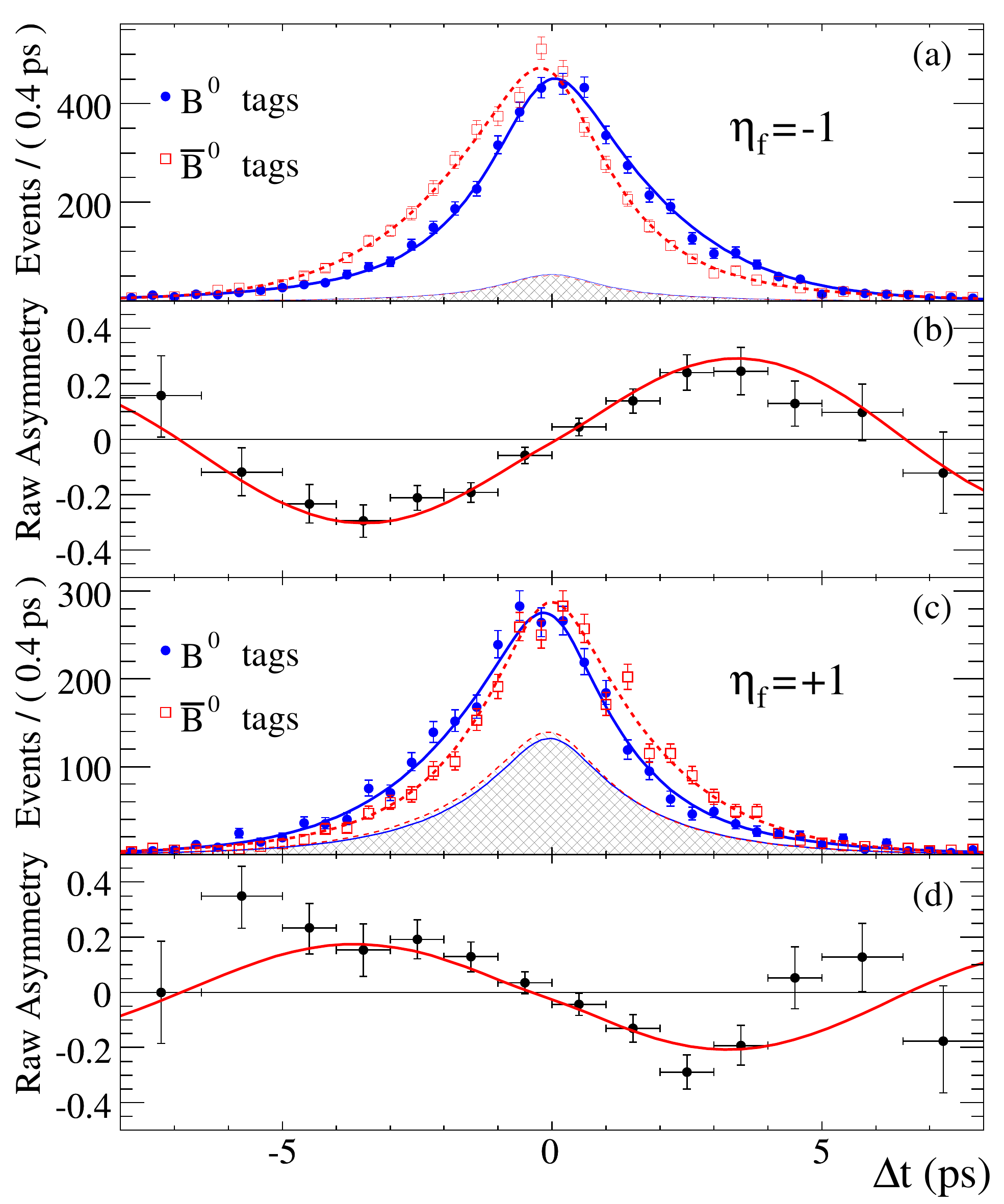}
\hspace{3mm}
\includegraphics[width=0.30\textwidth]{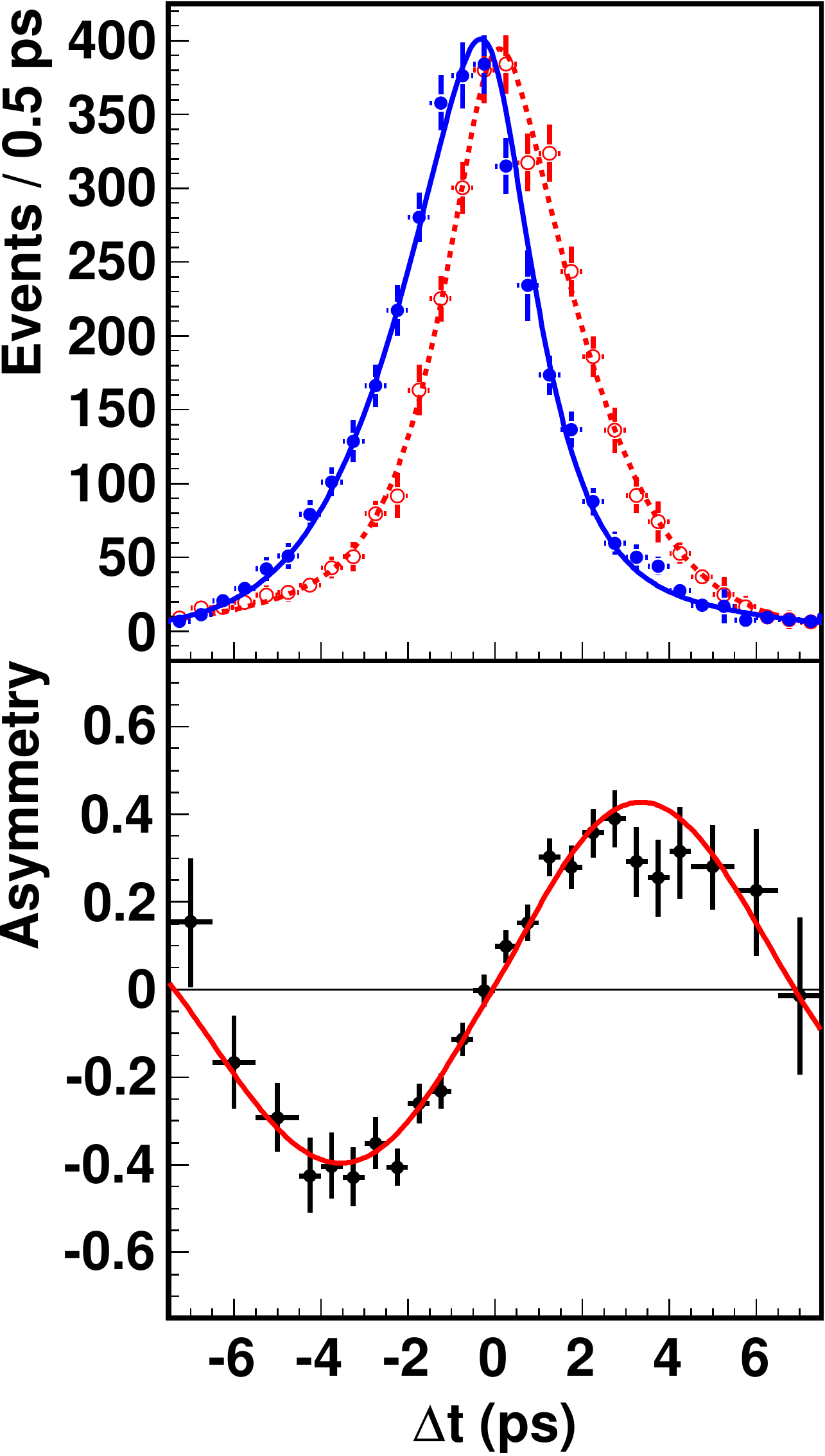}
\caption{\small
  Results from (left) \babar~\cite{Aubert:2009aw} and (right) \belle~\cite{Adachi:2012et} on the determination of $\sin2\beta$.  
  The \babar data are separated by the \CP-eigenvalue of the final state, while only \CP-odd modes from \belle are shown here.
}
\label{fig:sin2beta}
\end{figure}

Through the measurement of $\sin 2\beta$, \babar~\cite{Aubert:2001nu} and \belle~\cite{Abe:2001xe} were able to make the first observations of \CP violation outside the kaon sector, thus validating the Kobayashi-Maskawa mechanism.
The results based on the final data samples of \babar~\cite{Aubert:2009aw} and \belle~\cite{Adachi:2012et}, shown in Fig.~\ref{fig:sin2beta}, clearly show the large \CP violation effect.
These analyses include not only $\Bd \to \jpsi \KS$, but also decays to the \CP-odd final states $\psi(2S)\KS$, $\eta_c\KS$ and $\chi_{c1}\KS$ as well as the \CP-even final state $\jpsi\KL$.
As well as these, results have also been published on $\Bd\to\chi_{c0}\KS$~\cite{Aubert:2009me} and $\jpsi\KS\piz$~\cite{Aubert:2004cp,Itoh:2005ks} decays.
The latter are particularly interesting since the interference between the $K^{*0}$ resonance and the $K\pi$ S-wave can be used to measure $\cos(2\beta)$ and hence resolve an ambiguity in the solution for $\beta$ if only $\sin(2\beta)$ is known.
The results prefer the SM solution, but the precision is not sufficient to completely rule out the ambiguity.
Updates of the $\Bz\to\jpsi\KS\piz$ analysis with the full \babar and \belle statistics may be able to resolve the solutions definitively.

Results from LHCb also provide competitive precision on $\sin(2\beta)$ from $\Bd\to\jpsi\KS$.
The latest LHCb result is compared to those from \babar and \belle in Table~\ref{tab:sin2beta}.
For all experiments, the results are still statistically limited.
Moreover, the largest sources of systematic uncertainty differ between the $\epem$ and $pp$ collision environments. 
Among important sources of uncertainty for \babar and \belle are understanding of the vertexing and decay time resolution.  
For \lhcb, however, the dominant source is due to possible tagging asymmetries.
Since these effects are expected to scale with statistics, to some extent, there are good prospects for considerable further reduction in the uncertainty with larger data samples.

\begin{table}[!htb]
  \caption{
    Latest experimental results on $\sin(2\beta)$. 
    The first uncertainties are statistical and the second systematic.
 }
  \label{tab:sin2beta}
  \centering
\begin{tabular}{lcc}
  \hline
  & From $\jpsi\KS$ only & From all $c\bar{c}\KS$ \\
  \hline
  \babar~\cite{Aubert:2009aw} & $0.657 \pm 0.036 \pm 0.012$ & $0.687 \pm 0.028 \pm 0.012$ \\
  \belle~\cite{Adachi:2012et} & $0.670 \pm 0.029 \pm 0.013$ & $0.667 \pm 0.023 \pm 0.012$ \\
  \lhcb~\cite{LHCb-PAPER-2015-004} & $0.731 \pm 0.035 \pm 0.020$ & \\
  \hline
  World average~\cite{HFAG} & \multicolumn{2}{c}{$0.691 \pm 0.017$} \\
  \hline
\end{tabular}
\end{table}

The world average value, using determinations based on $b \to c\bar{c}s$ transitions, is~\cite{HFAG}
\begin{equation}
  \sin 2 \beta = 0.682 \pm 0.019 \ {\rm which \ gives} \ \beta = (21.5 \,^{+0.8}_{-0.7})^\circ \, .
\end{equation}
Here only the solution consistent with the SM is given.
The ambiguous value at $\pi/2 - \beta$ is disfavoured by several measurements (as discussed in this review); there is a further ambiguity at $\pi + \beta$ which cannot be resolved through any measurement of $2\beta$.

In addition to the well-studied charmonium-kaon modes discussed above, it is also possible to determine $\beta$ using $\Bd \to \DorDstarp\DorDstarm\KS$ decays.
These have been been proposed as providing potential to resolve ambiguities in the determination of $\beta$, since quantities proportional to both $\sin(2\beta)$ and $\cos(2\beta)$ can be measured~\cite{Browder:1999ng}.
Measurements have been performed by both \babar~\cite{Aubert:2006fh} and \belle~\cite{Dalseno:2007hx}, but are not yet sufficiently precise to resolve the sign of $\cos(2\beta)$.

\subsubsection{Measurements of $\beta_s$}
\label{sec:psiphiandbuddies}

Three $b \to c\bar{c}s$ transitions have been used to measure $\beta_s$ to date: $\Bs\to\jpsi K^+K^-$, $\Bs\to\jpsi \pi^+\pi^-$, and $\Bs\to\Dsp\Dsm$.
The latest results are collected in Table~\ref{tab:btoccbarqcpv}.\footnote{
  A recent result from LHCb with $\Bs\to\psitwos\phi$ decays~\cite{LHCb-PAPER-2016-027} is not included as it has worse precision compared to the other LHC results.
}
Note that the results are presented in terms of the experimentally observable \CP-violating phase \phis, which is equal to $-2\betas$ if the penguin contributions to these decays are small. 
All measurements agree well with each other and with the SM prediction~\cite{Charles:2011va} of $-2\betas = -0.0363 \pm 0.0013 \rad$, and give a world average~\cite{HFAG} of
\begin{equation}
  -2\beta_s = -0.034 \pm 0.033 \rad \, .
\end{equation}

\begin{table}[!htb]
  \caption{
    Latest experimental results on \CP violation in $\Bs\to\jpsi K^+K^-$, $\Bs\to\jpsi \pi^+\pi^-$ and $\Bs\to\Dsp\Dsm$ decays.
    Results are quoted in terms of $\phis$, in units of radians, which is equal to $-2\betas$ in the limit of vanishing penguin contributions.
    For the \cdf result the quoted interval combines statistical and systematic uncertainties at the $68\,\%$ confidence level, while \dzero report only the total uncertainty. 
    For all other results, the first uncertainty is statistical and the second systematic.
    The known sign of $\Delta\Gamma_s$~\cite{LHCb-PAPER-2011-028} has been used to break the ambiguity in the reported solutions. 
 }
  \label{tab:btoccbarqcpv}
  \centering
\begin{tabular}{lccc}
  \hline
  & $\jpsi K^+K^-$ & $\jpsi \pi^+\pi^-$ & $\Dsp\Dsm$ \\
  \hline
  \cdf~\cite{Aaltonen:2012ie} & $[-0.06,0.30]$ &  -- & -- \\ 
  \dzero~\cite{Abazov:2011ry} & $-0.55\,^{+0.38}_{-0.36}$ &  -- & -- \\ 
  \atlas~\cite{Aad:2016tdj} & $-0.098 \pm 0.084 \pm 0.040$ &  -- & -- \\ 
  \cms~\cite{Khachatryan:2015nza} & $-0.075 \pm 0.097 \pm 0.031$ &  -- & -- \\ 
  \lhcb~\cite{LHCb-PAPER-2014-059,LHCb-PAPER-2014-019,LHCb-PAPER-2014-051} & $-0.058 \pm 0.049 \pm 0.006$   & $0.070 \pm 0.068 \pm 0.008$   & $0.02 \pm 0.17 \pm 0.02$ \\ 
  \hline
\end{tabular}
\end{table}

\begin{figure}[!htb]
  \centering
  \includegraphics[width=0.95\textwidth,bb=80 620 545 755,clip=true]{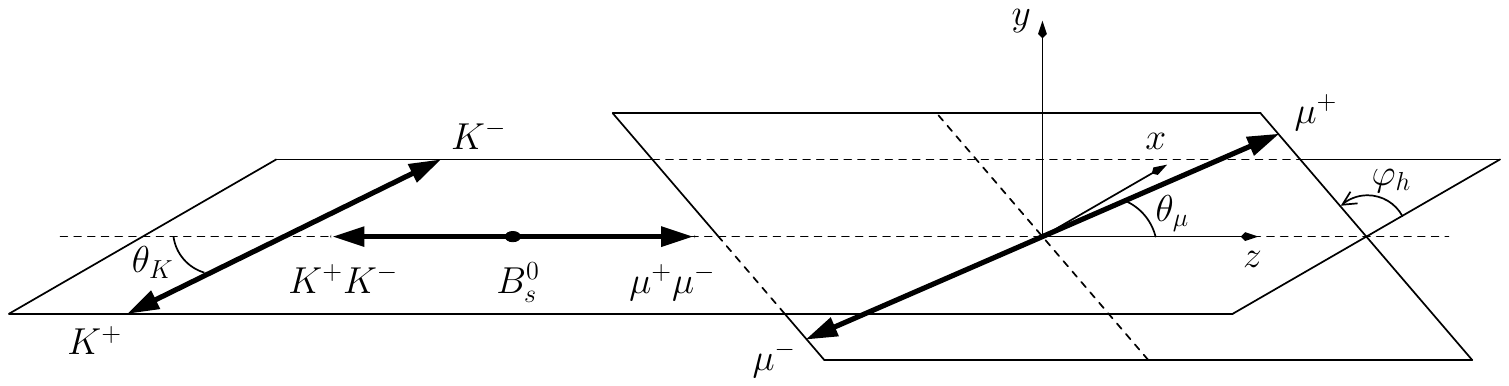}
  \caption{
    Angular (helicity) basis of the $\Bs\to\jpsi K^+K^-$, $\jpsi \to\mumu$ transition, reproduced from Ref.~\cite{LHCb-PAPER-2013-002}. 
\label{fig:psiphihel}
} 
\end{figure}

\begin{figure}[!htb]
  \centering
  \includegraphics[width=0.45\textwidth]{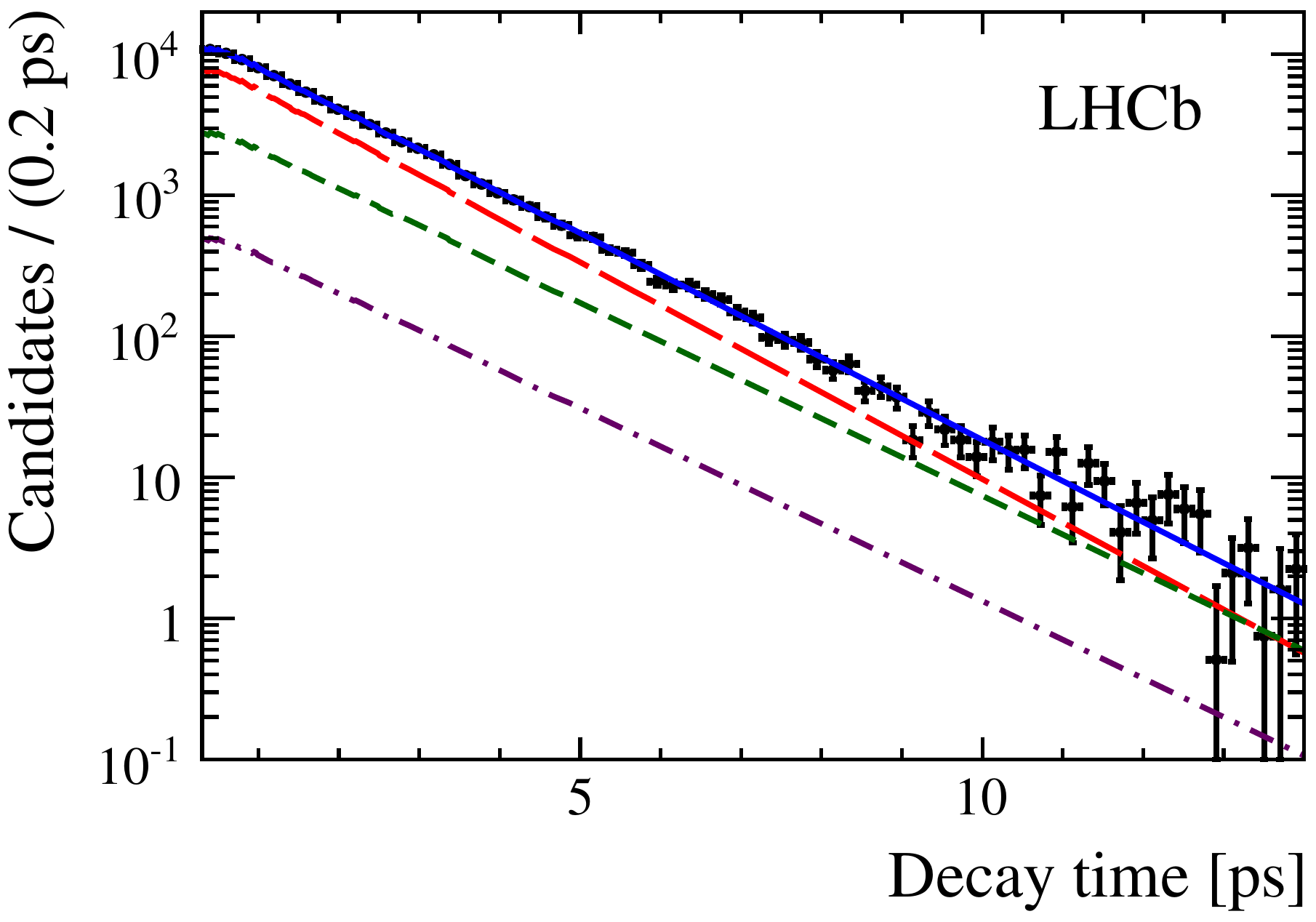}
  \includegraphics[width=0.45\textwidth]{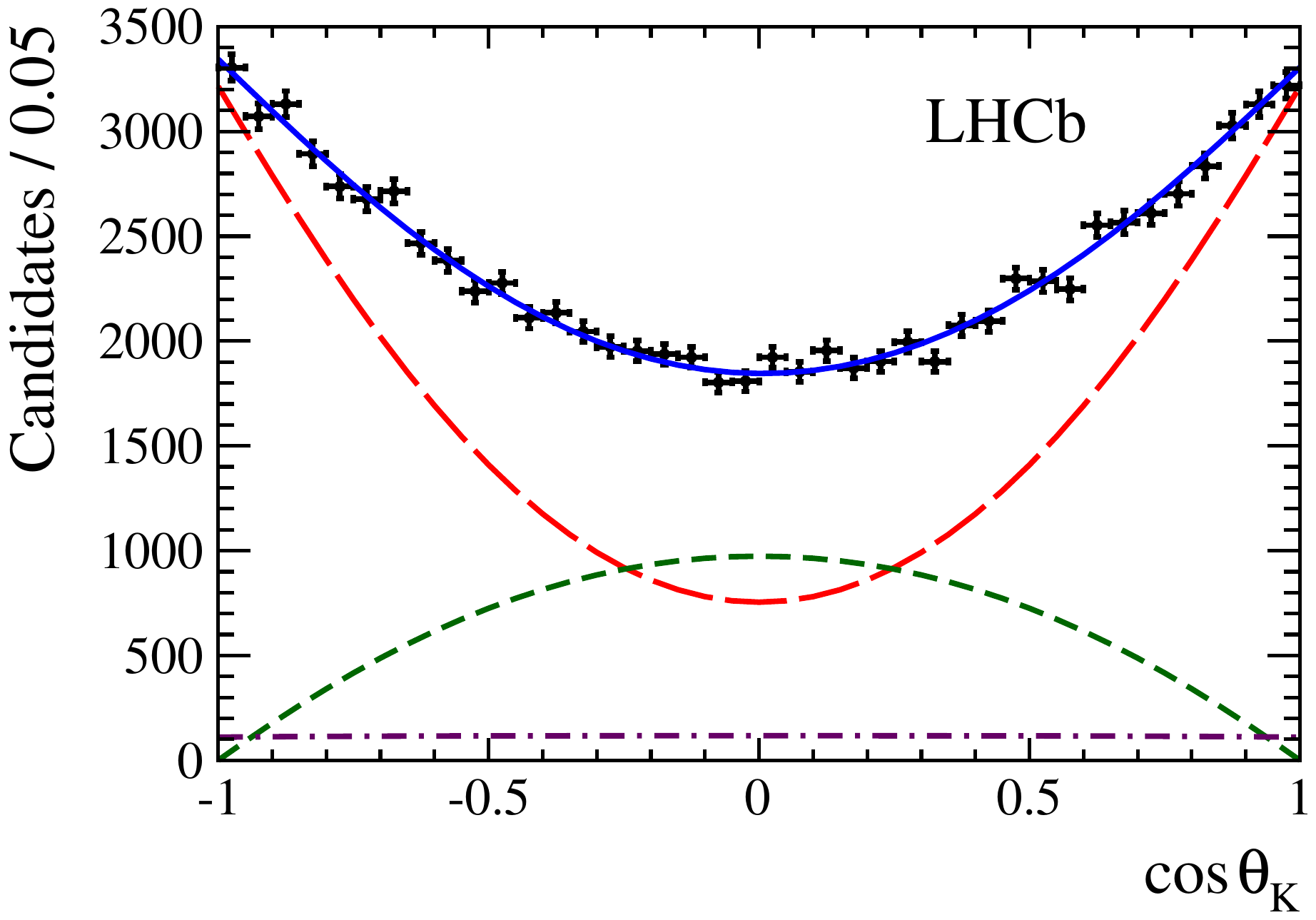}
  \includegraphics[width=0.45\textwidth]{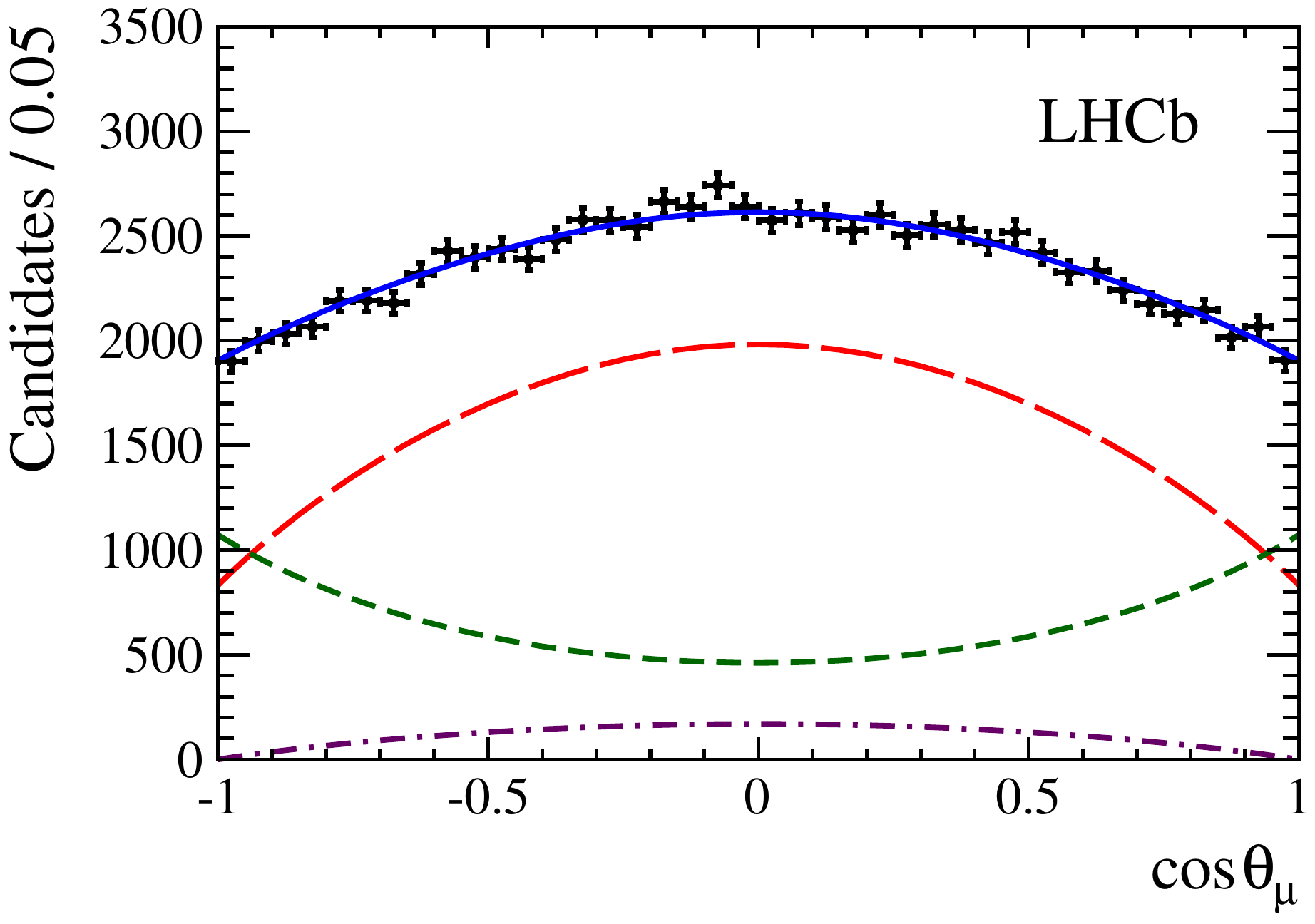}
  \includegraphics[width=0.45\textwidth]{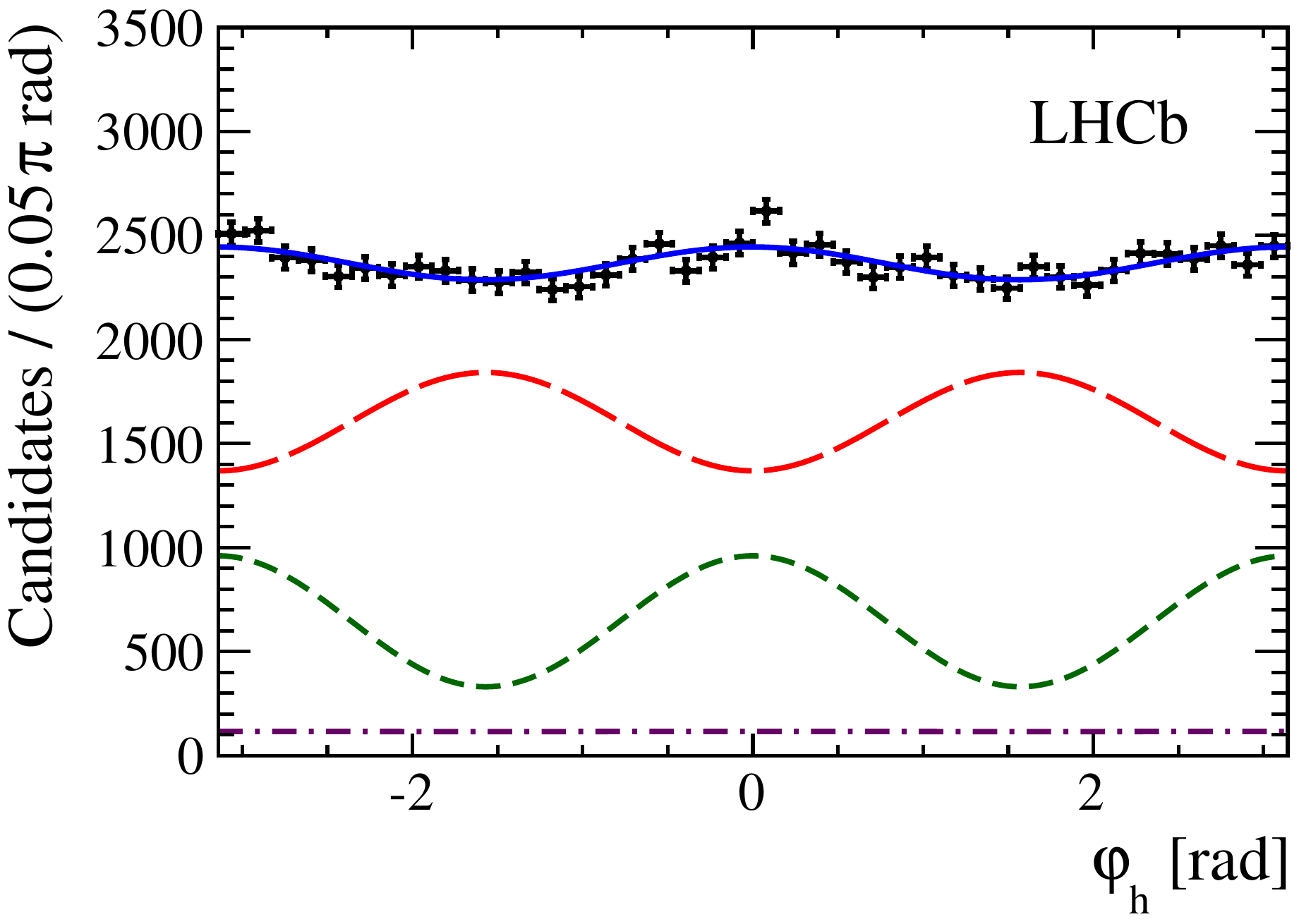}
\caption{
  Background-subtracted decay-time and decay-angle distributions for $\Bs \to \jpsi\Kp\Km$ decays (data points) in LHCb~\cite{LHCb-PAPER-2014-059}, with the one-dimensional fit projections overlaid. 
  The solid blue line shows the total signal contribution, which is composed of \CP-even (long-dashed red), \CP-odd (short-dashed green) and S-wave (dotted-dashed purple) contributions.
}
\label{fig:LHCb-PAPER-2014-059}
\end{figure}

The most widely used decay mode for the determination of $\beta_s$ is $\Bs \to \jpsi\Kp\Km$, which in the low $m(\Kp\Km)$ region is dominated by the $\phi$ resonance.
As $\Bs \to \jpsi\phi$ is a pseudoscalar to vector-vector transition, it contains a mixture of \CP-even and \CP-odd amplitudes, due to the different possible polarisations of the final state, which must be accounted for in the fit.
These terms are disentangled by performing a simultaneous fit to the decay-time and decay-angle distributions of the signal, where the relevant angles are shown in Fig.~\ref{fig:psiphihel} in the so-called ``helicity basis'' used in the \lhcb measurement~\cite{LHCb-PAPER-2014-059}. 
A further complication arises due to interference between the $\phi$ resonance and a broad S-wave $\Kp\Km$ component, which must be accounted for in the fit. 
However, these features can be turned to the benefit of the analysis, providing better sensitivity and allowing to resolve an ambiguity in the results.
For this reason, results are presented in terms of $\phi_s$ rather than $\sin(\phi_s)$ and $\cos(\phi_s)$ or the $C_f$, $S_f$ and $A_{f}^{\Delta \Gamma}$ quantities introduced in Sec.~\ref{sec:rates}. 
The separation of the different components in the simultaneous decay-time and decay-angle fit in the LHCb analysis is shown in Fig.~\ref{fig:LHCb-PAPER-2014-059}.
It should be noted that in principle there could be different weak phase differences in each of the polarisation amplitudes; to date, only LHCb has taken this into account in the analysis~\cite{LHCb-PAPER-2014-059}.

All current measurements are statistically limited.
The dominant systematic uncertainty for the most precise measurements is the knowledge of the variation of the selection efficiency with the decay angles.
This is taken from simulation samples reweighted to match the data distributions, and therefore significant further reduction in the uncertainty is expected to be possible with larger data samples.
Moreover, all results until now use only the low $m(\Kp\Km)$ region of $\Bs \to \jpsi\Kp\Km$ decays, and further sensitivity to $\phi_s$ can be achieved from contributions such as $\Bs \to \jpsi f_2^\prime(1525)$ at higher mass~\cite{LHCb-PAPER-2012-040}.

\begin{figure}[!htb]
  \centering
  \includegraphics[width=0.65\textwidth]{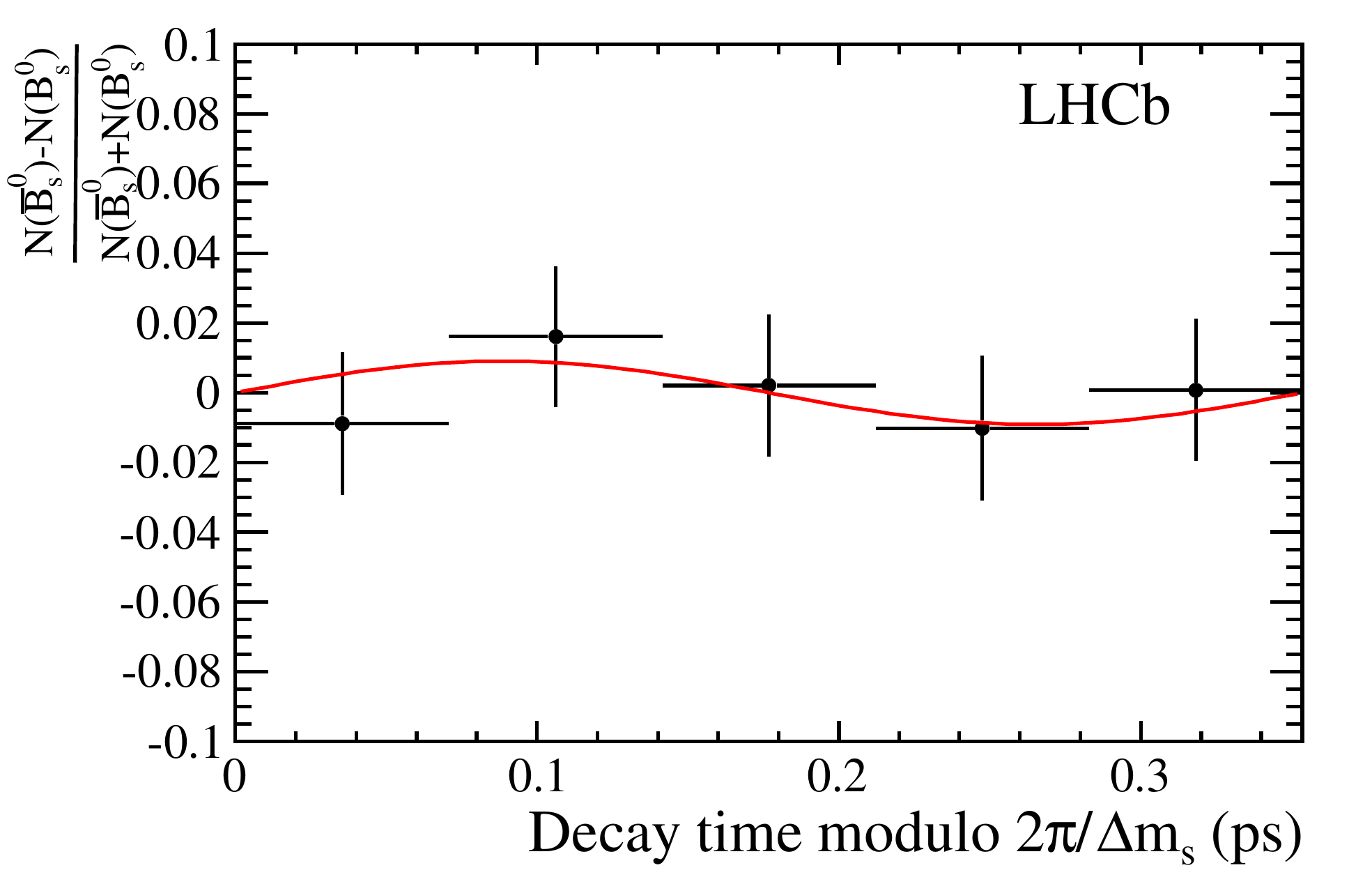}
\caption{
  Decay-time asymmetry for $\Bs \to \jpsi\pip\pim$ decays (data points) in LHCb~\cite{LHCb-PAPER-2014-019}, with the fit projection overlaid. 
  The data are folded into one oscillation period.
}
\label{fig:LHCb-PAPER-2014-019}
\end{figure}

\CP violation in the decay $\Bs\to\jpsi \pi^+\pi^-$ has to date only been studied by \lhcb~\cite{LHCb-PAPER-2014-019}. 
The decay proceeds predominantly through the $f_{0}(980)$ resonance, and has been shown to be an almost pure \CP-odd eigenstate~\cite{LHCb-PAPER-2013-069}.
In that limit, angular analysis of the final state would not be required; nevertheless, because the precision is quite similar to that of $\Bs\to\jpsi K^+K^-$, a decay-time-dependent amplitude analysis has been used to model both the dominant \CP-odd and subleading \CP-even components. 
The dominance of one \CP eigenvalue allows the decay-time-dependent asymmetry to be visualised, as shown in Fig.~\ref{fig:LHCb-PAPER-2014-019}.
The result is statistically limited, with the dominant systematic uncertainty coming from the precision of the amplitude model.

\CP violation in the decay $\Bs\to\Dsp\Dsm$ has also been studied only by LHCb~\cite{LHCb-PAPER-2014-051}. 
In this case the final state consists of two pseudoscalar mesons and there is no ambiguity about its \CP content and consequently no need for any angular analysis. 
However, since there are no muons in the final state, requirements imposed in the trigger (in particular, on the impact parameter with respect to the primary $pp$ collision vertex) tend to bias the decay-time distribution.
Precise knowledge of this effect is necessary, and causes the largest contribution to the systematic uncertainty, which nonetheless is negligible compared to the current statistical precision. 

It is worth noting that in the $\Bs\to\jpsi K^+K^-$ mode, all of \Gs, \DGs, \dms, and \phis can be measured simultaneously, together with parameters that describe the relative magnitudes and phases of the different polarisation amplitudes.
Indeed, this channel gives the most precise determination of $\Gs$ and $\DGs$, and also gives good precision on $\dms$ (though not competitive with the result based on $\Bs\to\Dsm\pip$ decays~\cite{LHCb-PAPER-2013-006}).
In some of the experimental analyses, however, these parameters are fixed in order to simplify the analysis. 
For the \CP-eigenstate modes, only a specific combination of $\Gs$ and $\DGs$ (corresponding to the effective lifetime; see Eq.~(\ref{eq:tau-eff})) can be determined, and it is common to fix these parameters, as well as $\dms$, in order to reduce the need for precise understanding of the variation of the efficiency with decay time.

\subsubsection{Investigation of penguin contributions with $b \to c\bar{c}d$ transitions} 
\label{sec:ccd}

As experimental precision improves, it becomes increasingly important to quantify the contribution of penguin diagrams to $b \to c\bar{c}s$ modes, in order to interpret the results in terms of $\beta$ and $\beta_s$ with minimal theoretical uncertainty. 
One consequence of a significant penguin contribution could be a non-zero parameter of \CP violation in decay in $b \to c\bar{c}s$ transitions.
Results for $C_f$ in $\Bd$ decay modes are consistent with zero within a few percent uncertainty~\cite{HFAG}; 
results in $\Bs$ decays are also consistent with no \CP violation in decay.
The most precise limit on \CP violation in this type of transition is however~\cite{PDG2016,Abazov:2013sqa,Sakai:2010ch}
\begin{equation}
  {\cal A}_{\CP}(\Bp\to\jpsi\Kp) = 0.003 \pm 0.006 \, .
\end{equation}
While these results indicate that the penguin contribution is likely to be small, it is possible that effects of \CP violation in decay are suppressed by small strong phase differences.
Thus, additional information is needed to estimate the size of the penguin effect on $\beta$ and $\beta_s$.

\begin{figure}[!htb]
  \centering
  \includegraphics[width=0.60\textwidth]{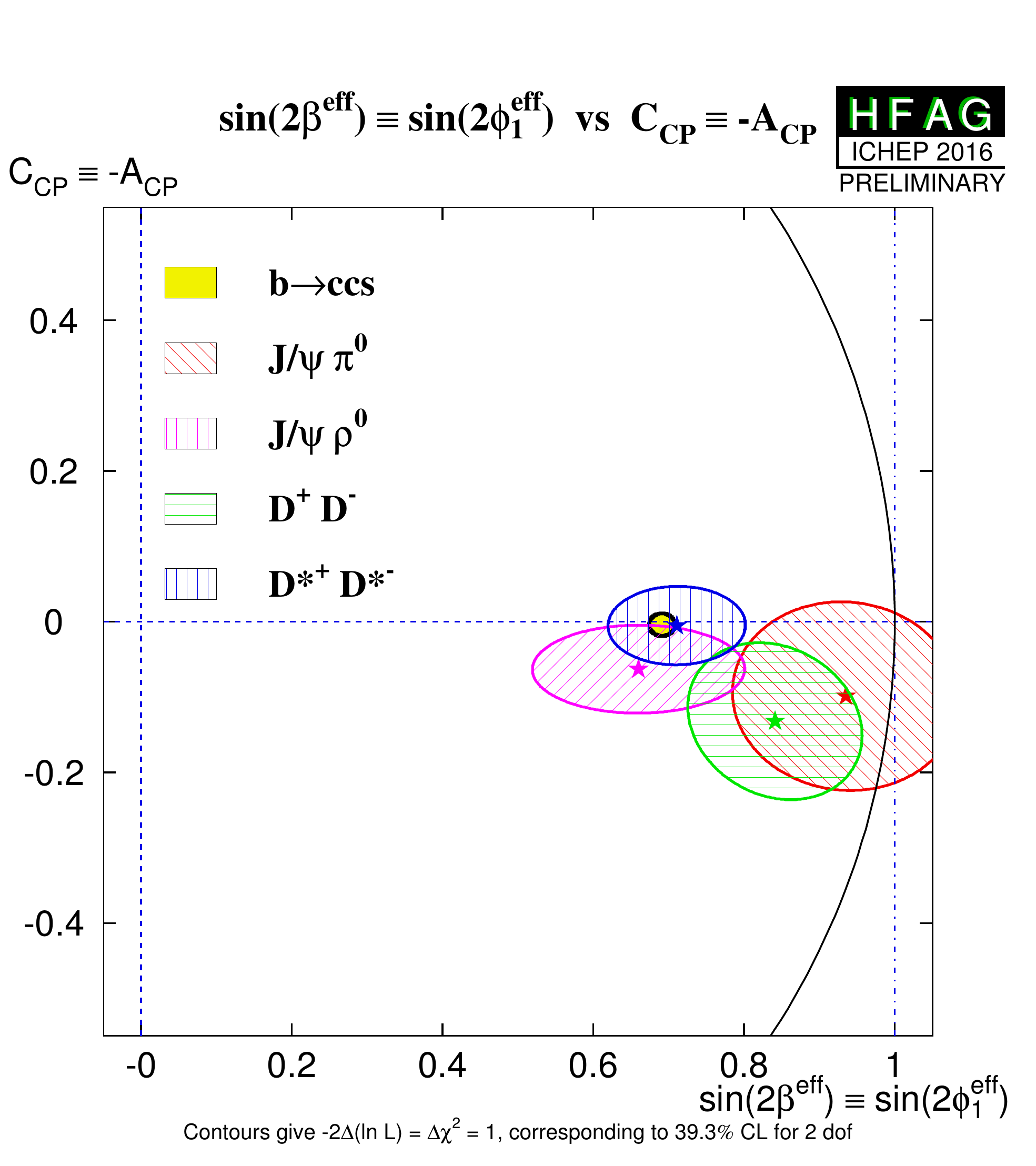}
\caption{
  Summary of measurements of $-\eta_f S_f$ and $C_f$ in $\Bd$ decays dominated by the $b \to c\bar{c}d$ transition~\cite{HFAG}.
  The yellow point shows the reference point ($-\eta_f S_f = \sin(2\beta)$) determined from $b \to c\bar{c}s$ transitions.
}
\label{fig:HFAG-ccd}
\end{figure}

As mentioned in the introduction to Sec.~\ref{sec:ccq}, one way to investigate further the penguin transitions is to search for their effects in $b \to c\bar{c}d$ transitions (for detailed discussions, see for example Refs.~\cite{Faller:2008gt,Jung:2012mp,DeBruyn:2014oga,Jung:2014jfa,Frings:2015eva,Ligeti:2015yma}).  
Searches for \CP violation in decay, for example in $\Bp \to \jpsi\pip$~\cite{LHCb-PAPER-2011-024} and $\Bp \to \Dzb\Dp$ decays~\cite{Adachi:2008cj}, set limits at the level of a few percent, but as mentioned above there may be suppression due to small strong phase differences.
A more comprehensive approach is to search for deviations from the $b \to c\bar{c}s$ values of $\beta$ and $\beta_s$ in $\Bz$ and $\Bs$ decays to self-conjugate final states.  
This is particularly attractive in cases where decays can be related by the U-spin flavour-symmetry of the strong interaction, such as $\Bd \to \jpsi\KS \longleftrightarrow \Bs\to \jpsi \KS$ and $\Bs\to\Dsp\Dsm \longleftrightarrow \Bd \to \Dp\Dm$~\cite{Fleischer:1999nz}.
Measurements of mixing-induced \CP violation in $\Bd \to \Dp\Dm$ decays~\cite{Aubert:2008ah,Rohrken:2012ta,LHCb-PAPER-2016-037} are, in fact, in slight tension with the value of $\sin(2\beta)$ from $b \to c\bar{c}s$ decays, as shown in Fig.~\ref{fig:HFAG-ccd}.
However, this arises partly because the central value of the \belle result~\cite{Rohrken:2012ta} lies outside the physical region $S_f^2 + C_f^2 \leq 1$, and therefore any interpretation must be made with great care.
Improved measurements are needed. 
In the case of $\Bs\to \jpsi \KS$ decays, the available yield is sufficient only for a rather imprecise determination of the mixing-induced asymmetry parameters, as has recently been performed by LHCb~\cite{LHCb-PAPER-2015-005}.
ATLAS and CMS may be able to contribute measurements of these observables, which would be useful to reduce the overall uncertainty.

Better knowledge about the penguin effects can be obtained by using a larger basis of flavour-symmetries, such as SU(3).  
For example, measurements of \CP violation effects in $\Bd \to \jpsi \piz$ can be used to obtain useful constraints on the penguin contribution to $\Bd \to \jpsi \KS$~\cite{Ciuchini:2005mg,Jung:2012mp}.
Similarly, results on $\Bd \to \jpsi \rhoz$ decays provide information about possible penguin pollution in $\Bs \to \jpsi \phi$.
Branching fraction measurements are also an important component of the SU(3) analysis.
The latest results on \CP violation in $\Bd \to \jpsi \piz$~\cite{Aubert:2008bs,Lee:2007wd} and $\jpsi \rhoz$~\cite{LHCb-PAPER-2014-058} decays are also shown in Fig.~\ref{fig:HFAG-ccd}; these currently provide the strongest experimental limits on penguin pollution in $\beta$ and $\beta_s$.
Note that $\Bd \to \jpsi \rhoz$ is a pseudoscalar to vector-vector decay, and therefore contains an admixture of \CP-even and \CP-odd in the final state; however results are presented in terms of $-\eta_f S_f$ and $C_f$ to enable a useful comparison with other measurements.

Another channel that can help to understand penguin effects is $\Bs \to \jpsi\Kstarzb$~\cite{Faller:2008gt,DeBruyn:2014oga,Frings:2015eva}.
A detailed analysis allowing for \CP violation effects that may differ between different polarisation amplitudes has recently been performed by LHCb~\cite{LHCb-PAPER-2015-034}.
No significant \CP violation effect is seen.

Current determinations of the possible penguin effects in the determinations of $2\beta$ and $2\beta_s$ limit the bias at less than about $1^\circ$.
This is less than the current experimental uncertainties, as given in Secs.~\ref{sec:sin2b} and~\ref{sec:psiphiandbuddies}, but not by much.
Further reduction should be possible with improved measurements in the channels discussed above.

%% file: cud.tex
\subsection{Measurement of $\beta_{(s)}$ using $b \to c\bar{u}d$ transitions}
\label{sec:cud} 

An alternative approach to reduce uncertainties from penguin contributions is to determine $\beta$ and $\beta_s$ from decay modes where there is no possibility of such amplitudes.
Decays mediated by the $b \to c\bar{u}d$ transition, for example $\Bz \to D \piz$ and $\Bs \to D \KS$,  provide such potential~\cite{Grossman:1996ke}.
When the neutral \D meson is reconstructed in a \CP eigenstate, the formalism of Eqs.~(\ref{eg:tdcpv1})--(\ref{eg:tdcpv-asym}) can be used.
(The additional light meson, $\piz$ or $\KS$, must also be reconstructed in a \CP eigenstate, as achieved with the $\piz \to \gamma\gamma$ and $\KS\to\pip\pim$ decays.)
This, however, allows amplitudes mediated by the $b \to u \bar{c}d$ transition to contribute.
The relative weak phase between $b \to u \bar{c}d$ and $b \to c\bar{u}d$ is $\gamma$, while the ratio of magnitudes is expected to be around 0.02.
Therefore, some small biases in the determination of $\beta_{(s)}$ can be expected.
The leading effects, however, have opposite signs for the cases where the \D meson is reconstructed in \CP-even and -odd final states~\cite{Fleischer:2003ai,Fleischer:2003aj}, and therefore if both are measured the theoretical uncertainty on $\beta_s$ can be kept under excellent control.
Note also that, in certain circumstances, the interference between $b \to u \bar{c}d$ and $b \to c\bar{u}d$ amplitudes can be used to make a competitive determination of $\gamma$, as discussed in Sec.~\ref{sec:gamma-tree-level:mix}.

An interesting modification of this method utilises multibody decays of the \D meson such as $\KS\pip\pim$~\cite{Bondar:2005gk}.
In this case, interference between amplitudes contributing to the \D decay provides sensitivity to $\cos(2\beta_{(s)})$ as well as $\sin(2\beta_{(s)})$.
Similarly to the methods discussed in Sec.~\ref{sec:ccq}, the determination of  $\cos(2\beta)$ can remove the ambiguity inherent in measurements of $\sin(2\beta)$ alone.

Measurements of \CP violation in the decay-time dependence of $\Bz$ decays mediated by the $b \to c\bar{u}d$ transition have been performed combining many final states of the form $D^{(*)} h^0$, where $h^0$ is a light meson (such as $\piz$ or $\eta$).
\babar and \belle have independently performed analyses with $\D \to \KS\pip\pim$~\cite{Krokovny:2006sv,Aubert:2007rp}, and a combined analysis of $D_{\CP}^{(*)} h^0$ decays has been performed~\cite{Abdesselam:2015gha}.
The latter gives the first observation of \CP violation in $b \to c\bar{u}d$ transitions, and a combination of all results~\cite{HFAG} gives 
\begin{equation}
  \sin(2\beta)^{b \to c\bar{u}d} = 0.63 \pm 0.11 \, .
\end{equation}
Thus, \CP violation in the interference between mixing and decay has been observed in these transitions.
The average for $\cos(2\beta)$ prefers the SM solution for $\beta$.
The comparison of the $b \to c\bar{u}d$ value of $\beta$ with that from $b \to c\bar{c}s$ processes will become more and more important as the precision improves.

All measurements of $b \to c\bar{u}d$ processes to date are statistically limited, and therefore further reduction of uncertainties can be anticipated with larger samples.
These modes are, however, challenging to reconstruct at hadron colliders when the decay products of the light meson $h^0$ include photons.
A more attractive channel for LHCb is therefore $\Bz \to \D \pip\pim$, including a significant contribution from the $\D\rho^0$ intermediate state.
The decay-time distribution of this decay provides sensitivity to both $\sin(2\beta)$ and $\cos(2\beta)$~\cite{Charles:1998vf,Latham:2008zs}.
A Dalitz plot analysis of $\Bz \to \Dzb \pip\pim$ with $\Dzb \to \Kp\pim$~\cite{LHCb-PAPER-2014-070} demonstrates the large yields available at LHCb and the potential for a future decay-time-dependent analysis with the \D meson reconstructed in \CP eigenstates.

It should be noted that in the case of the measurements with $\D \to \KS\pip\pim$, lack of knowledge of the correct \D decay Dalitz plot model leads to a significant source of systematic uncertainty.
As precision improves, it could therefore be of interest to explore model-independent approaches based on binning the Dalitz plot, as discussed for the determination of $\gamma$ from $B \to DK$ in Sec.~\ref{sec:gamma-tree-level:decay}.
First results with the model-independent approach have been reported by \belle~\cite{Vorobyev:2016npn}.

While good experimental progress has been made on $\Bz$ decays mediated by the $b \to c\bar{u}d$ transition, the same cannot be said for the $\Bs$ modes.
The $\Bs \to \D \KS$ decay has recently been observed~\cite{LHCb-PAPER-2015-050}, but rather modest yields are available at LHCb, due to the relatively low efficiency to reconstruct the long-lived \KS meson.
Prospects may be somewhat better for $\Bs$ decays mediated by the $b \to c\bar{u}s$ transition.
The phenomenology in this case is similar to that for the $b \to c\bar{u}d$ case, but the interference effects between $b \to c\bar{u}s$ and $b \to u\bar{c}s$ are larger.
This leads to greater sensitivity to $\gamma$, and therefore these modes will be discussed in that context in Sec.~\ref{sec:gamma-tree-level:mix}.
However, it is germane to the discussion here to note that both the quasi-two-body $\Bs \to \Dzb\phi$~\cite{LHCb-PAPER-2013-035}, and the three-body $\Bs \to \Dzb \Kp\Km$~\cite{LHCb-PAPER-2012-018} decays have been measured at LHCb.
With larger data samples, these modes (in cases where the \D meson is reconstructed in a final state accessible to both \Dz and \Dzb decay) can be used to probe $\beta_s$ without uncertainty due to penguin contributions~\cite{Nandi:2011uw}.

%% file: gamma.tex
\subsection{Measurement of $\gamma$ exploiting interference between $b \to c\bar{u}s$ and $b \to u\bar{c}s$ transitions}
\label{sec:gamma-tree-level}

\subsubsection{Methods based on \CP violation in decay}
\label{sec:gamma-tree-level:decay}

\begin{figure}[!htb]
  \centering
  \includegraphics[width=0.495\textwidth]{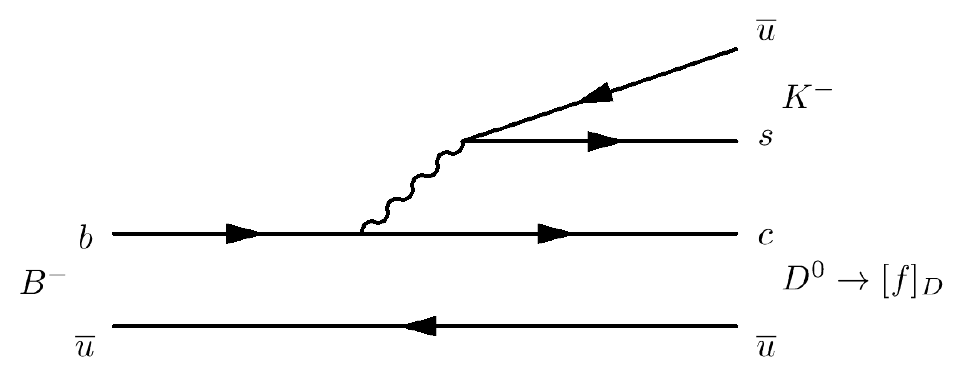}
  \includegraphics[width=0.495\textwidth]{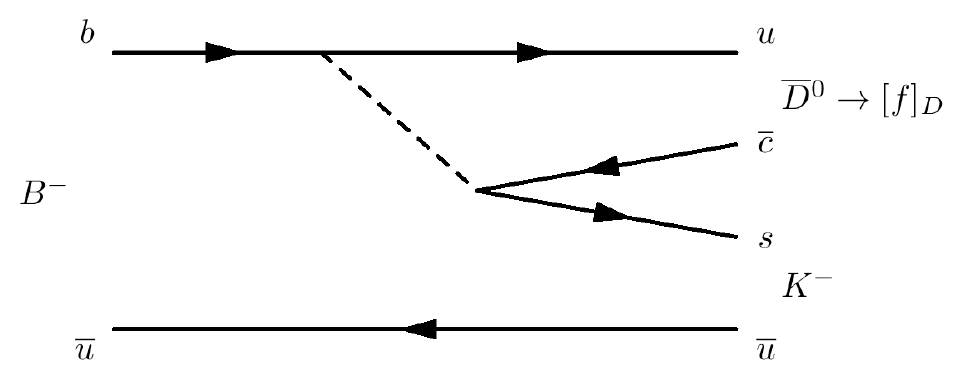}
  \caption{
    Leading diagrams for $\Bm \to \D\Km$ decays: (left) $b \to c\bar{u}s$ and (right) $b \to u\bar{c}s$ transitions.
  }
  \label{fig:gamma}
\end{figure}

Interference between the tree-level $b \to c\bar{u}s$ and $b \to u\bar{c}s$ transitions allows the angle $\gamma$ to be determined from \CP violation in decay.
This is possible since, although the two transitions give different final states, for example $\Bm \to \Dz\Km$ and $\Bm \to \Dzb\Km$, interference occurs when the neutral \D meson is reconstructed in a decay mode that is accessible to both, as illustrated in Fig.~\ref{fig:gamma}.
It is important to note that, since only tree-level amplitudes are involved, the determination of $\gamma$ provides a SM benchmark measurement of \CP violation, with essentially negligible theoretical uncertainty from higher-order electroweak amplitudes~\cite{Brod:2013sga}.
Small effects from mixing and \CP violation in the $\Dz$--$\Dzb$ system can be included in the analysis, and are neglected in this discussion.

In the simplest case, where the \D meson is reconstructed in a \CP eigenstate, the asymmetry of Eq.~(\ref{eq:acp-decay}) becomes 
\begin{equation}
  A_{\CP} = \frac{\pm 2 \, r_B \sin(\delta_B) \sin(\gamma)} {1 + r_B^2 \pm 2 \, r_B \cos(\delta_B) \cos(\gamma)} \, ,
\label{eq:acp-gamma}
\end{equation}
where the $+$ ($-$) sign corresponds to \CP-even (-odd), $r_B$ is the ratio of the magnitudes of the $b \to u\bar{c}s$ and $b \to c\bar{u}s$ amplitudes and $\delta_B$ is their relative strong phase.
The parameter $r_B$ governs the possible size of interference effects between the two amplitudes.
In $\Bm \to \D\Km$ decays, the expectation (confirmed by experiment) is $r_B \sim 0.10$ due to the magnitudes of relevant CKM matrix elements and the ``colour-suppression'' of the $b \to u\bar{c}s$ amplitude which arises as the $s$~quark produced from the internal virtual $W$~boson must form a colour neutral object with the spectator $\bar{u}$~quark, as seen in Fig.~\ref{fig:gamma}(right).

The crucial feature of $\Bm \to D\Km$ (and similar) decays is that the neutral \D meson can be reconstructed in different final states -- this provides the unique potential to determine a CKM phase from \CP violation in decay with low theoretical uncertainty.
A two-body \D decay final state $f$ introduces two new hadronic parameters ($r_D$ and $\delta_D$, which are, respectively, the ratio of the magnitudes and the relative phase of the amplitudes for $\Dzb$ and $\Dz$ decay to $f$), leading to a modified version of Eq.~(\ref{eq:acp-gamma}):
\begin{equation}
  A_{\CP} = \frac{2 \, r_D r_B \sin(\delta_B + \delta_D) \sin(\gamma)} {r_D^2 + r_B^2 + 2 \, r_B r_D \cos(\delta_B + \delta_D) \cos(\gamma)} \, .
\label{eq:acp-gamma2}
\end{equation}
The hadronic parameters describing the \D decay can be independently determined, either from samples of quantum-correlated $\psi(3770) \to \D\Dbar$ decays~\cite{Gronau:2001nr,Atwood:2002ak,Atwood:2003mj,Asner:2005wf} or from studies of charm mixing~\cite{Harnew:2013wea,Harnew:2014zla,HFAG}.
By combining this information with measurements of asymmetries and also rates in $\Bm \to \D\Km$ processes with various different \D decays, sufficient independent constraints can be obtained to determined the three parameters $\gamma$, $r_B$ and $\delta_B$.
Multibody \D decays can also be used, in a similar way.

It has become conventional to refer to analyses of $B \to DK$ decays with different \D meson final states by the initials of authors of theory papers discussing their use.
\begin{itemize}[leftmargin=*]
\item The use of \D decays to \CP eigenstates is referred to as the GLW method~\cite{Gronau:1990ra,Gronau:1991dp}.
  Reasonably high yields can be obtained with the \CP-even final states $\Kp\Km$ and $\pip\pim$, and results are available from several experiments~\cite{delAmoSanchez:2010ji,Abe:2006hc,Aaltonen:2009hz,LHCb-PAPER-2016-003}.
  The \CP-odd final states, such as $\KS\piz$, are challenging to reconstruct in a hadronic collision environment, and therefore results are only available from the $\epem$ \B factory experiments \babar and \belle~\cite{delAmoSanchez:2010ji,Abe:2006hc}.
  The world averages are~\cite{HFAG}
  \begin{eqnarray}
    A_{\CP}(\Bm \to \D_{\CP+}\Km) & = & +0.111 \pm 0.018\,, \nonumber \\
    A_{\CP}(\Bm \to \D_{\CP-}\Km) & = & -0.10 \pm 0.07\,.
  \end{eqnarray}
  The first average demonstrates \CP violation in decay in these modes.
  The most precise of the available results is that from LHCb~\cite{LHCb-PAPER-2016-003}, illustrated in Fig.~\ref{fig:LHCb-PAPER-2016-003}.

\begin{figure}[!htb]
  \centering
  \includegraphics[width=0.95\textwidth]{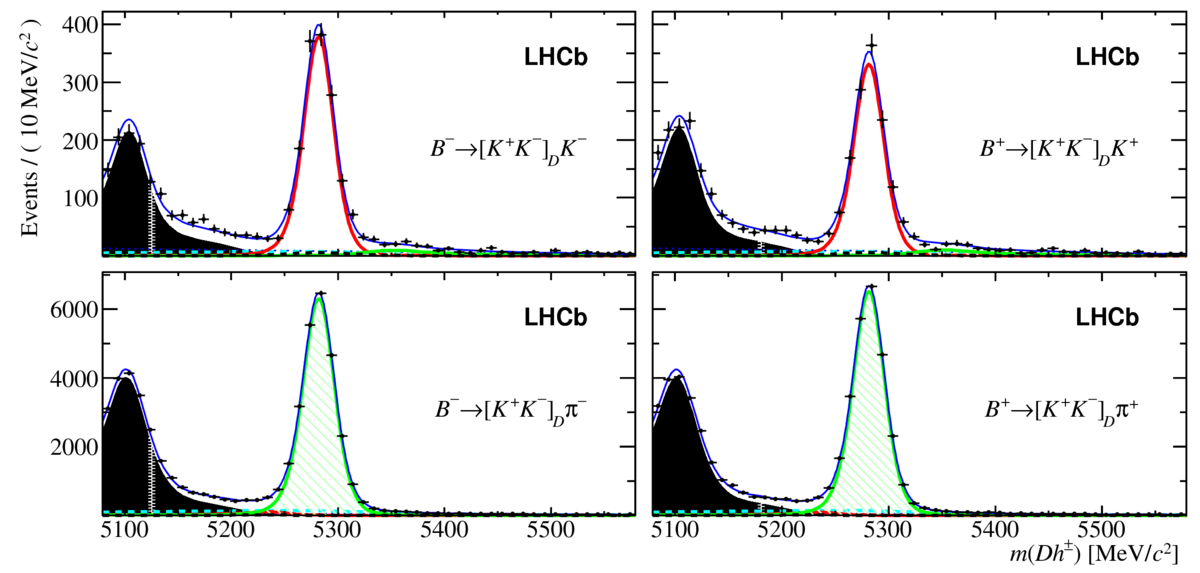}
\caption{
  Yields of (top) $\Bmp \to \D \Kmp$ and (bottom) $\Bmp \to \D \pimp$ candidates in LHCb~\cite{LHCb-PAPER-2016-003}, with fit projections overlaid. 
  The \D mesons are reconstructed in the \CP-even $\Kp\Km$ final state.
  The \CP asymmetry, measured to be $A_{\CP} = +0.097 \pm 0.018 \pm 0.009$ (including also $\D \to \pip\pim$ decays), can be seen as an excess in the $B \to DK$ signal peak for (left) \Bm compared to (right) \Bp candidates. 
}
\label{fig:LHCb-PAPER-2016-003}
\end{figure} 

\item Some multibody decays have been found to be dominantly \CP-even, although they are not {\it a priori} \CP eigenstates.  
  This can be quantified in terms of the fractional \CP-even content $F_+$.  
  This quantity has been measured using CLEOc $\psi(3770) \to \D\Dbar$ data for three modes: $F_+(\Dz \to \pip\pim\piz) = 0.973 \pm 0.017$, $F_+(\Dz \to \Kp\Km\piz) = 0.732 \pm 0.055$ and $F_+(\Dz \to \pip\pim\pip\pim) = 0.737 \pm 0.028$ ~\cite{Malde:2015mha}.
  It has been shown that such decays can be used to obtain information about $\gamma$ in a ``quasi-GLW'' analysis~\cite{Nayak:2014tea}, in which Eq~(\ref{eq:acp-gamma}) is modified to
\begin{equation}
  A_{\CP} = \frac{2 \,(2F_+-1)\, r_B \sin(\delta_B) \sin(\gamma)} {1 + r_B^2 + 2 \,(2F_+-1)\, r_B \cos(\delta_B) \cos(\gamma)} \, .
\label{eq:acp-gamma3}
\end{equation}
Such analyses have been performed by LHCb with the $\D \to \pip\pim\piz$, $\Kp\Km\piz$~\cite{LHCb-PAPER-2015-014} and $\pip\pim\pip\pim$~\cite{LHCb-PAPER-2016-003} decays.
An earlier analysis with $\Bpm \to \left[ \pip\pim\piz \right]_\D \Kpm$ decays by \babar~\cite{Aubert:2007ii} had noted the dominantly \CP-even nature of the final state and measured $A_{\CP}$, whilst also using this information in an amplitude analysis.

\item The asymmetry of Eq.~(\ref{eq:acp-gamma2}) can be largest when $r_D \sim r_B$. 
  Therefore, it is particularly interesting to use doubly-Cabibbo-suppressed \D decays, such as $\Dz \to \Kp\pim$ for which $r_D \sim 0.06$~\cite{HFAG}, which is known as the ADS method~\cite{Atwood:1996ci,Atwood:2000ck}.
  The fact that the $\Bm \to \left[ \Kp\pim \right]_\D \Km$ final state involves only charged tracks makes this channel accessible to several experiments~\cite{delAmoSanchez:2010dz,Belle:2011ac,Aaltonen:2011uu,LHCb-PAPER-2016-003}.
  The world average is~\cite{HFAG} 
  \begin{equation}
    A_{\CP}(\Bm \to \left[ \Kp\pim \right]_\D \Km) = -0.41 \pm 0.06 \, ,
  \end{equation}
  representing a significant \CP\ violation effect in this mode.

\item Multibody doubly-Cabibbo-suppressed \D decays can be used in an extension of the ADS method~\cite{Atwood:2000ck}.\footnote{
  Since Ref.~\cite{Atwood:2000ck} describes the use of multibody \D decays, this is considered part of the ADS method rather than a ``quasi-ADS'' approach.
}
  Similarly to the quasi-GLW method, additional hadronic parameters are introduced in a modified version of Eq.~(\ref{eq:acp-gamma2}),
  \begin{equation}
    A_{\CP} = \frac{2 \,\kappa\, r_D r_B \sin(\delta_B + \delta_D) \sin(\gamma)} {r_D^2 + r_B^2 + 2 \,\kappa\, r_B r_D \cos(\delta_B + \delta_D) \cos(\gamma)} \, .
    \label{eq:acp-gamma4}
  \end{equation}
  where the coherence factor $\kappa$ quantifies the dilution of the asymmetry due to interference between different resonances in the multibody final state ($0 < \kappa < 1$).
  The parameters $r_D$ and $\delta_D$ also become effective parameters, averaged over the phase space of the decay.
  Multibody ADS analyses have been performed by \babar, \belle and LHCb for $\Bm \to \left[ \Kp\pim\piz \right]_\D \Km$ decays~\cite{Lees:2011up,Nayak:2013tgg,LHCb-PAPER-2015-014} and by LHCb for $\Bm \to \left[ \Kp\pim\pip\pim \right]_\D \Km$ decays~\cite{LHCb-PAPER-2016-003}.

\item In the case of \D decay to a multibody self-conjugate final state that is not dominated by one \CP eigenstate, the distribution of decays over the phase space provides additional sensitivity to $\gamma$.
  The study of this distribution can be performed either with an amplitude model for the \D decay or model-independently; either approach is referred to as the GGSZ method~\cite{Giri:2003ty,Bondar}.
  In the former case, the choice of amplitude model results in a systematic uncertainty that is hard to quantify.  
  In the model-independent analysis the phase space (described by a Dalitz plot in the case of three-body decays) is binned, and the method requires knowledge of the average cosine and sine of the strong phase difference between amplitudes for $\Dz$ and $\Dzb$ decays to points within each bin~\cite{Giri:2003ty,Bondar:2005ki,Bondar:2008hh}.
  Such knowledge can be obtained from $\psi(3770) \to \D\Dbar$ samples, with results available from CLEOc~\cite{Briere:2009aa,Libby:2010nu}.
  The limited precision of these measurements leads to a systematic uncertainty on $\gamma$ which, however, often appears as part of the statistical error since it is experimentally convenient to constrain these hadronic parameters within uncertainties through a Gaussian penalty term in the likelihood function used in the fits.

  The GGSZ method gives good sensitivity to $\gamma$ as it combines the strong features of the GLW and ADS approaches in channels with relatively large yields available.
  For example, in $\Dz \to \KS\pip\pim$ decays there are contributions from the singly-Cabibbo-suppressed decay to the \CP eigenstate $\KS\rhoz$ and from the doubly-Cabibbo-suppressed decay to $\Kstarp\pim$, and interference between these resonances not only enhances the sensitivity, but allows ambiguities in the determination of $\gamma$ to be resolved.
  In fact, unlike most other methods, the value of $\gamma$ can be determined directly in the GGSZ method, but it has become conventional for experiments to fit instead for the parameters $x_\pm$, $y_\pm$ where 
  \begin{equation}
    x_\pm + i y_\pm = r_B \exp \left\{ i (\delta_B \pm \gamma)\right\}\,.
    \label{eq:xy}
  \end{equation}
  These parameters are statistically more robust, and allow for more straightforward combination with results from other methods.

  Model-dependent results for $\Bm \to \left[ \KS\pip\pim \right]_\D \Km$ are available from \babar, \belle and LHCb~\cite{delAmoSanchez:2010rq,Poluektov:2010wz,LHCb-PAPER-2014-017} (the \babar analysis also uses the $\D\to\KS\Kp\Km$ decay).
  The model-independent approach has been used by \belle and LHCb~\cite{Aihara:2012aw,LHCb-PAPER-2014-041}, in the former case using $\D\to\KS\pip\pim$ only, and in the latter including also the $\D\to\KS\Kp\Km$ decay.
  The experimental results are summarised in Table~\ref{tab:GGSZ} and illustrated in Fig.~\ref{fig:GGSZ}.
  The averages are more precise in the model-dependent case, and are~\cite{HFAG}
  \begin{equation}
    \begin{array}{c@{\qquad}c}
    x_+ = -0.098 \pm 0.024 \, , & y_+ = -0.036 \pm 0.030 \, , \\
    x_- = \phantom{-}0.070 \pm 0.025 \, , & y_- = \phantom{-}0.075 \pm 0.029 \, ,
    \end{array}
  \end{equation}
  where the effect of model dependence has been neglected since it is not known how much uncertainty should be applied to the average (the uncertainties assigned by \babar~\cite{delAmoSanchez:2010rq} and \lhcb~\cite{LHCb-PAPER-2014-017} are an order of magnitude smaller than those assigned by \belle~\cite{Poluektov:2010wz}).
  If this problem is overlooked, corresponding to negligible model dependence, the results demonstrate \CP violation in $\Bm \to D\Km$ decays.

\item A final category of hadronic \D decays contains singly-Cabibbo-suppressed transitions to non-self-conjugate final states, for example $\Dz \to \Kstarpm\Kmp$.
  These modes can be studied in the GLS approach~\cite{Grossman:2002aq}.
  One experimental result is available from LHCb~\cite{LHCb-PAPER-2013-068}.
  Until now, the yields available in the relevant channels make this method less statistically sensitive than the others.

\end{itemize}

\begin{table}[!htb]
  \centering
  \caption{
    Results from $\Bm \to \D\Km$ GGSZ analyses.
    For the model-dependent analyses the third uncertainty is due to the choice of \D decay model, while for the model-independent analyses the third uncertainty arises from the precision of the constraints of the hadronic parameters describing the average strong phase difference in each Dalitz plot bin.
    Note that the data samples used in the model-dependent and -independent analyses by each experiment are overlapping, and therefore the results are not statistically independent.
  }
  \resizebox{\textwidth}{!}{
  \begin{tabular}{lccc} 
    \hline
    & \babar~\cite{delAmoSanchez:2010rq} & \belle~\cite{Poluektov:2010wz,Aihara:2012aw} & LHCb~\cite{LHCb-PAPER-2014-017,LHCb-PAPER-2014-041} \\
    \hline
    & \multicolumn{3}{c}{Model-dependent} \\
$x_+$ & $-0.103 \pm 0.037 \pm 0.006 \pm 0.007$ & $-0.107 \pm 0.043 \pm 0.011 \pm 0.055$ & $-0.084 \pm 0.045 \pm 0.009 \pm 0.005$ \\
$y_+$ & $-0.021 \pm 0.048 \pm 0.004 \pm 0.009$ & $-0.067 \pm 0.059 \pm 0.018 \pm 0.063$ & $-0.032 \pm 0.048 \,^{+0.010}_{-0.009} \pm 0.008$ \\
$x_-$ & $\phantom{-}0.060 \pm 0.039 \pm 0.007 \pm 0.006$ & $\phantom{-}0.105 \pm 0.047 \pm 0.011 \pm 0.064$ & $\phantom{-}0.027 \pm 0.044 \,^{+0.010}_{-0.008} \pm 0.001$ \\
$y_-$ & $\phantom{-}0.062 \pm 0.045 \pm 0.004 \pm 0.006$ & $\phantom{-}0.177 \pm 0.060 \pm 0.018 \pm 0.054$ & $\phantom{-}0.013 \pm 0.048 \,^{+0.009}_{-0.007} \pm 0.003$ \\
    \hline
    & \multicolumn{3}{c}{Model-independent} \\
$x_+$ & & $-0.110 \pm 0.043 \pm 0.014 \pm 0.007$ & $-0.077 \pm 0.024 \pm 0.010 \pm 0.004$ \\
$y_+$ & & $-0.050 \,^{+0.052}_{-0.055} \pm 0.011 \pm 0.007$ & $-0.022 \pm 0.025 \pm 0.004 \pm 0.010$ \\
$x_-$ & & $\phantom{-}0.095 \pm 0.045 \pm 0.014 \pm 0.010$ & $\phantom{-}0.025 \pm 0.025 \pm 0.010 \pm 0.005$ \\
$y_-$ & & $\phantom{-}0.137 \,^{+0.053}_{-0.057} \pm 0.015 \pm 0.023$ & $\phantom{-}0.075 \pm 0.029 \pm 0.005 \pm 0.014$ \\
    \hline
  \end{tabular}
  }
  \label{tab:GGSZ}
\end{table}

\begin{figure}[!htb]
  \centering
  \includegraphics[width=0.46\textwidth]{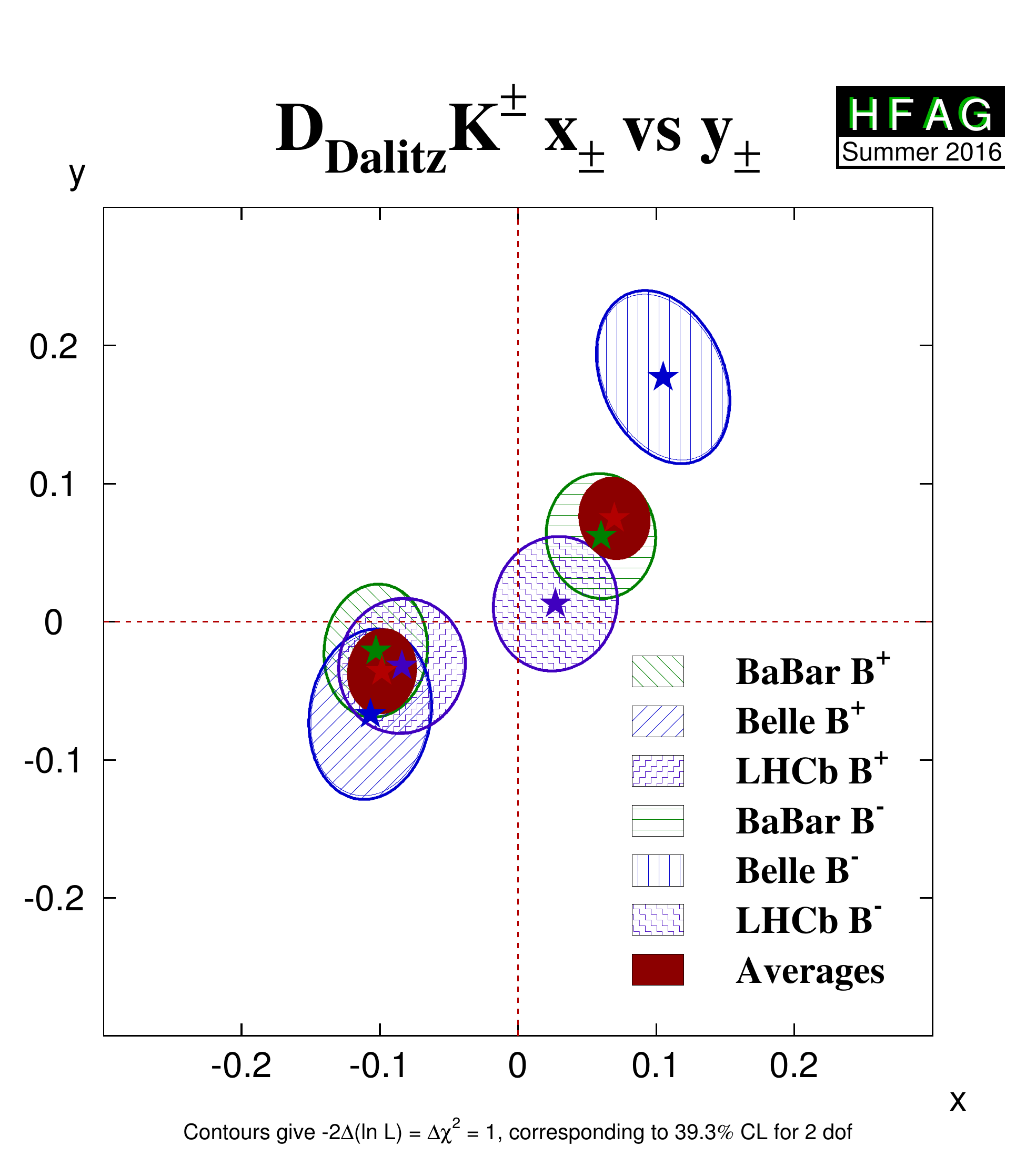}
  \includegraphics[width=0.46\textwidth]{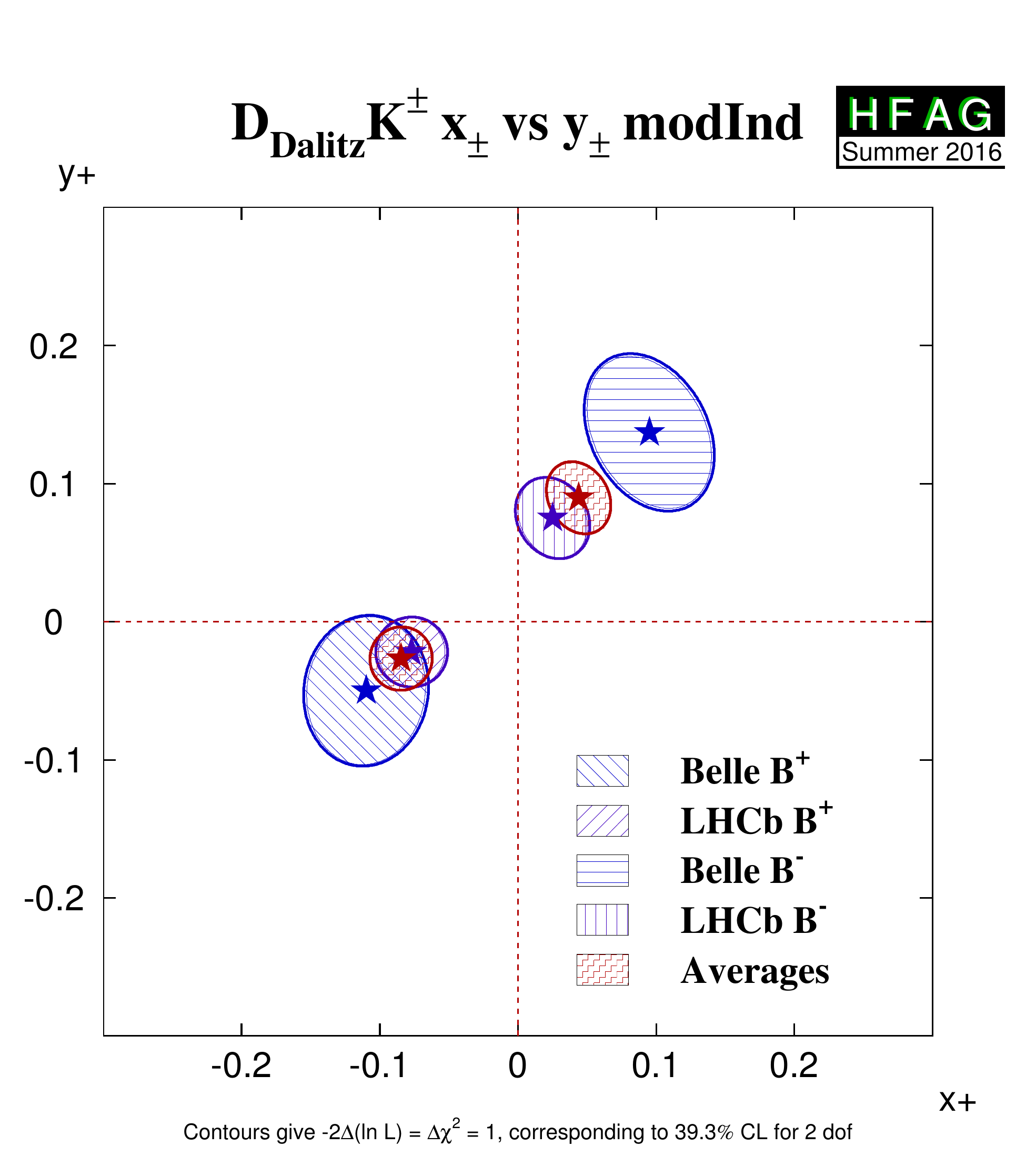}
  \caption{
    Compilation of (left) model-dependent and (right) model-independent results from GGSZ analyses of $\Bm \to \D\Km$ decays~\cite{HFAG}.
    The plotted variables $x_\pm$ and $y_\pm$ are defined in Eq.~(\ref{eq:xy});
    the angle between the two lines from the positions of $(x_\pm,y_\pm)$ to the origin is equal to twice the value of $\gamma$.
    Note that in the model-dependent case the average is performed without including the model uncertainty.
  }
  \label{fig:GGSZ}
\end{figure}

In addition to there being numerous \D decay final states to consider, there are also strong reasons to include additional \B decays.
Each \B decay has its own hadronic parameters $r_B$ and $\delta_B$, but as well as contributing to the reduction of statistical uncertainty, certain decays have particular advantageous features that can be exploited.

\begin{itemize}[leftmargin=*]
\item In $B \to D^*K$ decays, there is an effective strong phase shift of $\pi$ between the $D^*$ decays to $\D\piz$ and $\D\gamma$~\cite{Bondar:2004bi}.
  This must be accounted for in any analysis involving $D^*$ decays, but is particularly beneficial for the ADS method since it provides additional constraints on the phase $\gamma$.
  Results based on $B \to D^*K$ decays are available with the GLW, ADS and GGSZ methods from \babar and \belle~\cite{Aubert:2008ay,Lees:2011up,Abe:2006hc,delAmoSanchez:2010rq,Poluektov:2010wz}.
  These modes are challenging for experiments at hadron colliders.

\item The value of $r_B$ is expected to be larger in neutral \B meson decays to $DK$, compared to that for charged \B meson decays, as both amplitudes are colour-suppressed: the expectation is $r_B \sim 0.3$.
  In $\Bz \to \D\Kstarz$ decays, the charge of the kaon in the $\Kstarz\to\Kp\pim$ decay provides a tag of the flavour of the decaying \B meson~\cite{Dunietz:1991yd}.
  Thus initial state flavour tagging, which would lead to a reduction in sensitivity, is not required.
  However, the finite width of the $\Kstarz$ resonance means that other contributions to the $\Bz \to D\Kp\pim$ Dalitz plot can enter the selection window.
  Consequently an additional hadronic parameter must be included in the analysis in a similar way to the coherence factor of Eq.~(\ref{eq:acp-gamma4}), and the quantities $r_B$ and $\delta_B$ become effective parameters corresponding to the region of the $\D\Kp\pim$ Dalitz plot particular to the $\Kstarz$ selection requirements~\cite{Gronau:2002mu}.
  Results on $\Bz \to \D\Kstarz$ decays are available from \babar, \belle and LHCb with GLW, ADS and GGSZ methods~\cite{Aubert:2009ao,Negishi:2012uxa,LHCb-PAPER-2014-028,Aubert:2008yn,Negishi:2015vqa,LHCb-PAPER-2016-006,LHCb-PAPER-2016-007}.

\item The methodology of Ref.~\cite{Gronau:2002mu} can be extended to any $\B \to \D X_s$ decay, where $X_s$ is a charged or neutral hadronic system with unit strangeness. 
  Results on $\Bm \to \D\Kstarm$ have been presented by \babar and \belle using GLW, ADS and GGSZ methods~\cite{Aubert:2009yw,delAmoSanchez:2010rq,Poluektov:2006ia}, but the comparatively small yields available limit the sensitivity achievable.
  LHCb has presented results on $\Bm \to \D\Km\pip\pim$~\cite{LHCb-PAPER-2015-020}, which appears to be a promising channel.

\item The $\Bz \to \D\Kstarz$ analysis can be extended by using an amplitude model to fit the full $\Bz \to \D\Kp\pim$ Dalitz plot~\cite{Gershon:2008pe,Gershon:2009qc}.
  In this case interference between $\Bz \to D_2^*(2460)^-\Kp$ and $\Bz \to \D\Kstarz$ can be exploited to obtain additional sensitivity and to break ambiguities in the determination of $\gamma$.
  LHCb have obtained results with this method using favoured $D\to\Km\pip$ decays as well as the $D\to\Kp\Km$ and $\pip\pim$ modes~\cite{LHCb-PAPER-2015-017,LHCb-PAPER-2015-059}.
  The results of the analysis include values for the $x_\pm, y_\pm$ parameters for $\Bz \to \D\Kstarz$ decays as well as determinations of the hadronic parameters for this channel.

\item   In addition to GLW- and ADS-like analyses of the $\Bz \to \D\Kp\pim$ Dalitz plot distribution, a double Dalitz plot analysis of $\Bz \to \left[ \KS\pip\pim \right]_\D\Kp\pim$ (and similar) decays is possible~\cite{Gershon:2009qr}.
  In this case, it is possible to perform the analysis without model assumptions for both the \B and \D decay Dalitz plots.
  This approach, however, requires large samples and has not yet been pursued experimentally.

\item Interference between the $b \to c\bar{u}d$ and $b \to u\bar{c}d$ transitions contributing to $\Bm \to \D\pim$ decays can, in principle, be used to determine parameters sensitive to $\gamma$ in a similar way as for $\Bm \to \D\Km$ decays.
  However, the expected small size of the interference effect ($r_B \sim 0.01$) makes this approach in general less statistically sensitive, and more susceptible to systematic biases.
  An exception arises for the ADS method, as $r_D$ and $r_B$ are still of comparable magnitude.
  Such analyses have been carried out with $\Bm \to D^{(*)}\pim$ decays by several experiments~\cite{delAmoSanchez:2010dz,Belle:2011ac,Aaltonen:2011uu,LHCb-PAPER-2016-003,Nayak:2013tgg,LHCb-PAPER-2015-014}.
  In fact, it is becoming common for the experiments to report results with $\Bm \to \D\pim$ decays also for the GLW and GGSZ methods, and to include these channels in combinations to determine $\gamma$ (\eg in Refs.~\cite{LHCb-PAPER-2013-020,LHCb-PAPER-2016-032}).
  Even if $\Bm \to \D\pim$ decays have less sensitivity to $\gamma$, the inclusion of these modes is valuable to obtain the best possible precision and to ensure that subleading effects are correctly handled.

\end{itemize}

\subsubsection{Methods based on \CP violation in interference between mixing and decay}
\label{sec:gamma-tree-level:mix}

The \Bsb meson can decay to both $\Dsp\Km$ and $\Dsm\Kp$ final states, through $b \to c\bar{u}s$ and $b \to u\bar{c}s$ transitions respectively. 
Since different final states are involved, there is no \CP violation in decay, but instead the parameters of \CP violation in the interference between mixing in decay, given in Eq.~(\ref{eg:tdcpv1})--(\ref{eq:ACS-defs}), are sensitive to $\gamma - 2\beta_s$~\cite{Aleksan:1991nh,Dunietz:1997in,Fleischer:2003yb}.
If $\gamma$ were well-known from other processes, results on $\Bs\to\Dsmp\Kpm$ could therefore be used to obtain a determination of $\beta_s$ using only tree-level decays.  
Since this is not the case, but instead $\beta_s$ is well-known from $b \to c\bar{c}s$ transitions, the results are more commonly interpreted in terms of $\gamma$.
Similar determinations are possible in related processes -- for example \CP violation in interference between mixing and decay in $\Bz \to \DorDstarmp\pipm$ processes, which are mediated by $b \to c\bar{u}d$ and $b \to u\bar{c}d$ transitions, probes $2\beta+\gamma$.
Indeed, the $\Bs\to\Dsmp\Kpm$ and $\Bz\to\Dmp\pipm$ processes are related to U-spin, and therefore the measurements can combined in a joint analysis~\cite{Fleischer:2003yb} in order to better control the hadronic parameters. 
This, however, results in only a small~\cite{Gligorov:1106345} improvement on the knowledge of $\gamma$ which could be obtained from $\Bs\to\Dsmp\Kpm$ alone.

Due to the small value of $\Delta \Gamma_d$, the only relevant observables in $\Bz\to\DorDstarmp\pipm$ decays are $S_f$ and $S_{\bar{f}}$ which depend on $2 R_{\DorDstar\pi} \sin(2\beta+\gamma \pm \delta_{\DorDstar\pi})$.
The ratio of magnitudes of $b \to u\bar{c}d$ and $b \to c\bar{u}d$ amplitudes, $R_{\DorDstar\pi}$, has large CKM suppression and is expected to be about $0.02$, making the deviation of $\left| C_f \right|$ and $\left| C_{\bar{f}} \right|$ from unity unobservably small.  
In addition the strong phase difference between the two amplitudes, $\delta_{\DorDstar\pi}$ must be determined from the data.
Since there are only two observables that depend on three unknown quantities, it is necessary to use additional information, for example exploiting flavour symmetry relations to constrain $R_{\D\pi}$ from the measured branching fraction of $\Bz \to \Dsp\pim$~\cite{Dunietz:1997in}.
Such an approach however introduces theoretical uncertainty due to breaking of the SU(3) flavour symmetry.
Moreover, there are discrete ambiguities in the solutions for $2\beta+\gamma$.

Measurements have been made of the \CP violation parameters in $\Bz\to\Dmp\pipm$~\cite{Aubert:2006tw,Ronga:2006hv}, $\Dstarmp\pipm$~\cite{Aubert:2005yf,Aubert:2006tw,Ronga:2006hv,Bahinipati:2011yq} and $\Dmp\rhopm$~\cite{Aubert:2006tw} decays.
The samples of $\Bz\to\Dstarmp\pipm$ decays available are increased by use of a partial reconstruction technique~\cite{Aubert:2005yf,Bahinipati:2011yq}, and the average of results in this mode provides evidence for a significant \CP violation effect.
It is not, however, as yet possible to obtain strong constraints on $2\beta+\gamma$.

\begin{figure}[!htb]
  \centering
  \includegraphics[width=0.40\textwidth]{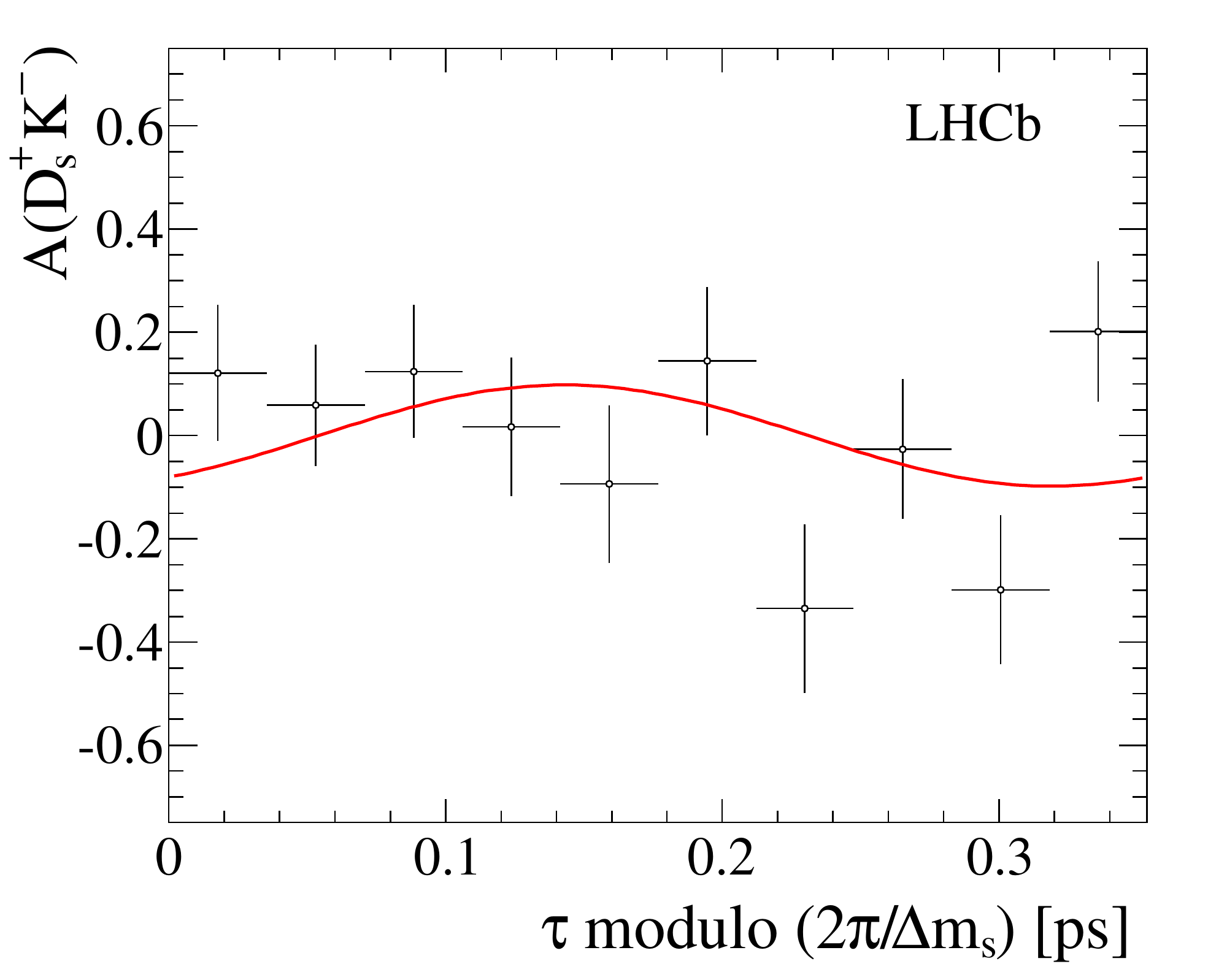}
  \includegraphics[width=0.40\textwidth]{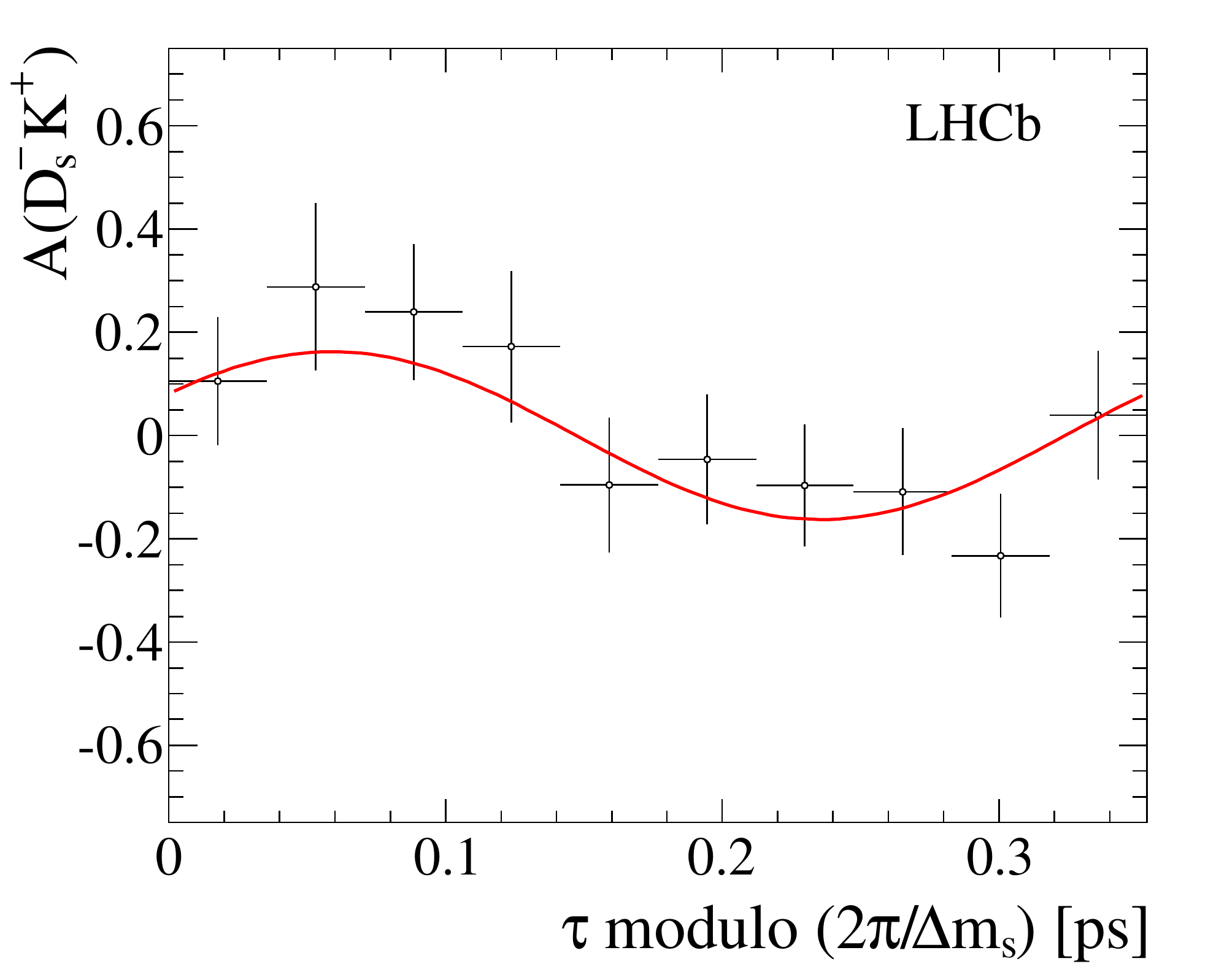}
  \caption{
    Decay-time-dependent asymmetries of Eq.~(\ref{eg:tdcpv-asym}) for (left) $\Bs \to \Dsp\Km$ and (right) $\Bs \to \Dsm\Kp$~\cite{LHCb-PAPER-2014-038}.
    To improve the visualisation the decay-time has been folded by $2\pi/\Delta m_s$.
  }
  \label{fig:PAPER-2014-038}
\end{figure}

In the case of $\Bs \to \Dsmp\Kpm$ decays, the ratio $R_{D_sK}$ is expected to be around $0.3$--$0.4$, allowing the full set of observables in Eq.~(\ref{eg:tdcpv1})--(\ref{eq:ACS-defs}) to be determined from the four decay-time-dependent decay rates.
They are related to the physics parameters $\gamma-2\beta_s$, $\delta_{D_sK}$ and $R_{D_sK}$ by
\begin{equation}
  \arraycolsep=1.4pt\def\arraystretch{1.5}
  \label{eq:dskobservables}
  \begin{array}{r@{\ }c@{\ }l@{\qquad}r@{\ }c@{\ }l}  
    C_{f} & = & - C_{\bar{f}} = \frac{1-R_{D_sK}^2}{1+R_{D_sK}^2} \,, \\
    A_{f}^{\Delta \Gamma}       & = & \frac{-2 R_{D_sK} \cos(\gamma-2\beta_s-\delta_{D_sK})}{1+R_{D_sK}^2} \,, &
    A_{\bar{f}}^{\Delta \Gamma} & = & \frac{-2 R_{D_sK} \cos(\gamma-2\beta_s+\delta_{D_sK})}{1+R_{D_sK}^2} \,, \\
    S_{f}       & = & \frac{-2 R_{D_sK}\sin(\gamma-2\beta_s-\delta_{D_sK})}{1+R_{D_sK}^2} \,, &
    S_{\bar{f}} & = & \frac{-2 R_{D_sK}\sin(\gamma-2\beta_s+\delta_{D_sK})}{1+R_{D_sK}^2} \,.
  \end{array}
\end{equation}
In this case the sinusoidal and hyperbolic observables result in staggered constraints in the $\delta_{D_sK}:(\gamma -2\beta_s)$ plane,
and consequently their combination results in only a twofold ambiguity in the measured value of $\gamma -2\beta_s$.
The observables have been measured by LHCb using the $1\invfb$ sample collected in 2011, as shown in Fig.~\ref{fig:PAPER-2014-038}, to be~\cite{LHCb-PAPER-2014-038}
\begin{equation}
  \begin{array}{r@{\ }c@{\ }l@{\qquad}r@{\ }c@{\ }l}  
    C_{f} & = & \phantom{-}0.53\pm0.25\pm0.04 \, , \\
    A_{f}^{\Delta \Gamma}       & = & \phantom{-}0.37\pm0.42\pm0.20 \, , & 
    A_{\bar{f}}^{\Delta \Gamma} & = & \phantom{-}0.20\pm0.41\pm0.20 \, ,\\ 
    S_{f}       & = & -1.09\pm0.33\pm0.08 \, , &
    S_{\bar{f}} & = & -0.36\pm0.34\pm0.08 \, ,
  \end{array}
\end{equation}
where the first uncertainty is statistical and the second systematic. 
From an experimental point of view, the key difficulty is the fact that the hyperbolic observables $A_{f}^{\Delta \Gamma}$ and $A_{\bar{f}}^{\Delta \Gamma}$ require precise knowledge of the variation with decay time of the selection efficiency, which is reflected in the larger systematic uncertainties on those observables.
This can, however, be reduced in future using the $\Bs\to\Dsm\pip$ decay as a control channel.
The results do not yet give strong constraints on $\gamma-2\beta_s$, but reduction in uncertainty is expected when larger data samples are analysed.  
It is anticipated that it will also be possible to use $\Bs\to\Dssmp\Kpm$ decays~\cite{LHCb-PAPER-2015-008} to help reduce the uncertainty. 

In addition to the two-body decays discussed above, similar methods can be applied to multibody final states.
In such cases additional interference effects arise due to the different resonances that contribute to the decay, potentially leading to enhanced sensitivity.
For example, a decay-time-dependent analysis of the Dalitz plot distribution of $\Bz\to\Dmp\KS\pipm$ decays can be used to determine $2\beta+\gamma$~\cite{Aleksan:2002mh}.
Unfortunately, the available yields do not allow a useful constraint~\cite{Aubert:2007qe}.
Similarly, the proposal to study \CP violation in $\Bs \to \D\Kp\Km$ decays~\cite{Nandi:2011uw} requires much larger samples than are available at present~\cite{LHCb-PAPER-2012-018}.

\subsubsection{Combined constraints on $\gamma$}
\label{sec:gamma-tree-level:combo}

For a specific \B decay, the methods discussed above are all sensitive to the same underlying parameters: $\gamma$ and the hadronic parameters $r_B$ and $\delta_B$.
The different methods introduce additional hadronic parameters related to the \D decay used, but these are known from independent measurements, typically from quantum-correlated $\psi(3770)\to \D\Db$ samples.
Therefore, the best sensitivity to $\gamma$, and the hadronic parameters, can be achieved by combining results from as many different methods as possible.
Similarly, it is desirable to include \B decays to different final states ($D\Km$, $D\Kstarz$, $\Dstar\Km$, \etc), as long as there are sufficient measurements in each mode to compensate for the additional hadronic parameters introduced.

\begin{figure}[!tb]
  \centering
  \includegraphics[width=0.49\textwidth]{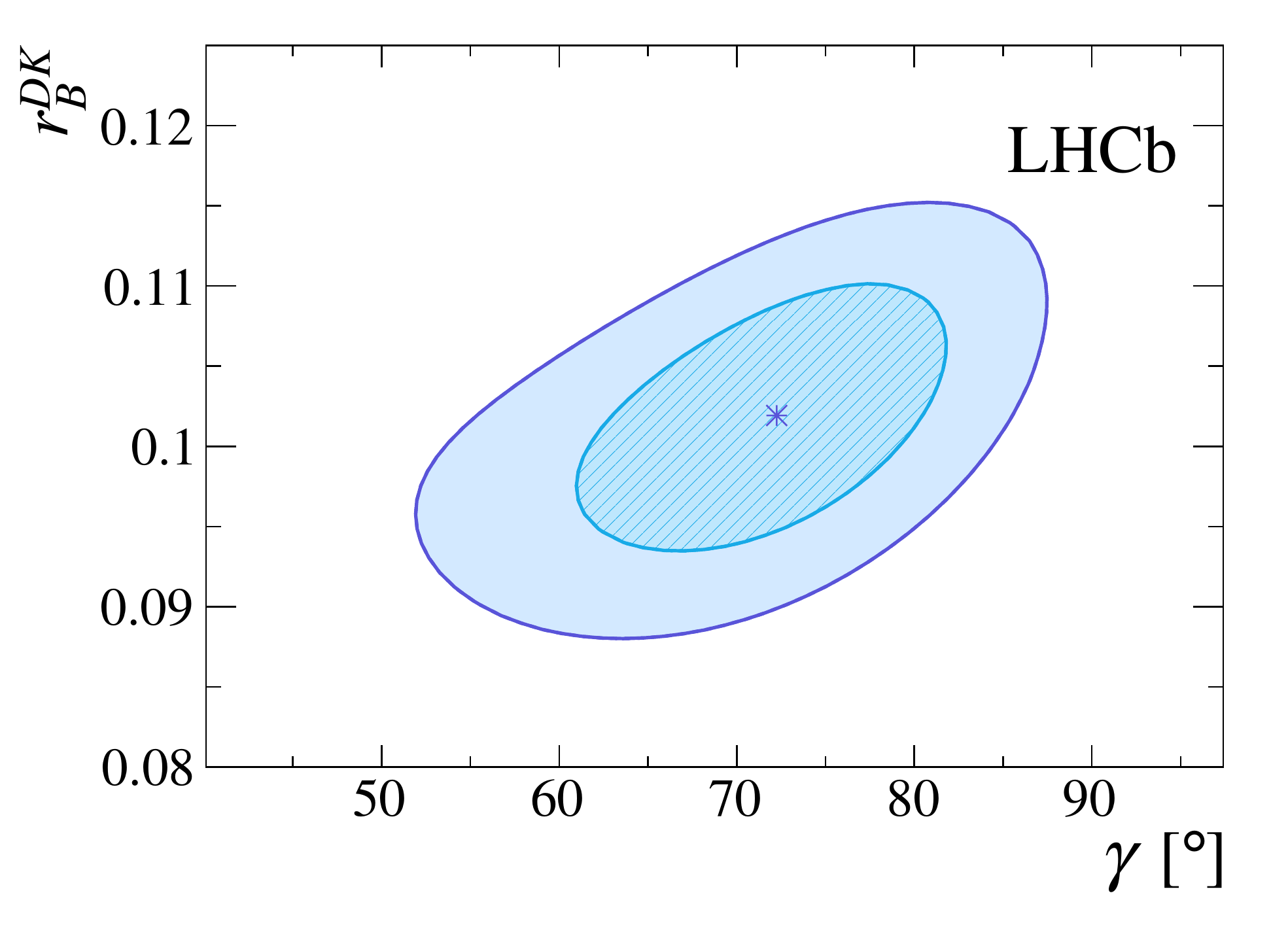}
  \includegraphics[width=0.49\textwidth]{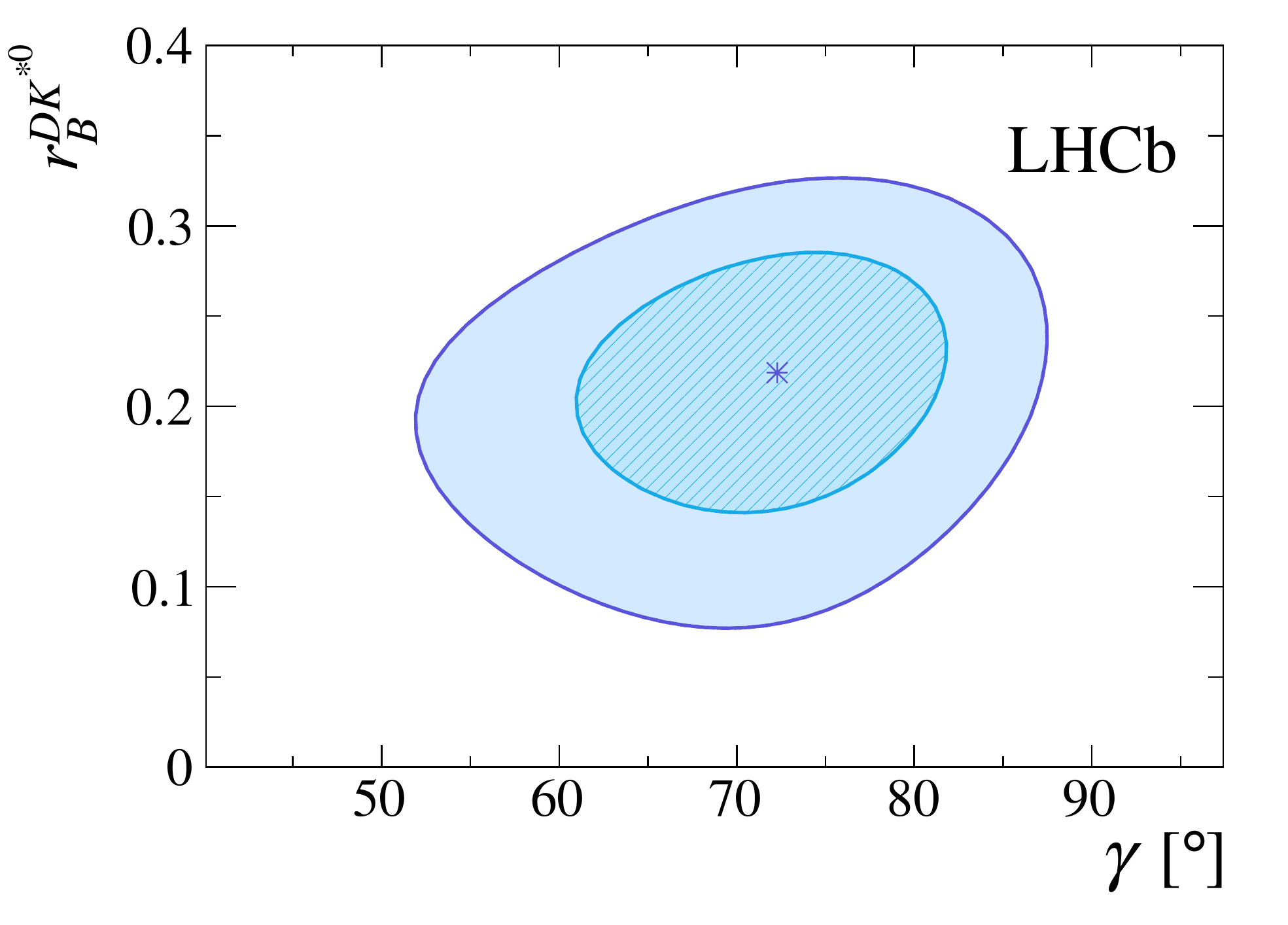}
  \caption{
    Two-dimensional contours for $\gamma$ versus the hadronic parameter $r_B$ in (left) $\Bm \to D\Km$ and (right) $\Bz \to D\Kstarz$ decays~\cite{LHCb-PAPER-2016-032}.
    The obtained constraints on the hadronic parameters are $r_B^{DK} = 0.1019 \pm 0.0056$ and $r_B^{D\Kstarz} = 0.218 \,^{+0.045}_{-0.047}$.
  }
  \label{fig:lhcbgamma}
\end{figure}

Such combinations have been performed by \babar~\cite{Lees:2013nha}, \belle~\cite{Trabelsi:2013uj} and \lhcb~\cite{LHCb-PAPER-2013-020,LHCb-PAPER-2016-032}, each using only their own results, together with auxiliary data on the hadronic parameters in charm decays, as inputs.
The results are 
\begin{equation}
  \label{eq:gammaAverages}
  \gamma = \left\{
  \begin{array}{c@{\quad}c}
    \left(69 \,^{+17}_{-16}\right)^\circ & \babar \, ,\\
    \left(68 \,^{+15}_{-14}\right)^\circ & \belle \, ,\\
    \left(72.2^{+6.8}_{-7.3}\right)^\circ & \lhcb \, .\\   
  \end{array}
  \right.
\end{equation}
Two-dimensional contours for $\gamma$ versus the hadronic parameter $r_B$ in $\Bm \to D\Km$ and $\Bz \to D\Kstarz$ decays, obtained by LHCb~\cite{LHCb-PAPER-2016-032}, are shown in Fig.~\ref{fig:lhcbgamma}.
The world average, combining results from all experiments, gives $\gamma = (71.3\,^{+5.7}_{-6.1})^\circ$~\cite{HFAG}.

As mentioned above, it is also possible to include results from $\Bm \to \D\pim$ decays.
In this case, however, the smaller value of $r_B$ makes the determination of $\gamma$ more sensitive to effects due to charm mixing~\cite{Silva:1999bd,Grossman:2005rp,Rama:2013voa}.
Biases are also possible due to \CP violation in the kaon or charm systems, although these are not important effects with the current precision.
Since the fractional uncertainty on $r_B$ in $\Bm\to\D\pim$ decays is larger compared to the $\Bm \to D\Km$ case, the analysis is less statistically robust, and more dependent on the choice of framework used for the combination (such as frequentist or Bayesian).
For these reasons, results from $\Bm \to \D\pim$ decays are often excluded from the combinations, as here.

%% file: penguin-dominated.tex
\section{\boldmath \CP violation in loop-dominated transitions}
\label{sec:penguin-dominated}

Decays that receive contributions from penguin diagrams only are of interest to test the SM, as effects of physics at high energy scales can contribute through loops.
This is true for the radiative and semileptonic decays $b \to (s,d)(\gamma,\ell^+\ell^-,\nu\bar{\nu})$, as well as for hadronic decays that proceed through $b \to s\bar{s}s$, $b \to d\bar{d}s$ and $b \to s\bar{s}d$ quark-level transitions.
However, there is some uncertainty in the SM predictions for \CP violating phenomena since, unlike the case for pure tree amplitudes, each penguin diagram can have three different SM quarks appearing in the loop, and therefore there may be contributions with different weak phases. 
For example, the $b \to s$ amplitude can be written
\begin{equation}
  P = P_u \, V^{}_{ub}V^*_{us} + P_c \, V^{}_{cb}V^*_{cs} + P_t \, V^{}_{tb}V^*_{ts} \, .
\end{equation}
CKM unitarity can be used to replace one of these terms, for example
\begin{equation}
  \label{eq:P-reparam}
  P = (P_c-P_u) \, V^{}_{cb}V^*_{cs} + (P_t-P_u) \, V^{}_{tb}V^*_{ts} \, .
\end{equation}
This choice cannot, of course, affect the physical observables, as emphasised in the concept of ``reparametrisation invariance''~\cite{Botella:2005ks}.
It does, however, affect the weak phase between the two remaining terms and hence the interpretation of the observables.

A further complication, due to hadronisation, affects particularly the decays mediated by $b \to s\bar{s}s$, $b \to d\bar{d}s$ and $b \to s\bar{s}d$ transitions.
If one of the final state mesons has a $u\bar{u}$ component in its wavefunction, then tree diagrams can also contribute.  
For the discussion in this section, the $\phi$ and $\etapr$ resonances will be considered as $s\bar{s}$ dominated; in some works this is also considered a good approximation for the $f_0(980)$ meson, however since the scalar sector is in general less well understood~\cite{GarciaMartin:2011jx,Pelaez:2015qba}, that is not done here.
In all cases it should be remembered that subleading amplitudes -- which can include long-distance rescattering contributions (see, for example, Ref.~\cite{Gronau:2012gs}) -- can have important effects.
For the same reasons, the $b \to d\bar{d}d$ transition is not considered in this section as a loop-dominated process.
Decays with both tree and loop contributions are discussed in Sec.~\ref{sec:tree-penguin}.

\subsection{Searches for \CP violation in decay in loop-dominated transitions to hadronic final states}
\label{sec:penguin-dominated:decay}

For $b \to s\bar{s}s$ transitions, the relative weak phase between the two terms of Eq.~(\ref{eq:P-reparam}) is $\beta_s$.
Since this is small, and since the $P_t$ term is expected to dominate, decays dominated by this transition are not expected to exhibit \CP violation in decay in the Standard Model.
As loop diagrams are sensitive to physics at high scales, such decays can be used to make powerful tests of the SM with, as usual, the caveat that a strong phase difference is also necessary for any manifestation of \CP violation in decay.

Large yields are available in the $\Bp \to \phi \Kp$ and $\Bp \to \etapr \Kp$ decays. 
Searches for \CP violation in decay in these modes~\cite{Lees:2012kxa,LHCb-PAPER-2013-048,Schumann:2006bg,Aubert:2009yx,LHCb-PAPER-2014-065} have not revealed any significant effect up to the precision of a few percent, as expected in the SM.
The world averages are 
\begin{equation}
  {\cal A}_{\CP}(\Bp \to \phi \Kp) = 0.016 \pm 0.013 \, , \quad
  {\cal A}_{\CP}(\Bp \to \etapr\Kp) = 0.004 \pm 0.011 \, .
\end{equation}
Possible evidence for \CP violation in $\Bp \to \phi \Kp$ decays, seen by \babar~\cite{Lees:2012kxa}, is not confirmed by other experiments and appears to be due to contributions from other structures in the $\Bp \to \Kp\Km\Kp$ Dalitz plot~\cite{Lees:2013ngt,LHCb-PAPER-2014-044}, as discussed further in Sec.~\ref{sec:tree-penguin}.
This demonstrates that, even for a relatively narrow resonance such as the $\phi$ meson ($\Gamma < 5 \mev$~\cite{PDG2016}), amplitude analyses of the full Dalitz plot structure of three-body decays are necessary to separate correctly the different contributions.
Such analyses have already been performed for several modes including $\Bp \to \Kp\Km\Kp$~\cite{Garmash:2004wa,Lees:2012kxa}, but more detailed investigations with larger data samples are needed.

Another interesting channel to search for \CP violation in decay is $\Bz \to \phi \Kstarz$, where the $\Kstarz \to \Kp\pim$ mode is used to tag the flavour of the \B meson, so decay-time-dependent analysis is not required.
As the final state contains two vector mesons, it is possible to search for \CP violation in the polarisation amplitudes in addition to the rates, leading to a wider range of tests of the SM.
However, all measurements of \CP violation effects~\cite{Aubert:2008zza,LHCb-PAPER-2014-005,Prim:2013nmy} are consistent with zero, as expected in the SM.

The relative weak phase in $b \to d\bar{d}s$ transitions is also $\beta_s$, and so effects of \CP violation in decay are expected to be small.
Results to date, for example in $\Bp \to \KS \pip$ decays~\cite{Aubert:2006gm,Duh:2012ie,LHCb-PAPER-2013-034}, are consistent with this expectation.
The $\Bs \to \Kstarz\Kstarzb$ decay provides a vector-vector final state, and therefore additional observables, but is not self-tagging.
Nonetheless, \CP violation in decay can in principle still be observed through asymmetries in the distribution of final state particles~\cite{Gardner:2002bb,Gardner:2003su,Durieux:2015zwa}, but such effects are consistent with zero in the currently available data~\cite{LHCb-PAPER-2014-068}. 
This decay can also be used to probe \CP violation in the interference of mixing and decay, as discussed in subsequent sections.
It should also be noted that if any \CP violation is observed, the large width of the $\Kstarz$ state means that
a full amplitude analysis of $\Bs\to\Km\pip\Kp\pim$ will be required to correctly interpret the observables.

Within the SM, larger effects of \CP violation in decay may occur in $b \to s\bar{s}d$ transitions, since the relative weak phase involved is $\beta$.
However, the yields of the relevant decays are smaller due to the additional CKM suppression of the amplitudes.
Searches have been made, for example in the $\Bp \to \etapr\pip$~\cite{Schumann:2006bg,Aubert:2009yx} and $\Bp \to \KS\Kp$~\cite{Aubert:2006gm,Duh:2012ie,LHCb-PAPER-2013-034} modes, with results consistent with zero.
The available yields of the vector-vector decays $\Bz\to \Kstarz\Kstarzb$~\cite{Aubert:2007xc,Chiang:2010ga,LHCb-PAPER-2014-068} and $\Bs \to \phi\Kstarz$~\cite{LHCb-PAPER-2013-012} are not yet sufficient to obtain useful constraints on \CP violating parameters.

\subsection{Searches for \CP violation in mixing/decay interference in loop-dominated transitions}
\label{sec:penguin-dominated:mix}

Values of $\beta$ and $\beta_s$ can be determined from the decay-time-dependence of the decay rate of $b \to s\bar{s}s$ transitions, in the same way as described for $b \to c\bar{c}s$ transitions in Sec.~\ref{sec:ccq}.
For the loop-dominated transitions, the measured values may be affected by contributions from new particles contributing to the penguin amplitudes, and therefore a difference between the measured value of $\beta_{(s)}$ from that obtained in $b \to c\bar{c}s$ transitions could be a signature of physics beyond the SM.
However, since small deviations are possible in the SM, the observables are often denoted as $\beta^{\rm eff}_{(s)}$.
Various theoretical methods have been used to try to evaluate the deviation in the SM~\cite{Grossman:2003qp,Gronau:2003ep,Gronau:2004hp,Cheng:2005bg,Gronau:2005gz,Buchalla:2005us,Beneke:2005pu,Cheng:2005ug,Engelhard:2005ky,Gronau:2006qh,Silvestrini:2007yf}, with the conclusion that the shifts are expected to be at or below the few percent level for the cleanest modes involving $\phi$ or $\etapr$ mesons.
The situation for modes involving $f_0(980)$ mesons is less clear, due to uncertainties concerning the quark-level substructure of this state~\cite{Dutta:2008xw}.

For $b \to d\bar{d}s$ transitions, significant contributions with different weak phases are possible in the SM, making a straightforward interpretation of results in terms of $\beta_{(s)}$ impossible.
Nonetheless, interesting tests of the SM are possible, most notably for $\Bs \to \KorKstarz \KorKstarzb$ decays, for example by exploiting flavour symmetries~\cite{Fleischer:2004vu,Datta:2006af,DescotesGenon:2006wc}.

\subsubsection{Measurements of $\beta^{\rm eff}$}

\begin{figure}[!htb]
\centering
\includegraphics[width=0.58\textwidth]{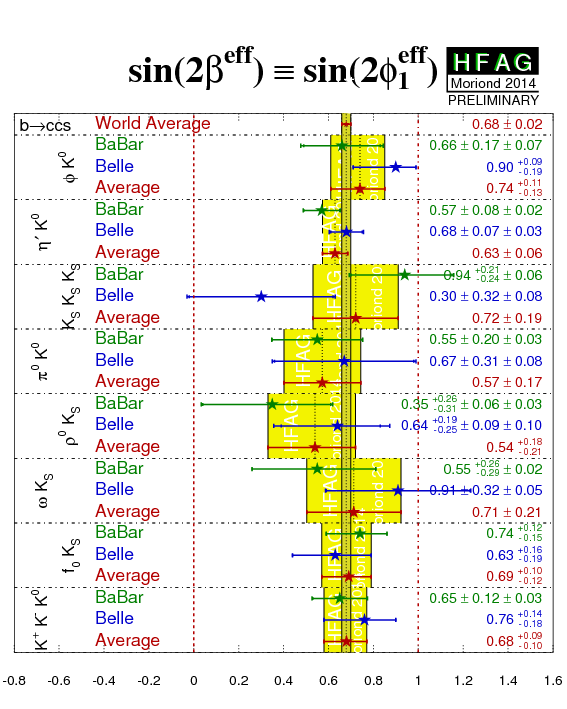}
\caption{
  Measurements and world averages of $\sin(2\beta^{\rm eff})$ in $b \to s$ transitions, compared to the world average value of $\sin(2\beta)$ in $b \to c\bar{c}s$ decays~\cite{HFAG}.
  Note that the measurements labelled $\Kp\Km\Kz$ exclude the contribution with an intermediate $\phi$ meson.
}
\label{fig:bsssbetaeff}
\end{figure}

\begin{figure}[!htb]
\centering
\includegraphics[width=0.29\textwidth]{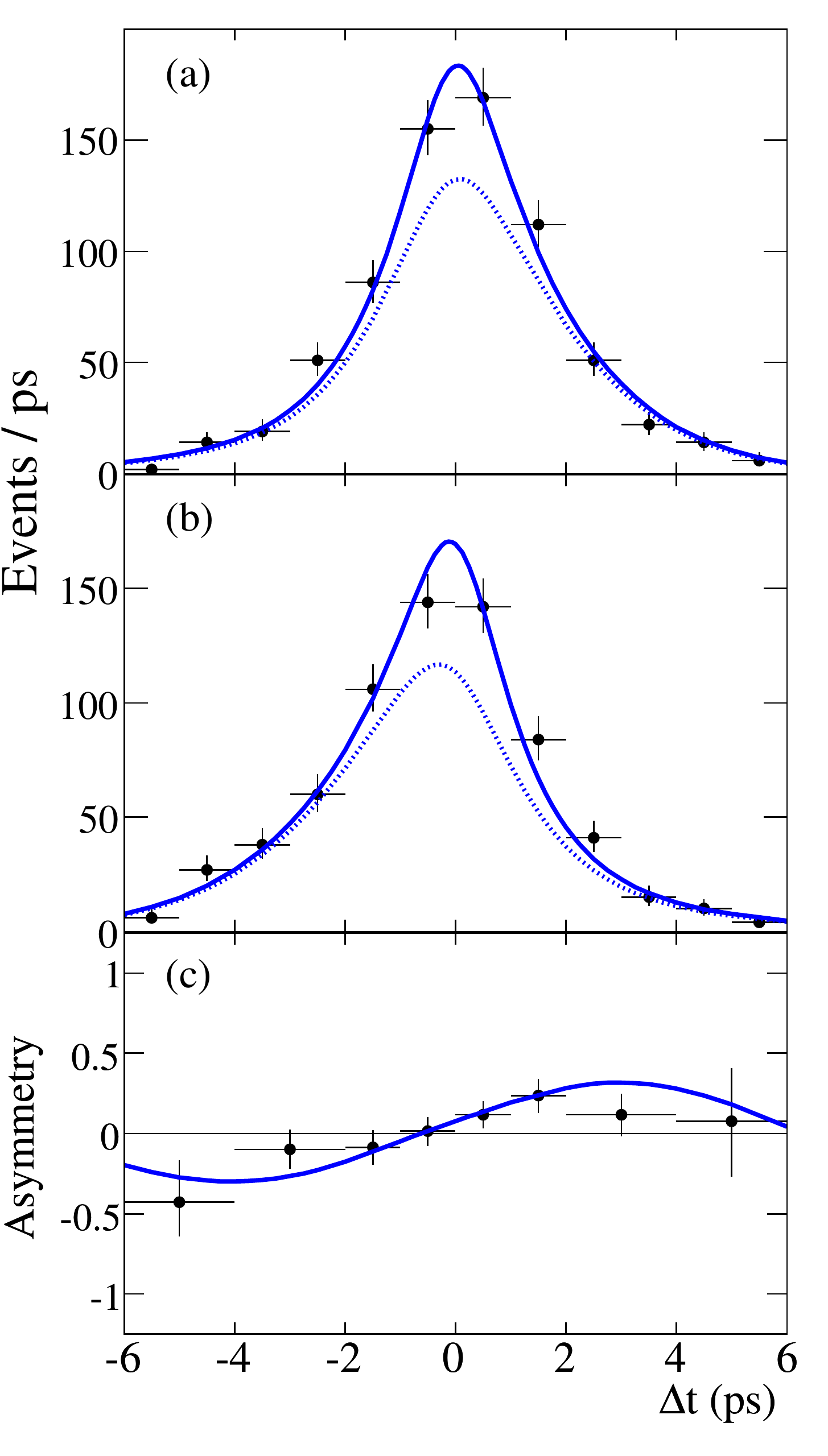}
\hspace{5mm}
\includegraphics[width=0.50\textwidth]{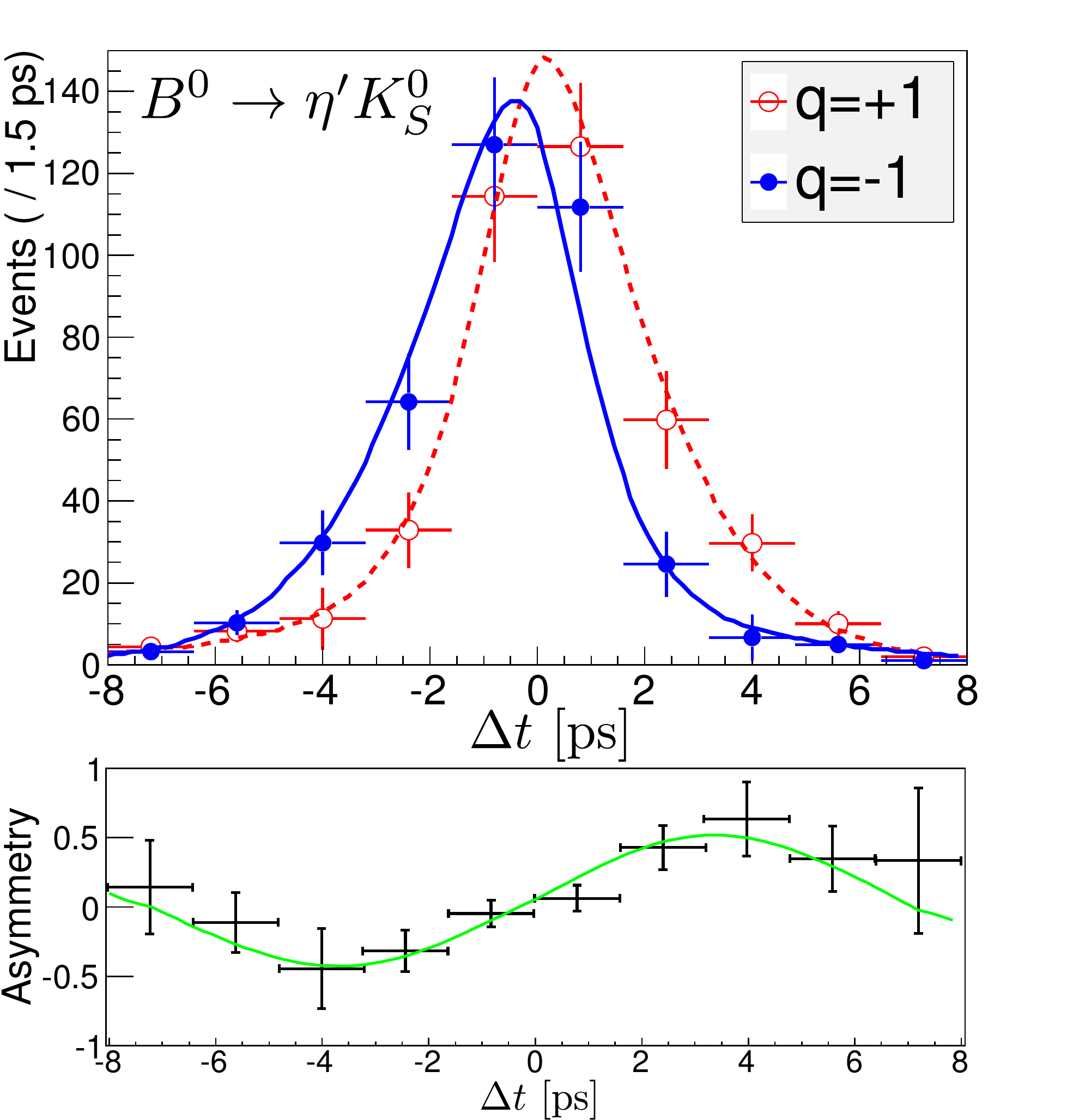}
\caption{
  Decay time distributions and asymmetries in $\Bz \to \etapr\KS$ decays from (left) \babar~\cite{Aubert:2008ad} and (right) \belle~\cite{Santelj:2014sja}.
  In the \babar plot, the solid (dotted) line displays the total fit function (signal component), separated into (a) \Bz and (b) \Bzb tags.
  In the \belle plot, background-subtracted data are shown, with $q = +1~(-1)$ indicating a \Bz~(\Bzb) tag. 
}
\label{fig:etaprKS}
\end{figure}

Measurements of $\beta^{\rm eff}$ have been made in many $b \to s$ transitions at the $\epem$ \B factories, as summarised in Fig.~\ref{fig:bsssbetaeff}.
The results are generally reported in terms of 
\begin{equation}
  S_{f_{\CP}} = -\eta_{f_{\CP}}\sin\left(2\beta^{\rm eff}\right)\,,
\end{equation}
where $\eta_{f_{\CP}}$ is the \CP eigenvalue of the final state $f_{\CP}$,
but are converted to $\sin\left(2\beta^{\rm eff}\right)$ in Fig.~\ref{fig:bsssbetaeff}.
The cosine coefficient, $C_{\CP}$ is also measured, with all values consistent with zero.
No significant departure from the SM expectation is observed. 
All individual measurements are statistically limited, with systematic uncertainties of typically a few percent. 
The most precise results are for the $\Bz\to\etapr\Kz$~\cite{Aubert:2008ad,Santelj:2014sja} (as shown in Fig.~\ref{fig:etaprKS}) and $\Bz \to \phi\Kz$~\cite{Lees:2012kxa,Nakahama:2010nj} modes, where the latter are determined from decay-time-dependent Dalitz plot analyses of $\Bz \to \Kp\Km\Kz$ decays.
There are no results yet available from LHCb on the \CP violation parameters in these modes, although large yields have been reported~\cite{LHCb-PAPER-2013-042}.

Among $b \to d\bar{s}s$ transitions, decay-time-dependent analyses have been performed only for $\Bz \to\KS\KS$ decays~\cite{Aubert:2006gm,Nakahama:2007dg}.
The available yields are too small to provide significant constraints on $S_{f_{\CP}}$.  

\subsubsection{Measurements of $\beta_s^{\rm eff}$}

The $\Bs \to \phi\phi$ channel is of particular interest among the relevant \Bs meson decays.
In the approximation that the $\phi$ meson is a pure $s\bar{s}$ state, the \CP-violation phases from mixing and decay cancel, so that the relation $S_{f_{\CP}} \approx 0$ holds in the SM~\cite{Beneke:2006hg,Bartsch:2008ps,Cheng:2009mu} (for consistency, the determination of $S_{f_{\CP}}$ is nonetheless considered here as a measurement of $\beta_s^{\rm eff}$).
Moreover, the small width of the $\phi$ meson makes this channel experimentally attractive when reconstructed in the $\Kp\Km$ final state, and ensures the contribution in the signal window from $\Kp\Km$ pairs in a S-wave configuration is small.
Since the vector-vector final state contains a mixture of \CP-even and \CP-odd terms, a combined fit to the \Bs decay-time and angular distributions is necessary, similarly as for $\Bs\to J/\psi\phi$ discussed in Sec.~\ref{sec:psiphiandbuddies}.
Among current and planned experiments, \lhcb\ is uniquely able to study the $\Bs \to \phi\phi$ decay mode, and has measured~\cite{LHCb-PAPER-2014-026}
\begin{equation}
  \phi_s^{\rm eff} = -2\beta_s^{\rm eff} = -0.17 \pm 0.15 \pm 0.03 \ \rad \, .
\end{equation}
A number of other parameters are determined, including untagged asymmetries; there is no significant \CP violation effect seen in any of them.

The $b \to d\bar{d}s$ transition $\Bs\to\Kstarz\Kstarzb$ also has great potential to test the SM, for example through exploitation of the U-spin relation with \Bz decays to the same final state~\cite{DescotesGenon:2006wc,Ciuchini:2007hx,Bhattacharya:2012hh}.
No decay-time-dependent analysis of this modes has yet been carried out, but the yields available at LHCb are sufficient for initial studies~\cite{LHCb-PAPER-2014-068}.
As mentioned in the context of measurements of \CP violation in decay,
the non-negligible width of the $\Kstarz$ resonance means that a full amplitude analysis of $\Bs\to\Km\pip\Kp\pim$ final state will ultimately be necessary in order to separate $\Bs\to\Kstarz\Kstarzb$ decays from contributions with a $K\pi$ pair in S-wave, or from reflections from other decays such as $\Bs \to \phi \pip\pim$.

\input{rare}

%% file: rare.tex
\subsection{Studies of \CP violation in $b \to q\gamma$ and $b \to q\ell^+\ell^-$ transitions}
\label{sec:penguin-dominated:rad}

A large number of observables in $b \to q\gamma$ and $b \to q\ell^+\ell^-$ transitions are sensitive to physics beyond the SM (see, for example, Ref.~\cite{Blake:2015tda} for a review).  
In general, since the final states contain particles that do not undergo strong interactions, the hadronic uncertainties in the SM predictions for these observables are reduced compared to fully hadronic final states.
These arguments are relevant for the \CP asymmetries discussed in this section.

Searches for \CP violation effects in several different $b \to q\gamma$ and $b \to q\ell^+\ell^-$ transitions have been carried out.
In addition to studies of particular exclusive final states, inclusive measurements have been performed at the $\epem$ $B$ factory experiments.
Such measurements can be made either by summing together as many different exclusive states as possible, or by exploiting the potential to reconstruct fully the $\Upsilon(4S) \to \B\Bbar$ event so that the hadronic system (denoted $X_s$ or $X_d$ depending on the flavour of the produced quark, or $X_{s+d}$ for their sum) need not be reconstructed at all.
The latest results on \CP violation in decay in inclusive processes are~\cite{HFAG,Lees:2014uoa,Lees:2013nxa}
\begin{equation}
  {\cal A}_{\CP}(B \to X_s \gamma) = 0.015 \pm 0.020 \, , \quad
  {\cal A}_{\CP}(B \to X_s \ell^+\ell^-) = 0.04 \pm 0.11 \, ,
\end{equation}
where both are obtained with the sum-of-exclusive method.

\begin{figure}[!htb]
  \centering
  \includegraphics[width=0.48\textwidth]{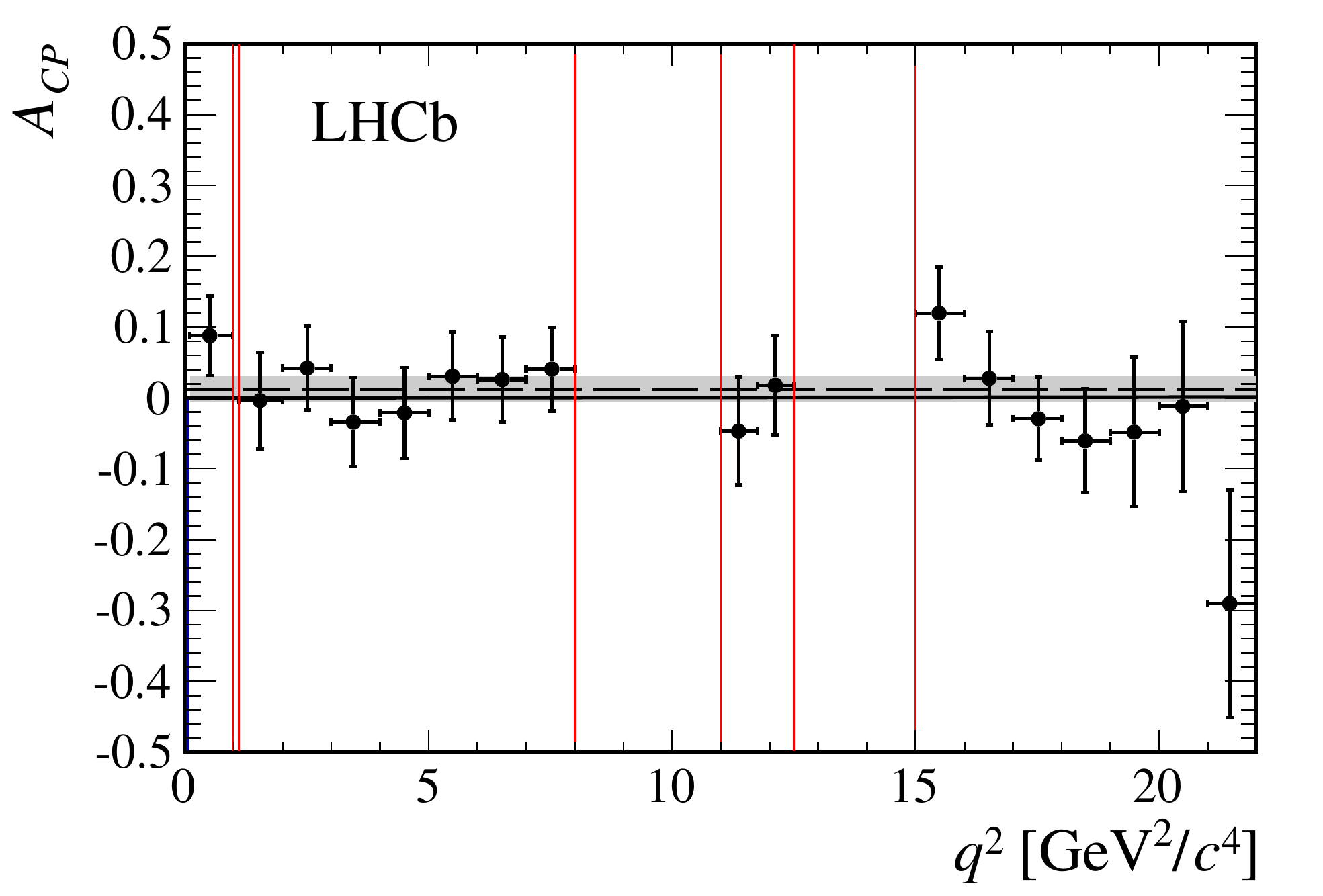}
  \includegraphics[width=0.48\textwidth]{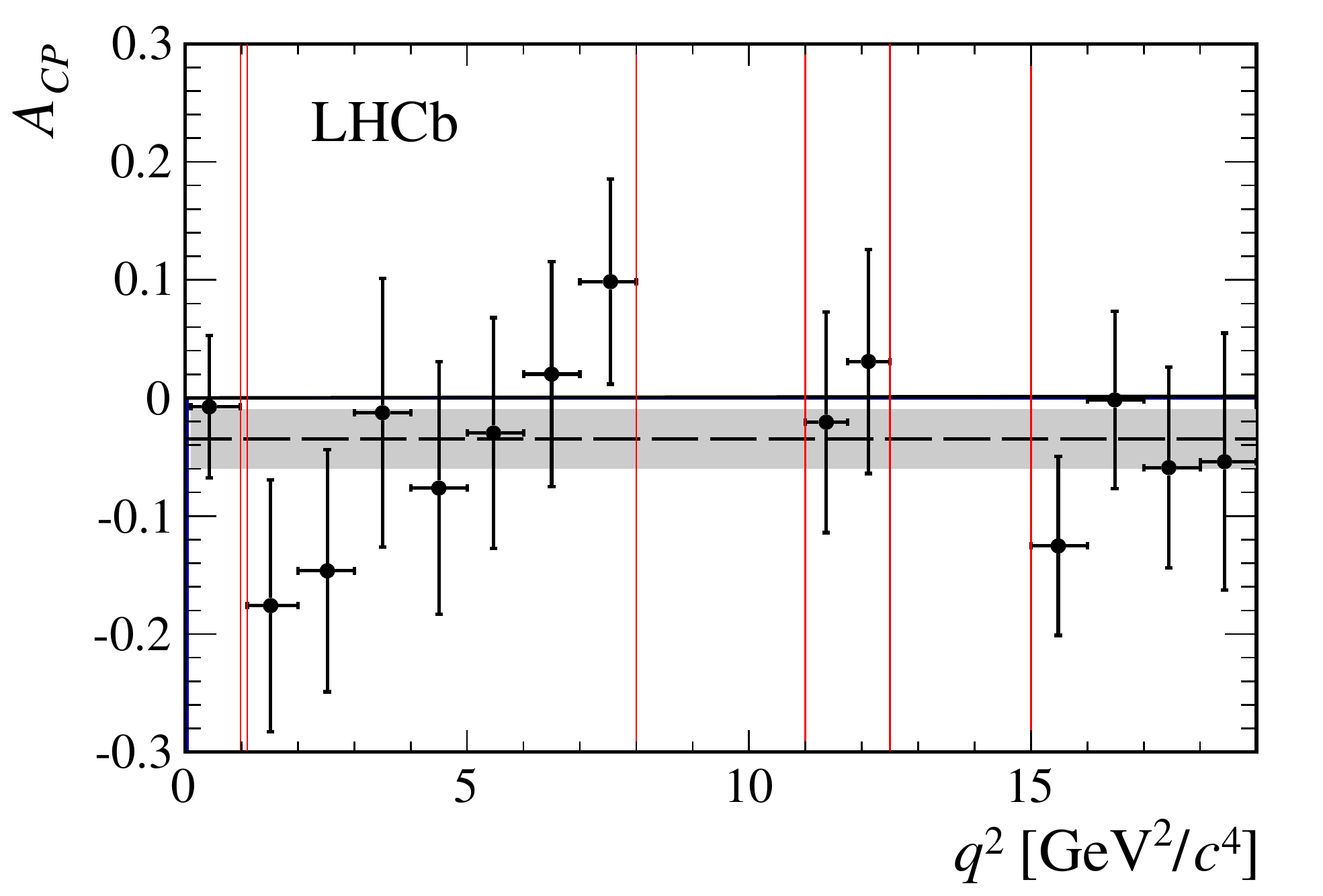}
  \caption{
    \CP asymmetries as a function of $q^2$ in (left) $\Bp \to \Kp\mumu$ and (right) $\Bz \to \Kstarz\mumu$ decays~\cite{LHCb-PAPER-2014-032}.
    The horizontal grey bands show the average over all $q^2$ bins.
    Vertical red lines indicate regions vetoed due to contributions from the $\phi$, $\jpsi$ and $\psi(2S)$ resonances.
  }
  \label{fig:acp-btosmumu}
\end{figure}

More precise results are available for exclusive decays, in particular in channels where large yields are available at LHCb.  
Among $b \to s\gamma$ decays, the most precise result is~\cite{HFAG,Aubert:2009ak,LHCb-PAPER-2012-019}
\begin{equation}
  {\cal A}_{\CP}(\Bz \to \Kstarz \gamma) = -0.002 \pm 0.015 \, .
\end{equation}
With the large yields available in some $b \to s\ell^+\ell^-$ modes, it is possible to study asymmetries as a function of $q^2$, the dilepton invariant mass squared.
This has been done, for example, for the rate asymmetries in $\Bp \to \Kp\mumu$ and $\Bz \to \Kstarz\mumu$ decays~\cite{LHCb-PAPER-2014-032}, as shown in Fig.~\ref{fig:acp-btosmumu}.
All results are consistent with the SM prediction of small \CP violation.
The yields available in the $\Bp \to \pip\mumu$ channel, which is mediated by the $b \to d\ell^+\ell^-$ transition and therefore may exhibit larger \CP violation in the SM, now allow a similar analysis to be performed~\cite{LHCb-PAPER-2015-035}.
The precision achieved is however not sufficient to observe a non-zero asymmetry.
Asymmetries in angular observables have also been measured in $\Bz \to \Kstarz\mumu$~\cite{LHCb-PAPER-2015-051} and $\Bs \to \phi\mumu$ decays~\cite{LHCb-PAPER-2015-023}, where in the latter case the analysis is currently limited to untagged asymmetries; all measured angular asymmetries are also consistent with zero.

The photon emitted in $b \to q \gamma$ transitions is expected to be almost fully polarised in the SM.
This feature suppresses the observability of \CP violation in mixing/decay interference, since the photons produced in, for example, $\Bzb \to \Kstarzb\gamma \to \KS\piz\gamma$ and $\Bz \to \Kstarz\gamma \to \KS\piz\gamma$ decays have predominantly left- and right-handed polarisations respectively, and therefore interference cannot occur.
This, however, provides an interesting possibility to test the SM prediction for the polarisation, since the magnitude of the parameter $S_f$ is reduced from that of Eq.~(\ref{eq:ACS-defs}) by a factor of $\sin 2 \Psi$, where $\tan \Psi$ is the relative magnitude of the suppressed and favoured polarisation amplitudes~\cite{Atwood:1997zr,Atwood:2004jj}.

Decays of the $\Bd$ meson by the $b \to s\gamma$ transition are ideal for such measurements, as the relatively large value of $\sin 2\beta$ would allow an observable effect in case $\sin 2 \Psi$ is non-zero.
Several measurements have been made, with the most precise results for $\Bd \to \KS\piz\gamma$~\cite{Ushiroda:2006fi,Aubert:2008gy} and $\Bz\to\KS\rhoz\gamma$~\cite{Li:2008qma,Sanchez:2015pxu}. 
All results to date are consistent with zero \CP violation and therefore with the SM prediction of strong polarisation of the emitted photon.

For \Bd meson decays through the $b \to d\gamma$ transition, the SM \CP violating phase is small and the oscillations are further suppressed by the photon polarisation, giving a vanishing value of $S_f$.
The $\Bd \to \rhoz\gamma$ mode therefore provides a null test of the SM, that is sensitive to models that introduce new sources of both right-handed currents and \CP violation.
The experimental sensitivity is, however, rather limited~\cite{Ushiroda:2007jf}.
A similar argument applies for $\Bs \to \phi \gamma$ decays, although in this case the $A^{\Delta\Gamma}$ observable provides additional sensitivity~\cite{Muheim:2008vu}.
Large $\Bs \to \phi \gamma$ yields are available~\cite{LHCb-PAPER-2012-019}, but no decay-time-dependent analysis has yet been performed.

As the available data samples increase, searches for \CP violation in mixing/decay interference, from decay-time-dependent analyses, will be possible in an increasing number of $b \to q\gamma$ and $b \to q\ell^+\ell^-$ transitions.
One particularly interesting possibility is to probe for non-SM \CP violation in the recently observed $\Bs\to\mumu$ decays~\cite{Chatrchyan:2013bka,LHCb-PAPER-2013-046,LHCb-PAPER-2014-049,Aaboud:2016ire}.
This will allow to probe several extensions of the SM that do not significantly change the value of the branching fraction~\cite{DeBruyn:2012wk,CERN-THESIS-2013-249}.

%% file: tree-penguin.tex
\section{\boldmath \CP violation in interference between tree and loop amplitudes}
\label{sec:tree-penguin}

There are several types of \B decays to hadronic final states where both tree and loop amplitudes are expected to contribute with comparable magnitudes.
These include decays mediated by the $b \to u\bar{u}s$ and $b \to u\bar{u}d$ quark-level transitions. 
The interference between tree and loop amplitudes causes \CP violation in decay to be expected in the SM, but its interpretation in terms of fundamental parameters is challenging due to hadronic uncertainties.
As will be shown, several approaches, often involving flavour symmetries, have been proposed to overcome this problem.

As mentioned in Sec.~\ref{sec:penguin-dominated}, decays mediated by $b \to d\bar{d}s$ and $b \to d\bar{d}d$ loop transitions often contribute to the same final states due to hadronisation effects, and therefore modes involving $\piz$, $\eta$, $\rhoz$ and $\omega$ mesons are relevant to this Section.
There is consequently an enormous number of different final states, of which only a selection will be discussed.
Full listings of all measurements, including several where the \CP\ asymmetries are reaching an interesting level of significance (\eg\ ${\cal A}_{\CP}(\Bp \to \eta\Kp)$) can be found in Refs.~\cite{HFAG,PDG2016}.

\subsection{Studies of \CP violation in decay in $b \to u\bar{u}q$ transitions}

Ever since the relative branching fractions of $\Bz \to \Kp\pim$ and $\pip\pim$ were first measured~\cite{Godang:1997we,CroninHennessy:2000kg} it was known that loop diagrams make sizable contributions to these decays.
However, it was not known if \CP violation effects would be large, as it remained possible that small strong phase differences could cause suppression; a range of values were predicted in explicit models of QCD effects in these decays~\cite{Keum:2000wi,Beneke:2001ev}.
The first measurement of \CP violation in \B meson decays, in the $\Bz \to \Kp\pim$ channel~\cite{Aubert:2004qm,Chao:2004mn} was therefore of great importance, bearing in mind that it occurred only a few years after the first observation of direct \CP violation in the kaon system~\cite{Fanti:1999nm,AlaviHarati:1999xp}.
The latest world average is~\cite{HFAG,Lees:2012mma,Duh:2012ie,Aaltonen:2014vra,LHCb-PAPER-2013-018}
\begin{equation}
  {\cal A}_{\CP}(\Bz \to \Kp\pim) = -0.082 \pm 0.006 \, .
\end{equation}

\begin{figure}[!b]
  \centering
  \includegraphics[width=0.65\textwidth]{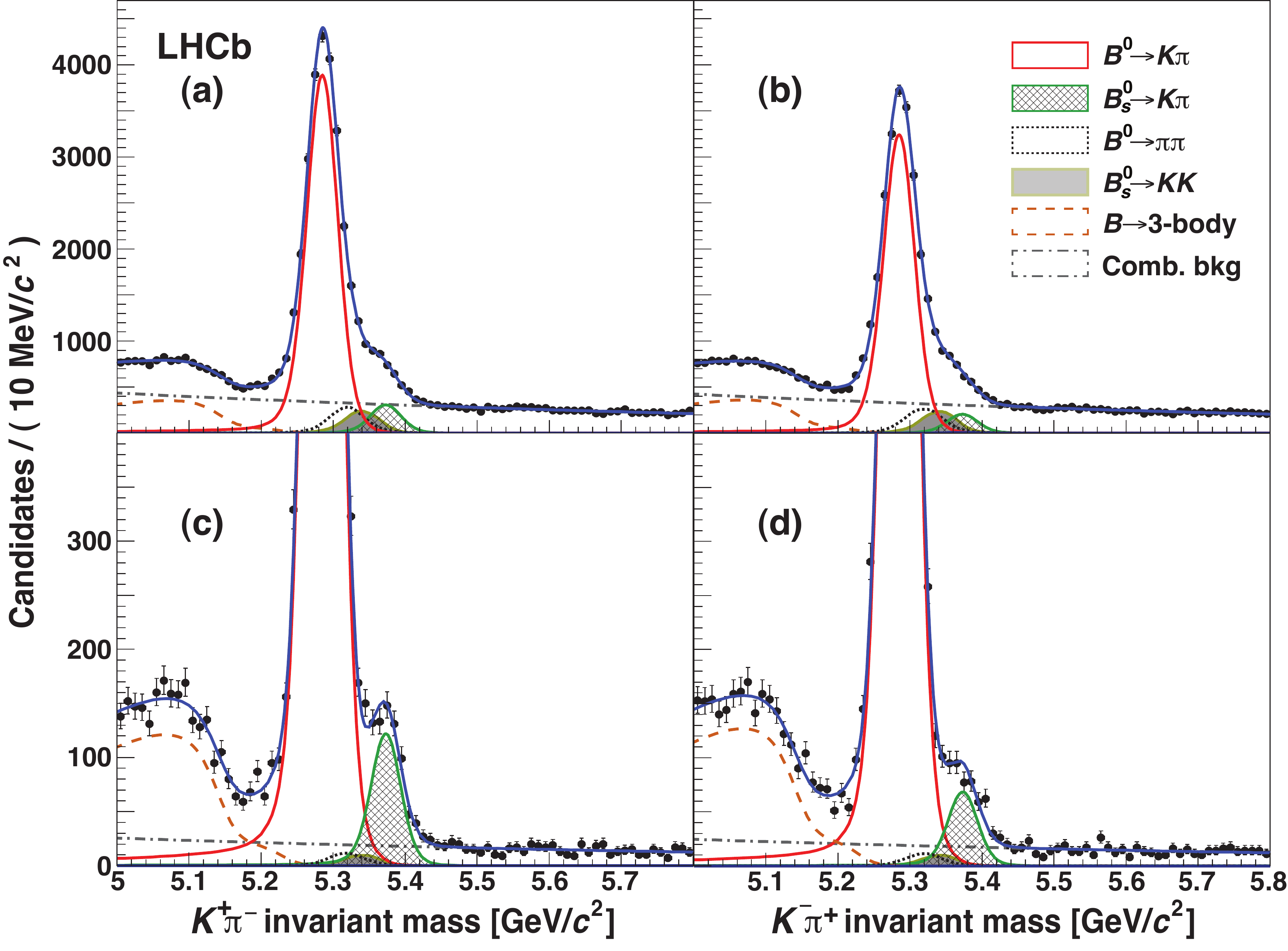}
  \caption{
    Measurements of ${\cal A}_{\CP}(\Bz \to \Kp\pim)$ and  ${\cal A}_{\CP}(\Bs \to \Km\pip)$~\cite{LHCb-PAPER-2013-018}.
    The final states are separated by charge: (a,c) $\Kp\pim$; (b,d) $\Km\pip$.
    The lower plots zoom in to allow better inspection of the \Bs region.
  }
  \label{fig:AcpKpi}
\end{figure}

The production of all types of \B meson in hadron collider experiments means that \CP violation effects can be studied simultaneously in \Bd and \Bs decays to the same final state.
Consequently, the first observation of \CP violation in \Bs decays~\cite{LHCb-PAPER-2013-018} was made together with the most precise determination of ${\cal A}_{\CP}(\Bz \to \Kp\pim)$, as illustrated in Fig.~\ref{fig:AcpKpi}.
Although in $pp$ collisions there may be production asymmetry effects that could differ for \Bd and \Bs mesons, these have been measured to be small~\cite{LHCb-PAPER-2014-042}.
The result is~\cite{HFAG,Aaltonen:2014vra,LHCb-PAPER-2013-018}
\begin{equation}
  {\cal A}_{\CP}(\Bs \to \Km\pip) = 0.26 \pm 0.04 \, .
\end{equation}
This is consistent with the SM prediction of approximately equal and opposite width differences between $\Bz \to \Kp\pim$ and $\Bs \to \Km\pip$~\cite{He:1998rq,Fleischer:1999pa,Gronau:2000md,Lipkin:2005pb}.

The difference $\Delta {\cal A}_{\CP}(\B \to K\pi) = {\cal A}_{\CP}(\Bz \to \Kp\pim)-{\cal A}_{\CP}(\Bp \to \Kp\piz)$ has been suggested as an interesting observable~\cite{Lin:2008zzaa}.
The measured value ${\cal A}_{\CP}(\Bp \to \Kp\piz) = 0.040 \pm 0.021$~\cite{Aubert:2007hh,Duh:2012ie} leads to 
\begin{equation}
  \Delta {\cal A}_{\CP}(\B \to K\pi) = -0.122 \pm 0.022 \, ,
\end{equation}
where possible correlations among systematic uncertainties have been neglected.
As seen in Fig.~\ref{fig:TandP}, the tree and penguin diagrams involved differ only by the spectator quark, and therefore if only these diagrams contribute then similar asymmetries, and $\Delta {\cal A}_{\CP}(\B \to K\pi) \approx 0$, would be expected.
Additional subleading amplitudes which contribute only to the $\Bp \to \Kp\piz$ decay and QCD corrections can, however, potentially explain the effect~\cite{Chiang:2004nm,Charng:2004ed,Chua:2007cm,Liu:2015upa}.

In order to investigate whether any non-SM physics causes the anomalous $\Delta {\cal A}_{\CP}(\B \to K\pi)$ value, a theoretically clean sum rule involving all four $\B \to K\pi$ decays has been proposed~\cite{Gronau:2008gu,Baek:2009hv}.
The precision of this approach is limited by the precision achieved with the $\Bz \to \KS\piz$ mode, where tagged analyses are necessary~\cite{Aubert:2008ad,Fujikawa:2008pk}.
Significant improvement can be expected with data from \belle~II.
Another interesting approach is to extend these measurements to final states involving pseudoscalar and vector mesons, \ie\ $B\to \Kstar \pi$ and $K \rho$.
Due to the non-negligible widths of the vector particles, such measurements must be made from Dalitz plot analyses, but large yields are available in most channels.
Several of the asymmetries have been measured~\cite{Garmash:2005rv,Aubert:2008bj,Aubert:2009me,Dalseno:2008wwa,BABAR:2011ae,BABAR:2011aaa,Lees:2015uun}, but the precision is not yet competitive with that in the $K\pi$ modes.

\begin{figure}[!tb]
  \centering
  \includegraphics[width=0.41\textwidth]{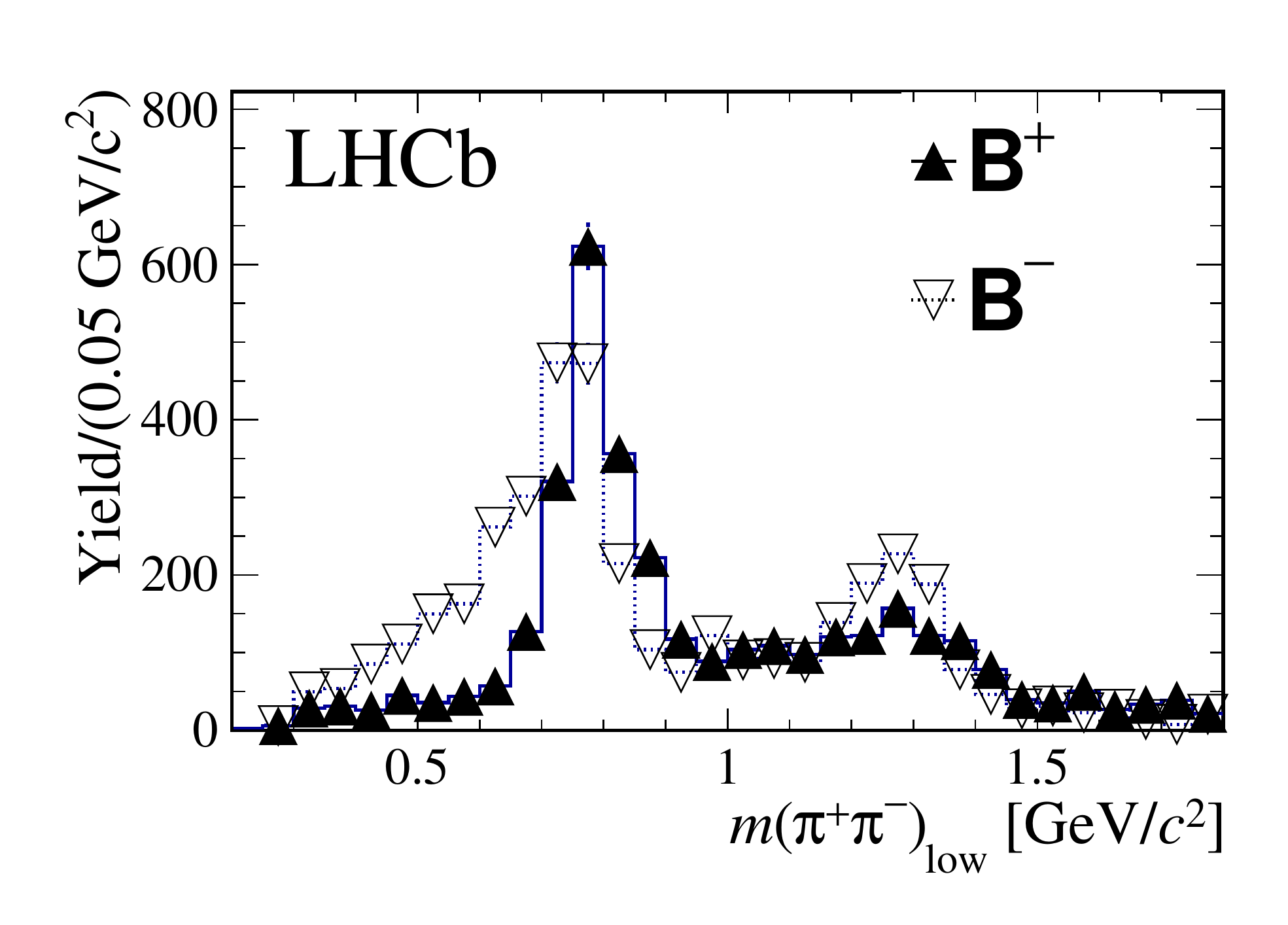}
  \includegraphics[width=0.41\textwidth]{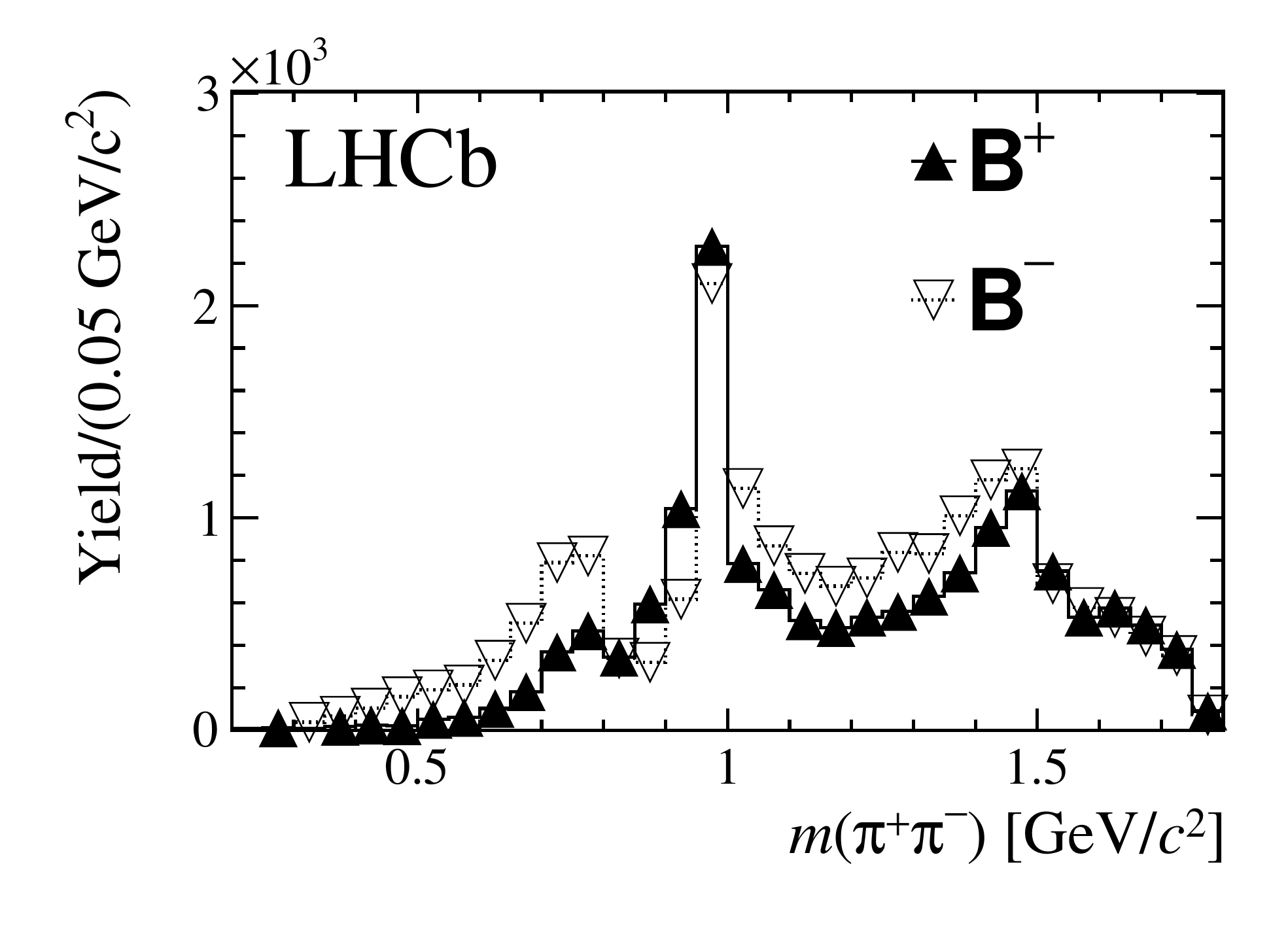}
  \includegraphics[width=0.41\textwidth]{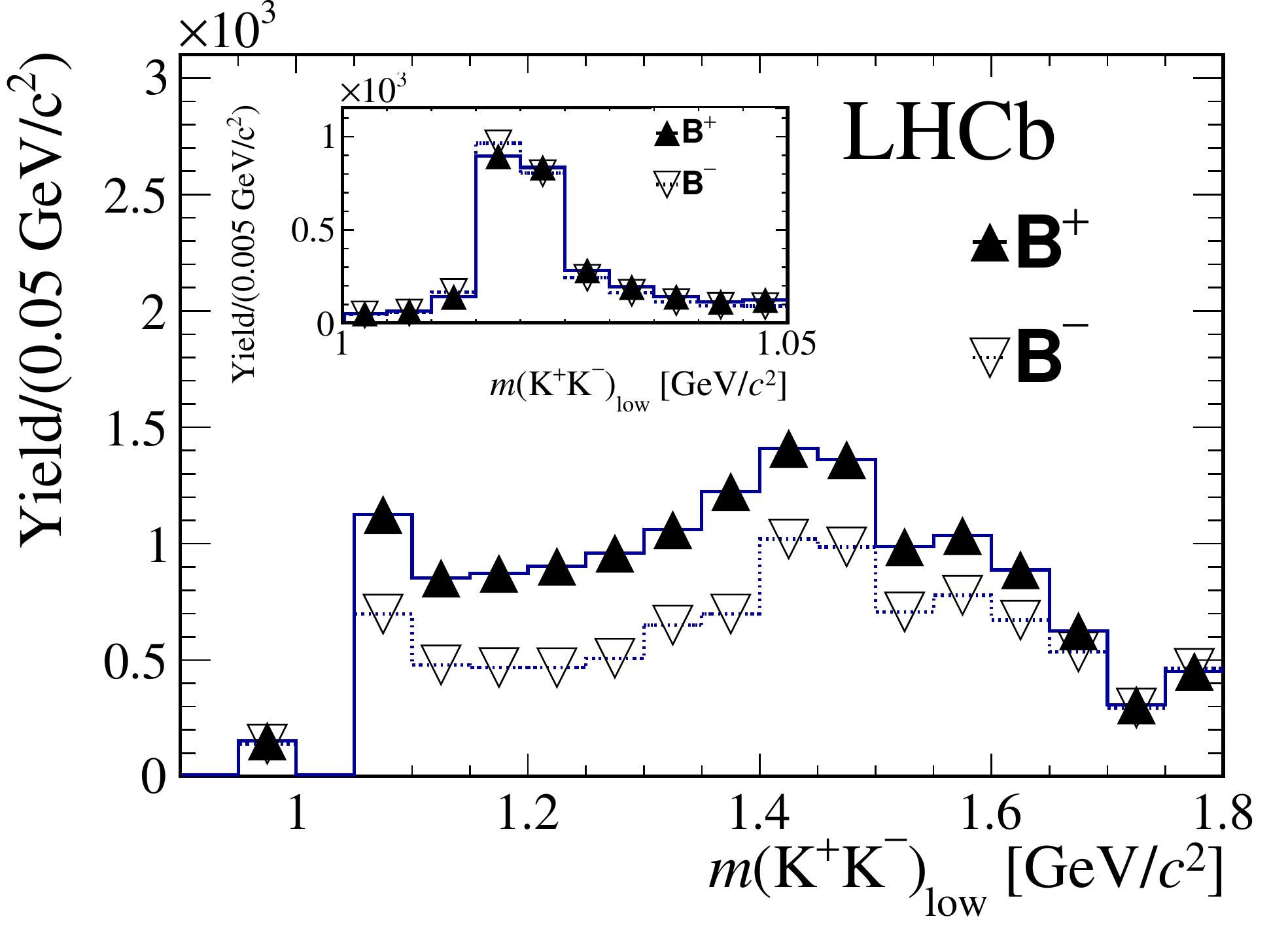}
  \includegraphics[width=0.41\textwidth]{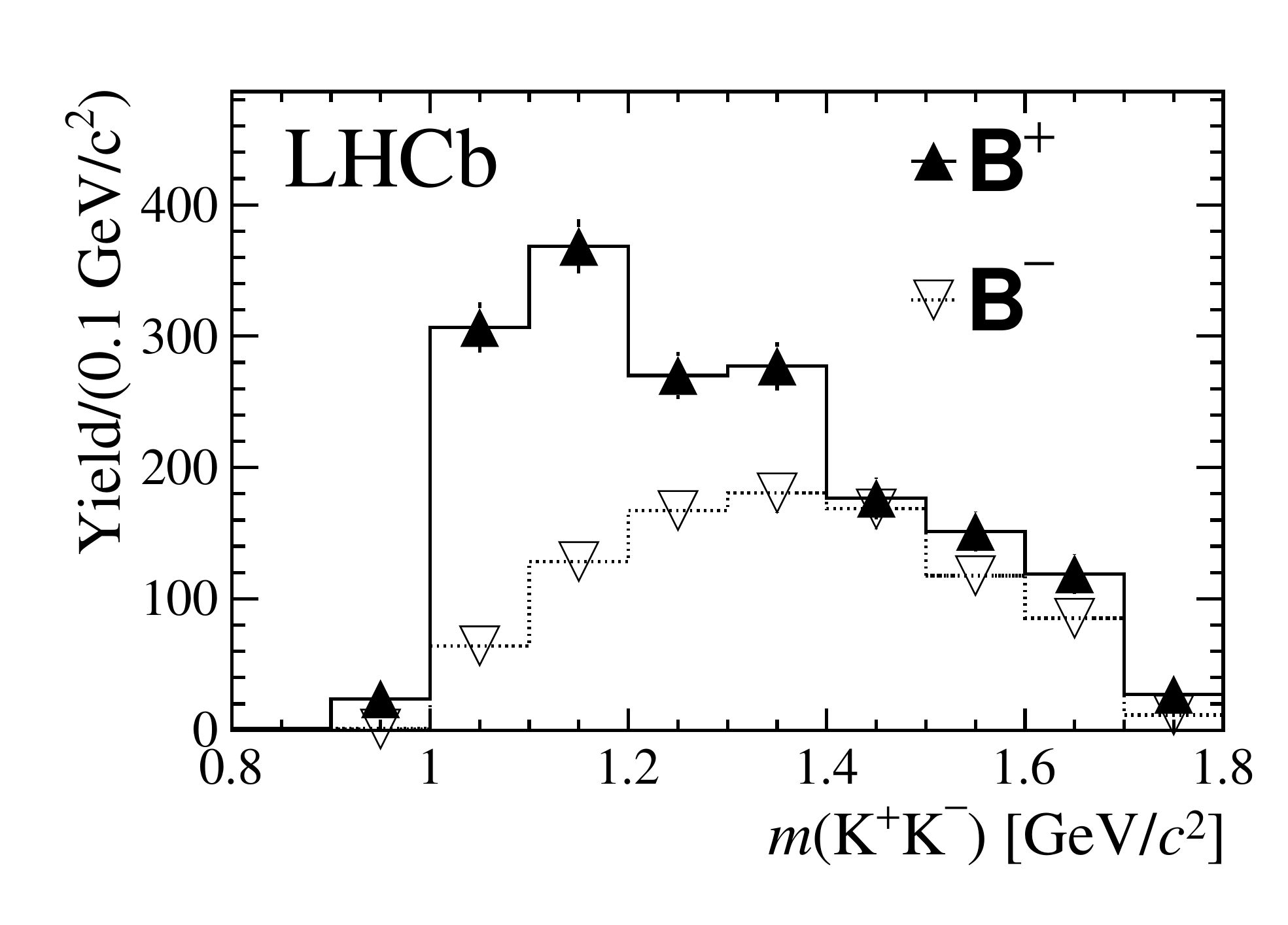}
  \caption{
    Yields of \Bp and \Bm decays as a function of two-body invariant mass in the (top left) $\pip\pim\pipm$, (top right) $\pip\pim\Kpm$, (bottom left) $\Kp\Km\Kpm$ and (bottom right) $\Kp\Km\pipm$ final states~\cite{LHCb-PAPER-2014-044}.
    In the first three plots, selections on the helicity angle have been imposed to enhance the observed \CP asymmetry.  
    The bottom left plot shows as an inset a zoom in the $\phi$ region.
  }
  \label{fig:Bto3h}
\end{figure}

To determine information about the \CP violation associated with particular resonant amplitudes, such as those for $B\to \Kstar \pi$ and $K \rho$ decays, it is necessary to perform a model-dependent amplitude analysis of the distribution of the three-body final state over the Dalitz plot. 
It is also possible to search for \CP violation model-independently, by comparing the binned distribution of events between \B and \Bbar decays~\cite{Aubert:2008yd,Bediaga:2009tr,Williams:2011cd,Bediaga:2012tm}.
(Similar methods have also been proposed for untagged decays~\cite{Burdman:1991vt,Gardner:2002bb,Gardner:2003su}.)
This has been done by LHCb for all of the $\Bpm \to \pip\pim\pipm$, $\pip\pim\Kpm$, $\Kp\Km\Kpm$ and $\Kp\Km\pipm$ decays~\cite{LHCb-PAPER-2013-027,LHCb-PAPER-2013-051,LHCb-PAPER-2014-044}, revealing very large \CP violation effects in some parts of the phase space as shown in Fig.~\ref{fig:Bto3h}.
The \CP violation is not localised around narrow resonances, but rather seems more pronounced in regions dominated by broad structures.
This presents a challenge for the interpretation, since the dynamical origin of the broad structures is not clear~\cite{Nogueira:2015tsa,Bediaga:2015mia}.
Detailed Dalitz plot studies will be necessary to gain deeper insights.
In addition to understanding the hadronic physics, it is of course interesting to investigate if the \CP violation is consistent with the SM, for which several methods based on flavour symmetries have been proposed~\cite{Ciuchini:2006kv,Ciuchini:2006st,Gronau:2006qn,Gronau:2007vr,Bediaga:2006jk,Gronau:2010dd,Gronau:2010kq,Imbeault:2010xg,Bhattacharya:2015uua}.

Similar analyses can also be performed for $\Bds \to \KS h^+h^{\prime -}$ decays, where $h^{(\prime)} = \pi,K$.
These provide additional interesting potential due to U-spin relations between \Bd and \Bs decays~\cite{Cheng:2014uga}, but the available yields are much smaller compared to modes with three charged tracks in the final state~\cite{LHCb-PAPER-2013-042}.
Ideally, such analyses should be complemented by results from the $\Bds \to h^+h^{\prime -}\piz$ and $\Bp \to \KS h^+\piz$ modes, but the reconstruction of neutral particles is challenging at a hadron collider and LHCb has not yet produced any results on these modes.
\belle~II, however, is expected to be able to study at least some of these channels.

\subsection{Studies of \CP violation in mixing/decay interference in $b \to u\bar{u}q$ transitions}

As noted above, hadronisation of states with $u\bar{u}$ content, produced from $b \to u\bar{u}s$ transitions, inevitably means that contributions from $b \to d\bar{d}s$ penguin amplitudes will also be present.  
Thus, results on modes such as $\Bd \to \KS\piz$, $\KS\rhoz$ and $\KS\omega$ are included in Fig.~\ref{fig:bsssbetaeff} in Sec.~\ref{sec:penguin-dominated:mix}.
The interpretation of these results in terms of CKM phases and potential non-SM amplitudes is not trivial due to the need to disentangle different SM contributions, but flavour symmetries can be exploited to reduce theoretical uncertainties~\cite{Fleischer:2008wb,Gronau:2008gu,Ciuchini:2008eh,Duraisamy:2009kb}.

In the case of $\Bs$ decays, $b \to u\bar{u}s$ transitions can lead to two types of final states: either $(s\bar{s})(u\bar{u})$, such as $\phi\piz$, or $(s\bar{u})(\bar{s}u)$, such as $\Kp\Km$.
In the former case the decays are isospin-violating and consequently both rare and sensitive to physics beyond the SM in electroweak penguin amplitudes~\cite{Hofer:2010ee}.
There are no experimental results on the $\Bs\to\phi\piz$ channel, but a recent amplitude analysis of $\Bs\to\phi\pip\pim$ decays by LHCb~\cite{LHCb-PAPER-2016-028} provides evidence for the $\Bs \to \phi\rhoz$ decay with a branching fraction of ${\cal O}(10^{-7})$.
Much larger data samples will be required for decay-time dependent analysis.

\begin{figure}[!tb]
  \centering
  \includegraphics[width=0.52\textwidth]{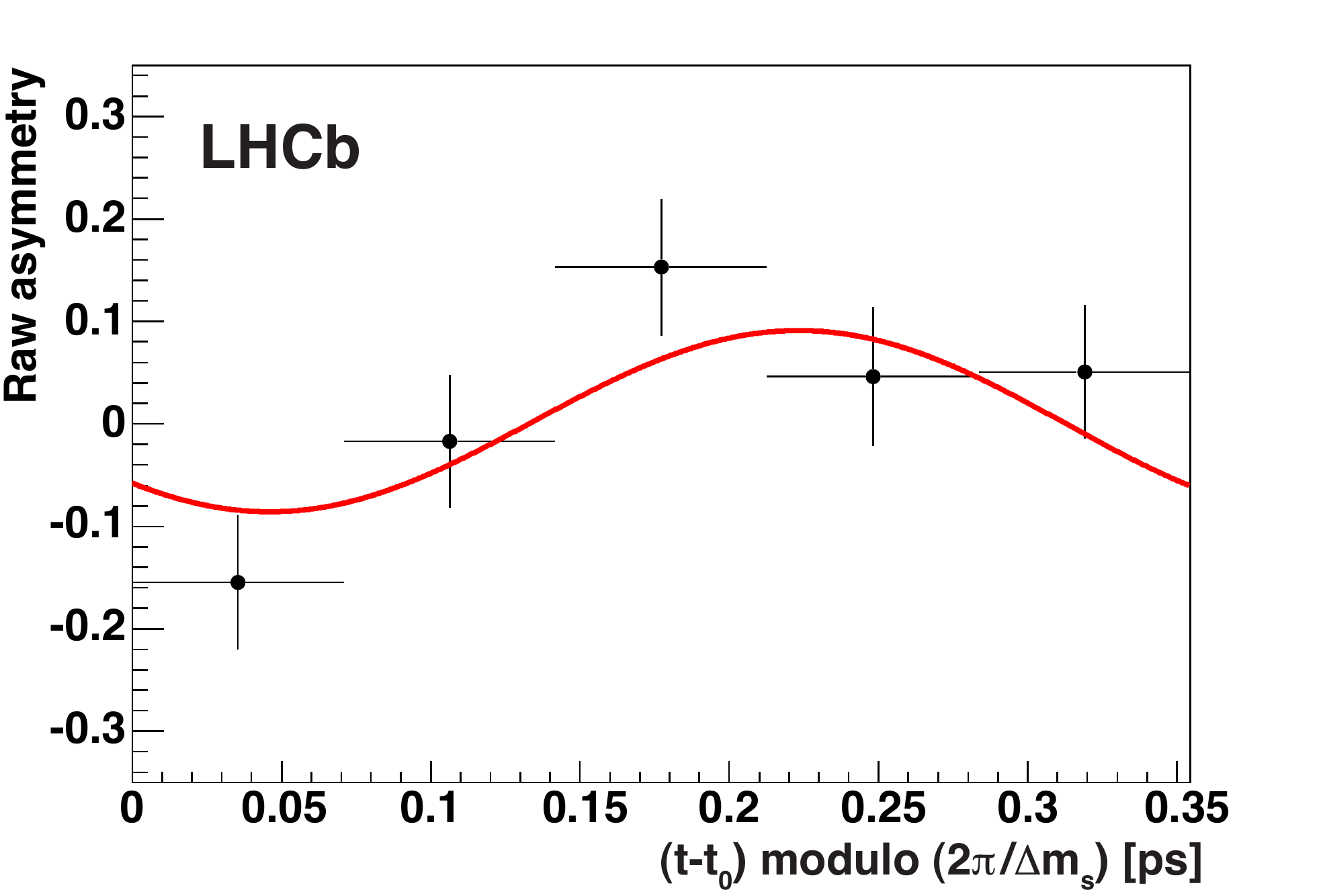} \\
  \includegraphics[width=0.42\textwidth]{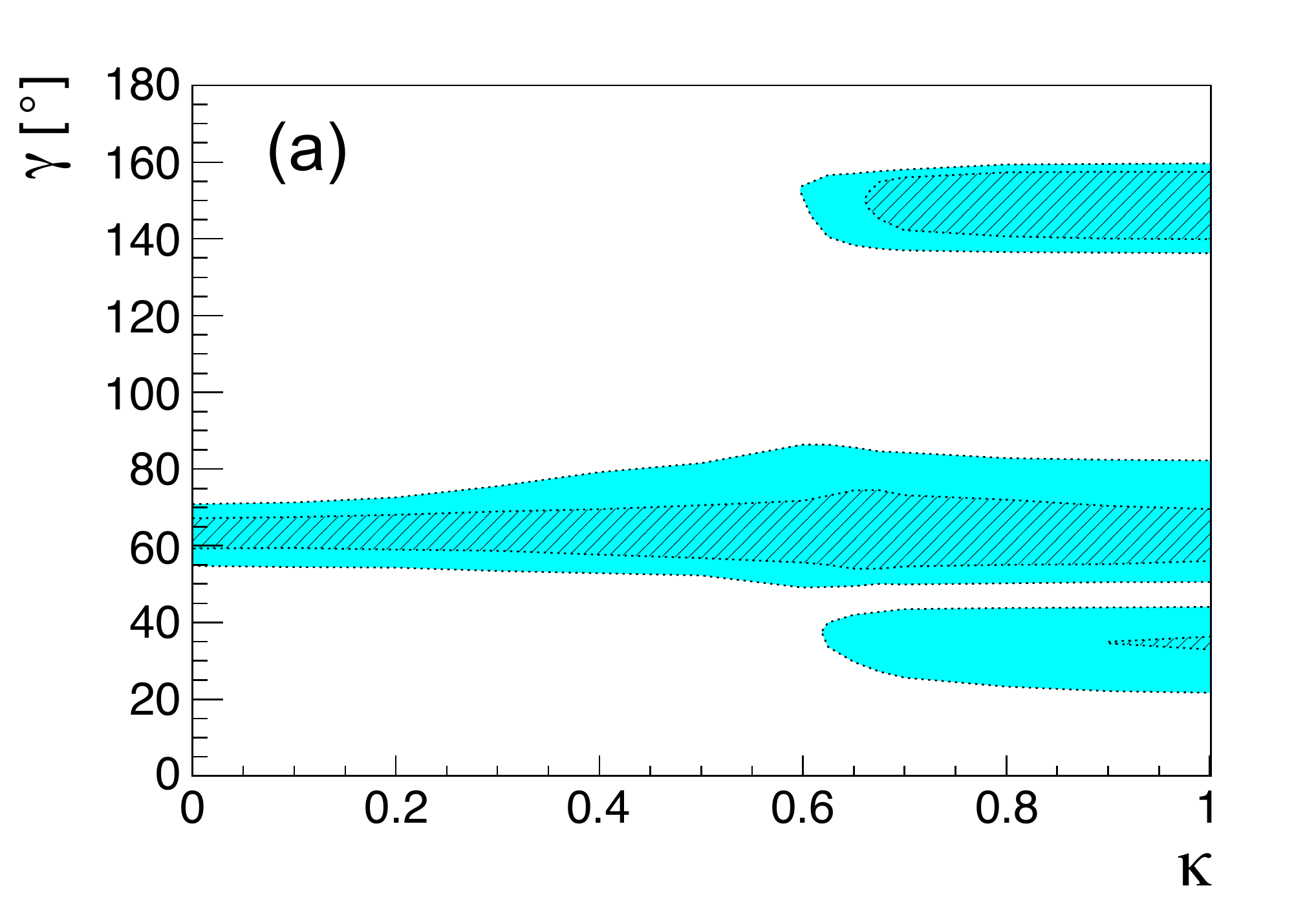}
  \includegraphics[width=0.42\textwidth]{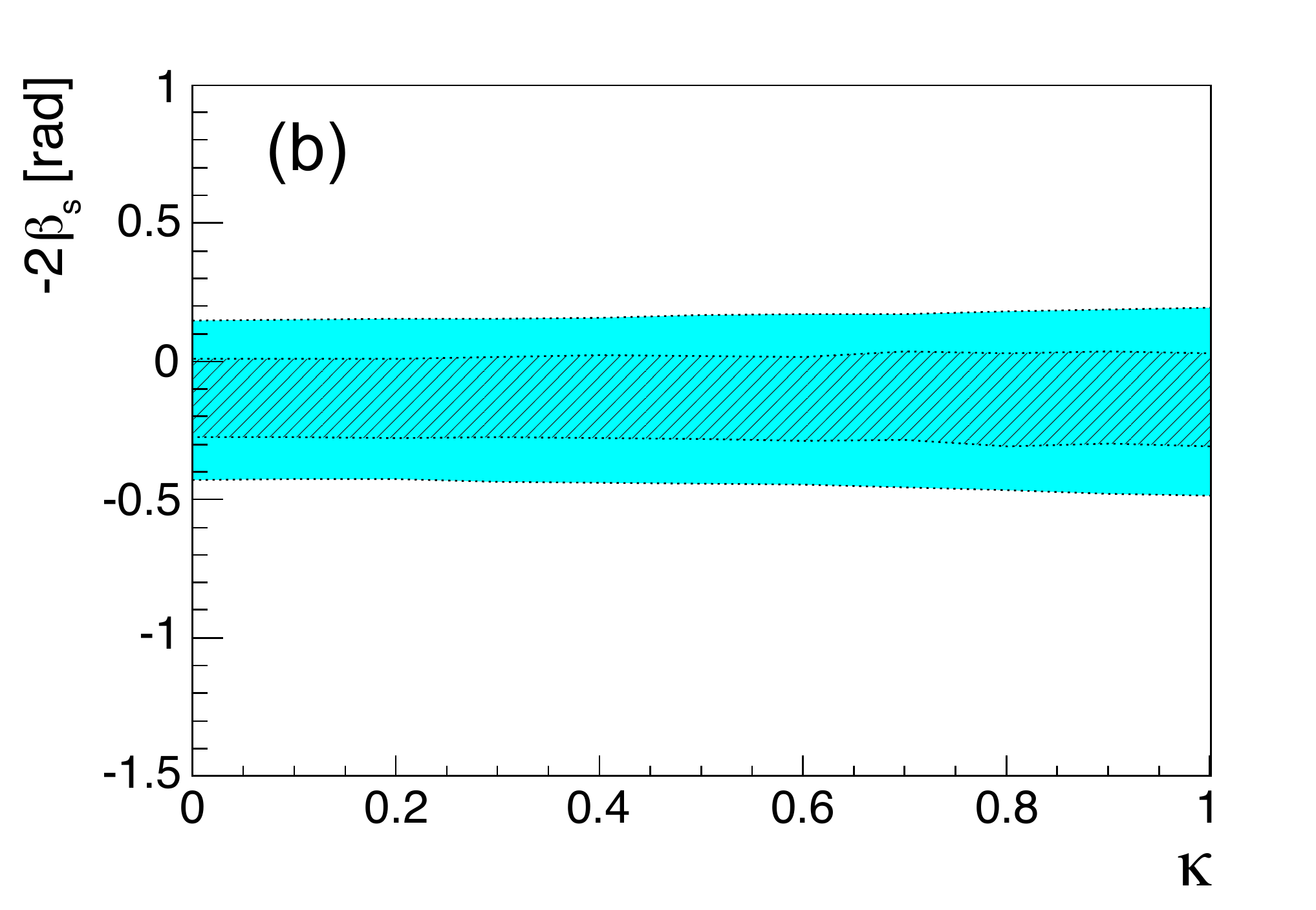}
  \caption{
    Asymmetries in the decay-time distributions, with fit results superimposed, for (top) $\Bs \to \Kp\Km$ decays~\cite{LHCb-PAPER-2013-040}, where the data have been folded to a single period of the oscillation.
    Constraints on (bottom left) $\gamma$ and (bottom right) $-2\beta_s$ obtained from $\Bs\to\Kp\Km$ and $\B \to \pi\pi$ decays as a function of the amount of non-factorisable U-spin breaking parametrised by $\kappa$~\cite{LHCb-PAPER-2014-045}.
  }
  \label{fig:KKpipi}
\end{figure}

The LHCb experiment is well-suited to study the $\Bs \to \Kp\Km$ decay.
The two-track final state allows high acceptance and trigger efficiency, leading to large yields.  
An analysis of the $1\invfb$ sample collected in 2011 gave the first measurement of the \CP violation parameters in this mode~\cite{LHCb-PAPER-2013-040}, as shown in Fig.~\ref{fig:KKpipi}: 
\begin{equation}
  C_{\Kp\Km} = 0.14 \pm 0.11 \pm 0.03 \, , \qquad
  S_{\Kp\Km} = 0.30 \pm 0.12 \pm 0.04 \, ,
\end{equation}
where the first uncertainties are statistical and the second systematic.
The constraint $|C_{\Kp\Km}|^2+|S_{\Kp\Km}|^2+|A_{\Kp\Km}^{\Delta \Gamma}|^2 = 1$ has been used.
An effective flavour tagging efficiency of $(2.45 \pm 0.25)\,\%$ was achieved.
It is likely that this can be improved in updated measurements based on larger data samples, and therefore there are good prospects for strong constraints to be obtained on $C_{\Kp\Km}$ and $S_{\Kp\Km}$, potentially leading to observations of \CP violation in decay and/or in mixing/decay interference in this channel.
In this context it is noteworthy that the largest sources of systematic uncertainty, due to the knowledge of the efficiency variation as a function of decay time and the decay time resolution function, are expected to be reducible.

If the contributions from the tree and penguin amplitudes to the $\Bs \to \Kp\Km$ decay can be disentangled, the results on $C_{\Kp\Km}$ and $S_{\Kp\Km}$ can be used to determine their relative weak phase, $\gamma$.
Alternatively, if the value of $\gamma$ from tree-dominated decays (Sec.~\ref{sec:gamma-tree-level}) is used as a constraint, a value of $-2\beta_s$ can be obtained.
There are not enough observables in the $\Bs \to \Kp\Km$ decay alone, but a U-spin relation with $\Bd \to \pip\pim$ decays that are discussed in the next subsection can be exploited~\cite{Fleischer:1999pa,Fleischer:2010ib,Ciuchini:2012gd}. 
Such an interpretation has been made by LHCb~\cite{LHCb-PAPER-2014-045}.
As shown in Fig.~\ref{fig:KKpipi}, it is found that the sensitivity to $\gamma$ depends strongly on the uncertainty assigned due to non-factorisable U-spin breaking effects.
The result for $-2\beta_s$, however, appears more robust to such effects, and therefore this method could be used to obtain a complementary and competitive determination to that from $\Bs \to \jpsi\phi$ and similar decays (Sec.~\ref{sec:psiphiandbuddies}).

As regards mixing/decay interference in $b \to u\bar{u}d$ transitions, the $\Bd$ decays are discussed in the next subsection.  
The $\Bs\to\KS\rhoz$ decay can in principle be studied at LHCb~\cite{LHCb-PAPER-2013-042}, but much larger data samples than currently available will be needed.
Other decays, such as $\Bs\to\KS\piz$, $\KS\eta^{(\prime)}$, $\KS\omega$, appear even less attractive.

\input{alpha}

%% file: alpha.tex
\subsection{Determination of CKM phases from isospin analysis of \CP violation in $b \to u\bar{u}d$ transitions}
\label{sec:alpha}

\begin{figure}[!tb]
  \centering
  \includegraphics[width=0.85\textwidth]{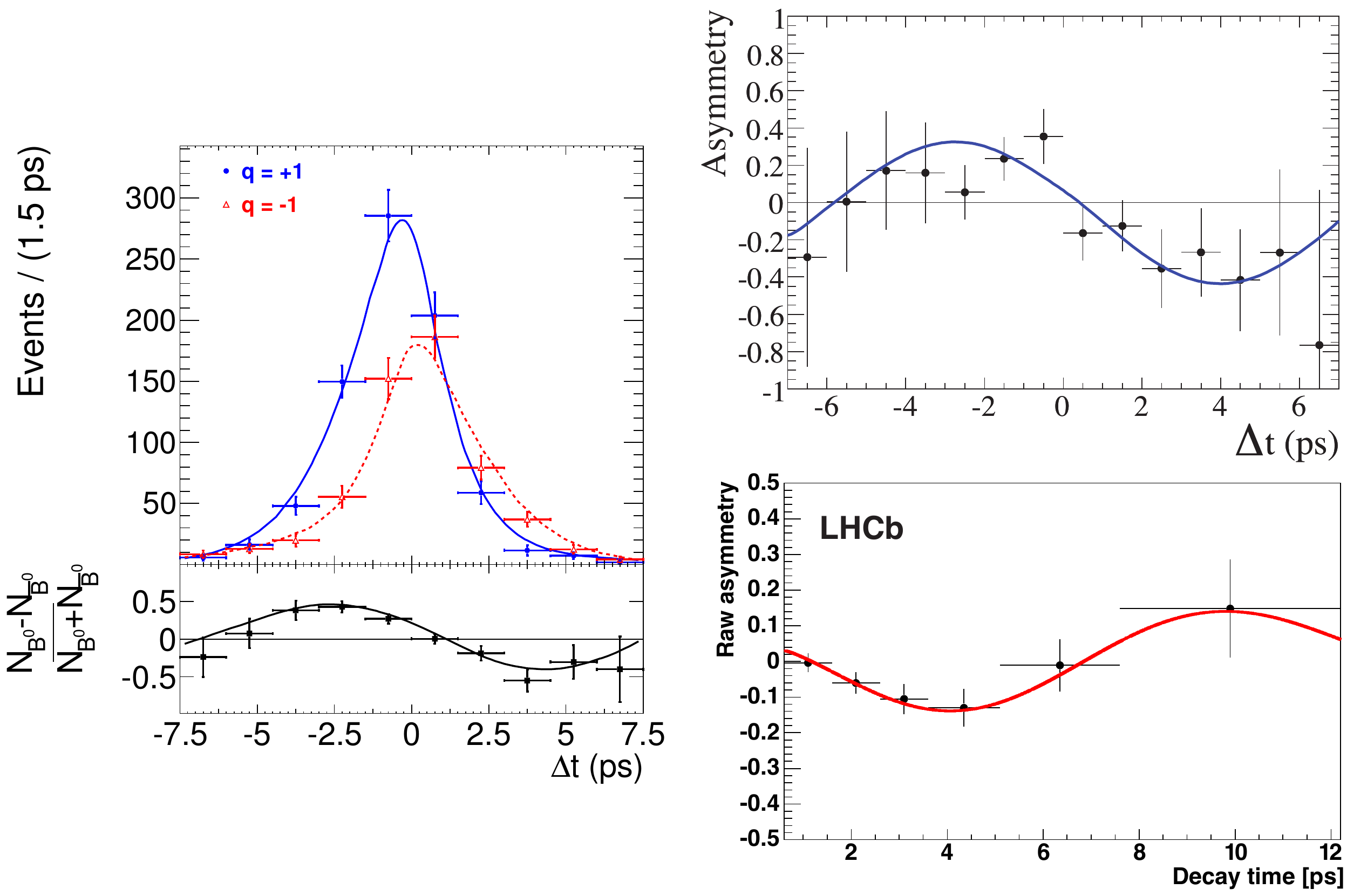}
  \caption{
    Asymmetries in the $\Bz \to \pip\pim$ decay-time distributions at (left) \belle~\cite{Adachi:2013mae}, (right top) \babar~\cite{Lees:2012mma}, and (right bottom) \lhcb~\cite{LHCb-PAPER-2013-040}, with fit results superimposed.
    The \belle plot includes also the background subtraction signal distributions separately for (blue) \Bz and (red) \Bzb tags.
  }
  \label{fig:pipi}
\end{figure}

Despite its low branching fraction, large yields of the $\Bz\to\pip\pim$ decay are available.
The \CP parameters have been measured from decay-time-dependent analyses by \babar, \belle and LHCb, as shown in Fig.~\ref{fig:pipi}.
The world averages for the observables are~\cite{HFAG,LHCb-PAPER-2013-040,Lees:2012mma,Adachi:2013mae}
\begin{equation}
  S_{\pip\pim} = -0.66 \pm 0.06 \, , \qquad
  C_{\pip\pim} = -0.31 \pm 0.05 \, .
\end{equation}
Thus, both \CP violation in decay and in mixing/decay interference have been observed in $\Bz\to\pip\pim$ decays.

If it were the case that $\Bz\to\pip\pim$ decays were tree-dominated, there would be no \CP violation in decay, hence $C_f = 0$, and the parameter of \CP violation in mixing/decay interference could be interpreted with low theoretical uncertainty as $S_f(\Bz\to\pip\pim) = +\eta_{\CP} \sin(2\alpha)$, where $\alpha$ is one of the Unitarity Triangle angles introduced in Sec.~\ref{sec:UT} and $\alpha \equiv \pi - \beta - \gamma$ by definition.
Loop diagrams, however, lead to so-called ``penguin pollution'' that makes such straightforward interpretation impossible.
The large measured magnitude of the value of $C_{\pip\pim}$ shows unambiguously that the penguin contribution is highly significant.
  
\begin{figure}[!tb]
  \centering
  \includegraphics[width=0.65\textwidth]{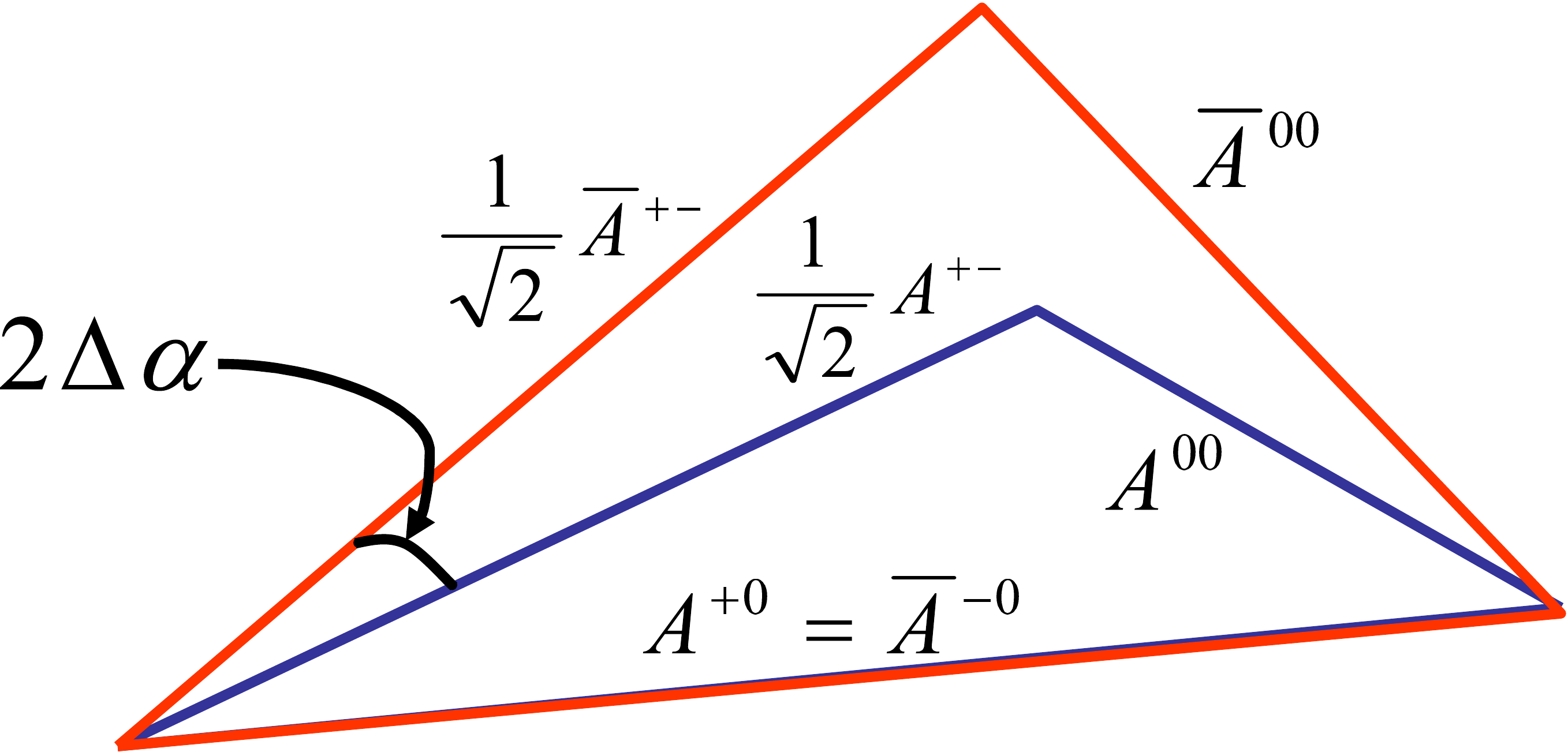}
  \caption{
    Isospin triangles for $\B\to\pi\pi$ decays~\cite{Gronau:1990ka}; reproduced from Ref.~\cite{Antonelli:2009ws}.
  }
  \label{fig:GLisospin}
\end{figure}

Isospin analysis of $\Bp \to \pip\piz$, $\Bz \to \pip\pim$ and $\Bz \to \piz\piz$ decays, and their charge conjugates, can be used to isolate, and correct for, the penguin pollution~\cite{Gronau:1990ka}.
The isospin decomposition leads to relations between the decay amplitudes (with obvious notation),
\begin{equation}
  \label{eq:GLisospin}
  A^{+0} = \frac{1}{\sqrt{2}} A^{+-} + A^{00} \, , \quad
  \bar{A}^{-0} = \frac{1}{\sqrt{2}} \bar{A}^{+-} + \bar{A}^{00} \, , \quad
  A^{+0} = \bar{A}^{-0} \, ,
\end{equation}
with small corrections possible due to isospin violation effects such as electroweak penguin contributions~\cite{Gronau:2005pq,Gardner:2005pq}.
These relations can be expressed as triangles that share a base, in which the phase between $A^{+-}$ and $\bar{A}^{+-}$, denoted $2\Delta \alpha$ in Fig.~\ref{fig:GLisospin}, is precisely the value needed to correct the value of $S_f$ and obtain $\alpha$,
\begin{equation}
  S_{\pip\pim} = \sqrt{1-C^2_{\pip\pim}} \sin \left(2\alpha - 2\Delta \alpha\right) \, .
\end{equation}

Thus, to determine $\alpha$ from $\B \to \pi\pi$ decays, the limiting factor becomes the knowledge of $\Delta \alpha$, which requires measurements of the branching fraction and \CP asymmetry parameters of the $\Bz\to\piz\piz$ decay. 
This mode is, unfortunately, difficult to study experimentally and its decay-time dependence cannot be determined due to the absence of a reconstructed vertex (see, however, Ref.~\cite{Ishino:2007pt}) which precludes a determination of $S_{\piz\piz}$.
The world averages~\cite{HFAG,Lees:2012mma,Vanhoefer:2014mfa} are 
\begin{equation}
  {\cal B}(\Bz \to \piz\piz) = (1.17 \pm 0.13) \times 10^{-6} \, , \ 
  {\cal A}_{\CP}(\Bz \to \piz\piz) = 0.43 \pm 0.27 \, .
\end{equation}
The quoted ${\cal A}_{\CP}(\Bz \to \piz\piz)$ average does not include results published by \belle in 2005~\cite{Abe:2004mp} since these are known to be affected by a source of background~\cite{Abe:2005bs} that was not accounted for in the analysis.
Nor is any scaling applied to the uncertainty due to the discrepancy between the \babar and \belle results for the branching fraction.

\begin{figure}[!tb]
  \centering
  \includegraphics[width=0.58\textwidth]{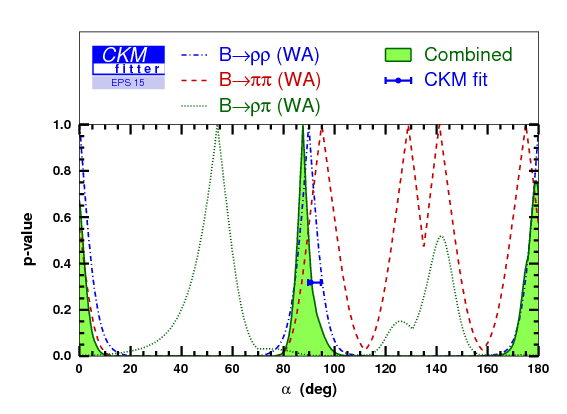}
  \caption{
    Constraints on the CKM angle $\alpha$ obtained from isospin analysis of (red) $\B\to\pi\pi$, (green) $\rho\pi$ and (blue) $\rho\rho$ decays~\cite{Charles:2004jd}.
    The combined result is $\alpha = (87.6\,^{+3.5}_{-3.3})^\circ$, consistent with the prediction from the CKM fit, $(90.6\,^{+3.9}_{-1.1})^\circ$, both of which are also shown.
  }
  \label{fig:alpha}
\end{figure}

Together with results on $\Bp\to\pip\piz$~\cite{Aubert:2007hh,Duh:2012ie}, constraints on $\alpha$ can be obtained, but with an eight-fold ambiguity due to the triangle relations.
This has been done by various groups~\cite{Charles:2004jd,Bona:2006ah,Bona:2007qta}; an example of the constraints obtained is shown in Fig.~\ref{fig:alpha}.
The ambiguities can be clearly seen, although it should be noted that the solution at $\alpha = 0$, which is allowed if only the isospin relations are used in the analysis, is excluded by the observation of \CP violation in decay in $\Bz \to \pip\pim$.
Imposing physical constraints on the hadronic parameters further restricts the allowed region near $\alpha = 0$~\cite{Antonelli:2009ws,Bona:2007qta}.

A similar isospin analysis can be performed for $\B \to \rho\rho$ decays.
In principle, matters are more complicated, as the final state involves two vector mesons, and the isospin triangles should be constructed for each polarisation state separately.  
However, it is observed that the decays are dominated by longitudinal polarisation~\cite{Aubert:2007nua,Vanhoefer:2015ijw}.
The finite width of the $\rho$ states leads to a further complication of the isospin analysis, but the effects are ${\cal O}\left(\Gamma/m\right)^2$ and hence small~\cite{Falk:2003uq}.
The world average values for the \CP violation parameters in $\Bz \to \rhop\rhom$ are~\cite{HFAG,Aubert:2007nua,Vanhoefer:2015ijw}
\begin{equation}
  S_{\rhop\rhom} = -0.14 \pm 0.13 \, , \qquad C_{\rhop\rhom} = 0.00 \pm 0.09 \, ,
\end{equation}
where the consistency of $C_{\rhop\rhom}$ with zero suggests smaller penguin pollution effects compared to the $\pi\pi$ case.
Indeed, the isospin triangles are found to be stretched and overlapping, which leads to good precision on $\alpha$, as seen in Fig.~\ref{fig:alpha}.
In this case, the isospin analysis benefits from the fact that the parameters of the $\Bz \to \rhoz\rhoz$ channel are more tractable for experimental study~\cite{Aubert:2008au,Adachi:2012cz,LHCb-PAPER-2015-006}, where existing results include a first measurement of $S_{\rhoz\rhoz}$, with large uncertainty~\cite{Aubert:2008au}.
However, improved determinations of the branching fraction and \CP asymmetry for the longitudinal component of $\Bp \to \rhop\rhoz$ decays~\cite{Zhang:2003up,Aubert:2009it} are needed in order to constrain further the $\B\to\rho\rho$ isospin triangles.  
Since the latest \belle result~\cite{Zhang:2003up} is based on analysis of around $10\,\%$ of the final \belle $\Upsilon(4S)$ sample, significant improvement should be possible.

Further channels involving only light non-strange mesons also have sensitivity to $\alpha$, but the isospin analysis tends to be more complicated~\cite{Lipkin:1991st}.
For example, results are available on decay-time-dependent \CP violation parameters in  $\Bz \to a_1^\pm \pimp$ decays~\cite{Aubert:2006gb,Dalseno:2012hp}, but these are often not used in global fits to obtain constraints on $\alpha$.
An exception occurs for $\B \to \rho\pi$ decays, where the interference of $\rhop\pim$, $\rhom\pip$ and $\rhoz\piz$ amplitudes in the $\Bz \to \pip\pim\piz$ Dalitz plot allows, in principle, penguin pollution to be corrected for and $2\alpha$ to be determined without ambiguity~\cite{Snyder:1993mx,Quinn:2000by}.
The relevant parameters have been measured in decay-time-dependent Dalitz plot analyses by \babar and \belle~\cite{Kusaka:2007dv,Kusaka:2007mj,Lees:2013nwa}; although the parameters associated with \CP violation in $\Bz \to \rhopm\pimp$ decays are determined quite precisely, giving evidence of \CP violation in decay, the obtained constraints on $\alpha$ are not yet competitive with those from other methods as seen in Fig.~\ref{fig:alpha}.

%% file: baryons.tex
\section{\boldmath \CP violation in decays of $b$ baryons}
\label{sec:baryons}

As baryons do not oscillate, only \CP violation in decay can occur, but no \CP violation has been observed in any $b$ baryon decay to date. 
Since the same classes of quark-level transitions as in the meson case are possible, it is likely that asymmetries of similar magnitude can arise. 
In search of a first observation of \CP violation in baryon decays, it is therefore of greatest interest to study charmless decays mediated by $b \to u\bar{u}s$ and $b \to u\bar{u}d$ transitions.
However, calculations of the expected effects are difficult due to hadronic uncertainties.

There has been much less experimental study of \CP violation in $b$ baryons than in $b$ mesons.
In part this is because many of the most accessible decays where \CP violation might be observed are experimentally more complex than their $b$-meson counterparts, and consequently there has been only limited exploitation of the significant samples of $b$-baryons available from the Tevatron and the LHC.
In particular, $b$-baryon production at the LHC is strongly biased towards low transverse momenta~\cite{LHCb-PAPER-2014-004}, which reduces the trigger and selection efficiencies compared to \B meson production. 

The lightest $b$ baryon is the \Lb ($udb$) state.
Searches for \CP violation in its decays to charmless two-body final states have been performed by CDF~\cite{Aaltonen:2014vra}, giving 
\begin{eqnarray*}
  A_{\CP}(\Lb \to p\pim) & = & +0.06 \pm 0.07 \stat \pm 0.03 \syst \, , \\
  A_{\CP}(\Lb \to p\Km)  & = & -0.10 \pm 0.08 \stat \pm 0.04 \syst \, .
\end{eqnarray*}
Within the current uncertainties, these results are compatible both with no asymmetry, and with the ${\cal O}(10\,\%)$ effects seen in charmless \B meson decays such as $\Bd \to \Kp\pim$, discussed in Sec.~\ref{sec:tree-penguin}.
LHCb should be able to improve on the precision, but the measurements require good knowledge of both \Lb production and $p/\bar{p}$ detection asymmetries and are therefore not trivial.

The search for \CP violation in $b$ baryons can be extended to three-body charmless hadronic decays.
These channels have the advantage that, if there is a component with an intermediate charmed hadron, it can be used to cancel the production and detection asymmetries.
Such searches have been carried out by LHCb with the $\Lb \to \KS p \pim$ $\Lz \Kp\Km$ and $\Lz\Kp\pim$ channels~\cite{LHCb-PAPER-2013-061,LHCb-PAPER-2016-004}, with the $\Lb \to \Lc \pim$, $\Lc \to \KS p$ and $\Lc \to \Lz\pip$ modes as reference.
The yields available are rather limited, due to the \KS\ and \Lz\ reconstruction efficiency at LHCb, and no significant asymmetry has yet been observed.

Since large \CP asymmetries have been seen in \B meson decays to three charged particles, it is natural to look for such effects in similar decays of $b$ baryons.
The weakly decaying charged $b$ baryons are the \Xibm ($bsd$) and \Omegabm ($bss$) states.
Charmless decays of these particles have not yet been observed, but if sufficient yields can be obtained they will provide good possibilities to search for \CP violation effects.

Larger yields are available for some four-body charmless hadronic \bquark-baryon decays such as $\Lb \to \proton\pim\pip\pim$ and $\proton\pim\Kp\Km$~\cite{LHCb-PAPER-2016-030}.
For these modes it is thus possible not only to determine phase-space integrated \CP asymmetries, but also to study the asymmetry in certain regions of the phase-space.
By comparing the yields in regions with the scalar triple product, constructed from the momenta of three final-state particles in the \Lb\ rest frame, either positive or negative, observables can be constructed that are robust against systematic uncertainties from production and detection asymmetries.
The measurement of such an observable for $\Lb \to \proton\pim\pip\pim$ decays reveals evidence, at the $3.3\,\sigma$ level, for a \CP violation effect~\cite{LHCb-PAPER-2016-030}.

Finally, $\Lb \to D\Lz$ decays can be used to determine the CKM angle $\gamma$ through \CP violation caused by the interference of the $b\to u$ and $b\to c$ tree-level transitions, in the same way as discussed in Sec.~\ref{sec:gamma-tree-level}.
This decay remains unobserved to date, so it is impossible to say whether it will make a useful contribution to the overall sensitivity to $\gamma$.
The related decay $\Lb\to \Dz p\Km$ has, however, been seen~\cite{LHCb-PAPER-2013-056}.
Further studies are needed to understand the contribution to the total rate from $p\Km$ resonances which could be used for a future measurement of $\gamma$.

%% file: fits.tex
\section{Global fits and future prospects}
\label{sec:fits}

As discussed throughout this review, studies of \CP violation in the \B sector can be carried out with decays mediated by many different quark transitions. 
A powerful way to probe the SM is through global fits which test the consistency of the different \CP violation measurements, both with each other and with \CP conserving quantities.
Since studies of \CP violation in the \B sector allow all three angles of the Unitarity Triangle to be measured, and the two non-trivial lengths of sides can also be determined, the apex of the triangle, located at $(\rhobar,\etabar)$ as defined in Eq.~(\ref{eq:rhoetabar}), can be over-constrained.
Each measurement defines an allowed area, at a given confidence level, in the $(\rhobar,\etabar)$ plane; within the SM all these areas must overlap at the true position of the apex of the Unitarity Triangle.
Any inconsistency between measurements would therefore indicate physics beyond the SM.

\subsection{Global fits to CKM matrix parameters}
\label{sec:fits-now}

There are two well-established collaborations who perform global fits to the CKM matrix parameters, differing in their preferred statistical treatment. 
(Some other analyses have also been performed, \eg\ those of Refs.~\cite{Lunghi:2008aa,Eigen:2013cv}, but will not be discussed in detail here.)
The CKMfitter~\cite{Charles:2004jd} collaboration uses a frequentist approach, while the UTFit~\cite{Bona:2006ah} collaboration uses a Bayesian method. 
Both collaborations include the latest measurements of individual observables, as well as theory input, for example from lattice QCD calculations~\cite{Aoki:2013ldr}, required to interpret the experimental observables in terms of the CKM matrix elements. 
The fit tests the overall consistency of the available results.

\begin{figure}[!tb]
\centering
\includegraphics[width=0.49\textwidth]{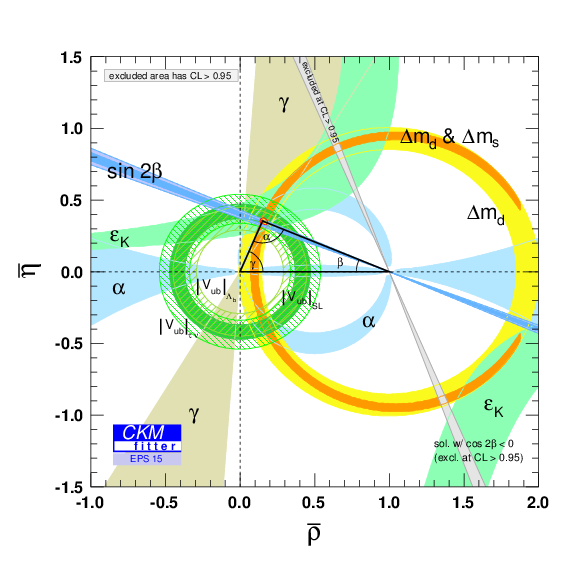}
\includegraphics[width=0.49\textwidth]{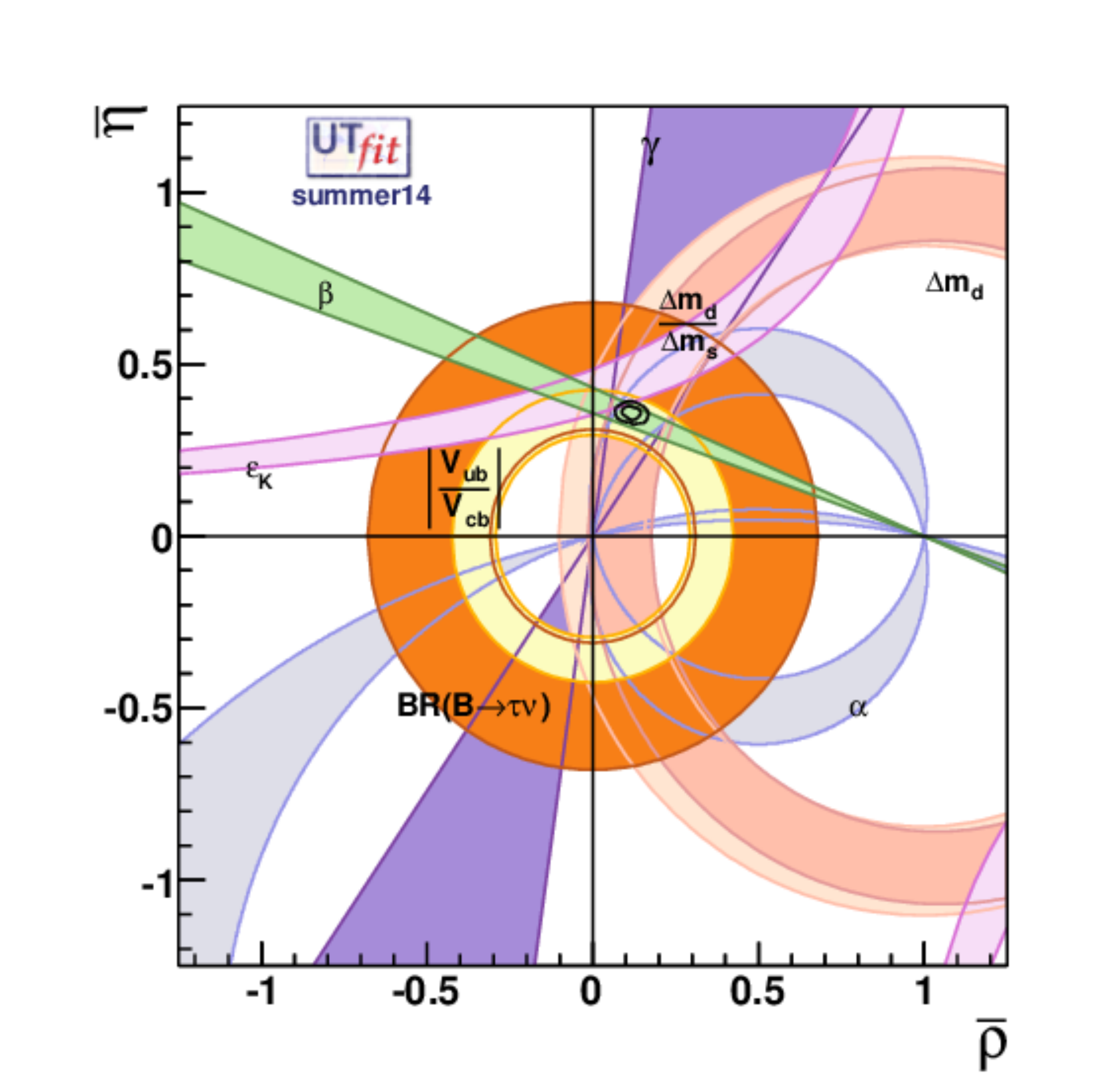}
\caption{
  Results of the latest (left) CKMfitter~\cite{Charles:2004jd} and (right) UTFit~\cite{Bona:2006ah} global fits to the CKM matrix parameters, showing good agreement with the SM picture of quark transitions. 
}
\label{fig:ckmfitterutfit}
\end{figure}

Typically, only the experimentally most precise and theoretically cleanest observables are included.
Among the observables related to \CP violation in the \B system, these include $\beta$ and $\beta_s$ from $b \to c\bar{c}s$ transitions (see Sec.~\ref{sec:ccq}), $\gamma$ from interference between $b \to c\bar{u}s$ and $b \to u\bar{c}s$ transitions (see Sec.~\ref{sec:gamma-tree-level}) and $\alpha$ from $b \to u\bar{u}d$ transitions (see Sec.~\ref{sec:alpha}).
The lengths of the two sides of the Unitarity Triangle are obtained from the relative magnitudes of CKM matrix elements $\left| V^{}_{ub} / V^{}_{cb} \right|$, determined from semileptonic \B decays~\cite{KowalewskiMannel}, and $\left| V^{}_{td} / V^{}_{ts} \right|$, determined from neutral \B meson oscillation frequencies~\cite{Schneider}; in both cases calculations of relevant hadronic quantities are also needed.
The parameter of \CP violation of $\Kz$--$\Kzb$ mixing, $\epsilon_K$, is also usually included and provides a constraint in the $(\rhobar,\etabar)$ plane in the shape of a hyperbola~\cite{Buchalla:1995vs}.
The results of the most recent analyses from the CKMfitter and UTfit collaborations are reproduced in Fig.~\ref{fig:ckmfitterutfit}; both show excellent consistency with the SM picture.
It is worth noting, however, that another analysis~\cite{Blanke:2016bhf} using more recent lattice QCD calculations of hadronic parameters relevant to $\Bds$--$\Bdsb$ mixing~\cite{Bazavov:2016nty}, indicates some tension in the fit.

The agreement with the SM can be quantified by excluding direct measurements of one of the parameters as inputs to the fit, and comparing the prediction for that quantity, obtained from the fit results, to the measured value.
Comparison of the predicted and measured values of the angle $\gamma$ is particularly interesting, as the fit is dominated by quantities determined from loop-level processes, while the measurement is obtained from tree-level decays.
Therefore, an inconsistency could be interpreted as a sign of physics beyond the SM. 
The latest predictions for $\gamma$ from the fits performed by the CKMfitter and UTFit collaborations are~\cite{Charles:2004jd,Bona:2006ah}
\begin{equation}
  \gamma^{\rm CKMfitter} = (66.9\,^{+0.9}_{-3.4})^\circ \, , \qquad
  \gamma^{\rm UTFit}  = (69.5 \pm 3.9)^\circ \, ,
\end{equation}
and are consistent with the measurements given in Eq.~(\ref{eq:gammaAverages}).

\subsection{Constraints on physics beyond the Standard Model}
\label{sec:BSMconstraints}

\begin{figure}[!tb]
  \centering
  \includegraphics[width=0.49\textwidth]{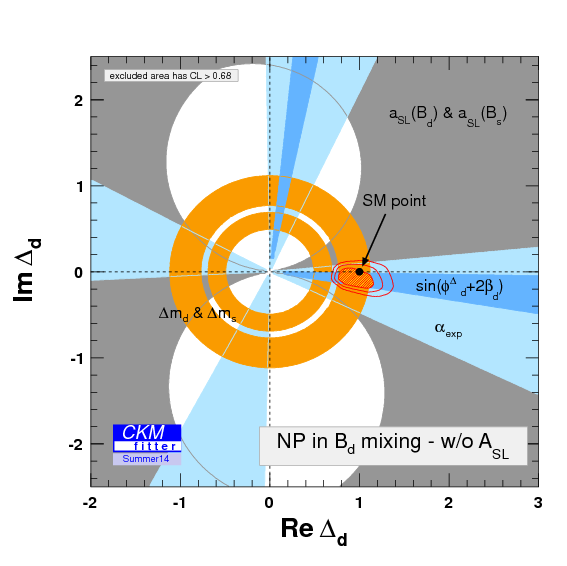}
  \includegraphics[width=0.49\textwidth]{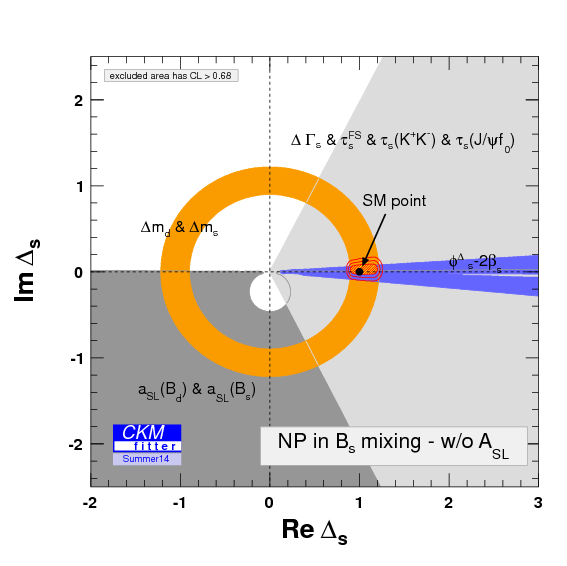}
  \caption{
    Constraints on contributions from physics beyond the SM to (left) $\Bz$--$\Bzb$ and (right) $\Bs$--$\Bsb$ mixing~\cite{Charles:2004jd}.
    The inclusive dimuon asymmetry result from the \dzero collaboration~\cite{Abazov:2013uma} (see Sec.~\ref{sec:SL}) is not included.
  }
  \label{fig:BSMconstraints}
\end{figure}

The large number of constraints on the parameters of the CKM matrix allows more sophisticated analyses to be performed.
For example, under the assumption that there is no contribution from physics beyond the SM to tree-level decay amplitudes, the consistency of the different measurements allows constraints to be put on possible beyond SM contributions~\cite{Buras:2000dm,Bona:2006sa,Bona:2007vi,Lenz:2010gu}.
The results of such an analysis~\cite{Lenz:2012az} are shown in Fig.~\ref{fig:BSMconstraints}.
The results are presented in terms of the real and imaginary parts of $\Delta_d$ and $\Delta_s$, which are the amplitudes for $\Bd$--$\Bdb$ and $\Bs$--$\Bsb$ mixing, respectively, normalised to their SM expectations; thus, the SM point is at $(1,0)$.
The consistency with the SM is again evident, and can now be interpreted as giving constraints on additional contributions to the amplitudes $\Delta_{d,s}$ at the $5$--$10\,\%$ level, with similar precision in both $\Bd$--$\Bdb$ and $\Bs$--$\Bsb$ systems.
Consequently, as discussed for example in Ref.~\cite{Isidori:2010kg}, if 
\begin{equation}
  \Delta_{d,s} = \Delta_{d,s}^{\rm SM} + \Delta_{d,s}^{\rm BSM} \, , 
  \ {\rm and} \ 
  \Delta_{d,s}^{\rm BSM} = c_{d,s}^{\rm BSM} / \left( \Lambda_{d,s}^{\rm BSM} \right)^2 \, ,
\end{equation}
then if the beyond SM coupling $c_{d,s}^{\rm BSM}$ is SM-like, \ie\ with loop and CKM-suppression as in scenarios referred to as ``minimal flavour violation''~\cite{D'Ambrosio:2002ex}, then the scale $\Lambda_{d,s}^{\rm BSM}$ of the additional contribution must be a factor of at least a few above the SM scale.  
On the other hand, in models where additional contributions enter with ${\cal O}(1)$ couplings, the scale must be ${\cal O}(10 \tev)$ or higher.

\subsection{Future prospects}
\label{sec:future}

Global fits to the CKM matrix parameters have established that the SM picture of \CP violation holds at the $5$--$10\,\%$ level, and the challenge in the coming years will be to test this picture at the percent level. 
This necessitates an order of magnitude improvement in experimental precision as well as strict control of theoretical uncertainties that are associated with the interpretation of experimental observables in terms of CKM matrix parameters. 
The already approved \belle~II~\cite{Aushev:2010bq,Abe:2010gxa} and upgraded \lhcb~\cite{LHCb-TDR-012,LHCb-PAPER-2012-031} experiments will deliver the necessary increase in recorded luminosity by the the mid-2020s, while \atlas and \cms are expected to continue to contribute a few important measurements, in particular with decays such as $\Bs\to\jpsi\phi$ which have high trigger efficiencies.
There are also good prospects for further reduction of uncertainties associated to quantities obtained from lattice QCD calculations~\cite{Pena:2016xww}.

\begin{figure}[!htb]
\centering
\includegraphics[width=0.325\textwidth]{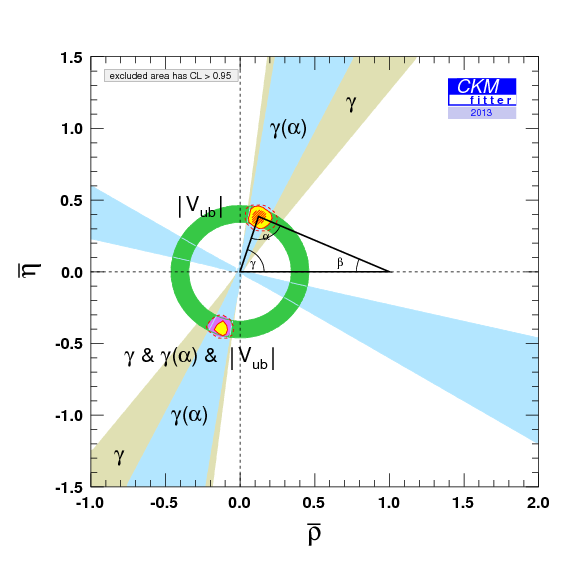}
\includegraphics[width=0.325\textwidth]{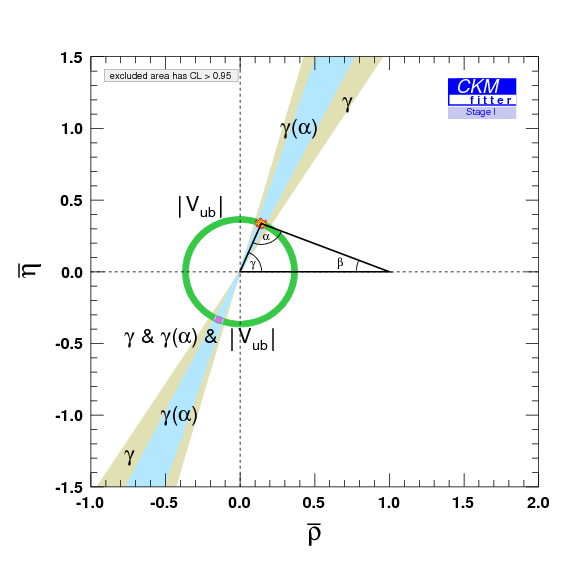}
\includegraphics[width=0.325\textwidth]{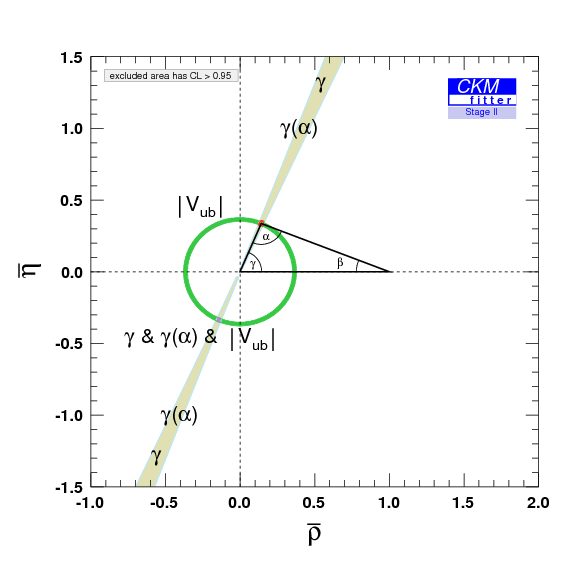}
\caption{
  Anticipated evolution of the fit to the CKM matrix using tree-level processes only, from (left) 2013 to projections for (middle) 2018 and (right) 2023~\cite{Charles:2013aka}.
  Constraints from $\left| V^{}_{ub} \right|$ from rates of semileptonic \B decays, $\gamma$ from interference between $b \to c\bar{u}s$ and $b \to u\bar{c}s$ transitions and $\alpha$ from $b \to u\bar{u}d$ transitions are shown, where the latter is interpreted as a constraint on $\gamma$ using $\alpha \equiv \pi - \beta - \gamma$ together with the measurement of $\beta$ from $b \to c\bar{c}s$ transitions.
  The limitations of each of these quantities is expected to come from different sources: for $\left| V^{}_{ub} \right|$ from the precision of lattice QCD calculations; for $\alpha$ from necessary approximations in the isospin analysis; for $\gamma$ from experimental uncertainty.
  }
\label{fig:ckmevol}
\end{figure}

The expected evolution of the constraint on the apex of the Unitarity Triangle determined from tree-level processes only~\cite{Charles:2013aka} is illustrated in Fig.~\ref{fig:ckmevol}. 
The high precision that will be achieved will not only lead to much greater sensitivity to effects of physics beyond the SM in $\Bd$--$\Bdb$ and $\Bs$--$\Bsb$ mixing, but the larger number of \CP violation parameters that will be measured will allow to search also for non-SM effects in the decay amplitudes.
Comparisons of the types of measurements mediated by different quark-level transitions, as outlined in Secs.~\ref{sec:tree-dominated}--\ref{sec:tree-penguin}, will enable the CKM paradigm to be tested at unprecedented precision.

While measurements of \CP violation in the \B sector form an important part of this programme, it must be stressed that there are many additional observables that provide complementary sensitivity to the parameters of the CKM matrix.
In addition to the rates of semileptonic decays and of neutral \B meson oscillations, mentioned above, important constraints can also be obtained from rare \B meson decays such as $\Bds \to \mumu$~\cite{Blake:2015tda}. 
The rare kaon decays, $\Kp \to \pip \nu\bar{\nu}$ and $\KL \to \piz \nu\bar{\nu}$, also provide constraints on the apex of the Unitarity Triangle~\cite{Bryman:2011zz,Cirigliano:2011ny}.
Results on both modes are anticipated to be forthcoming in the next few years~\cite{NA62-10-07,Yamanaka:2012yma}, and will greatly add to the global fit to the CKM matrix parameters.

%% file: summary.tex
\section{Summary}
\label{sec:summary}

The wealth of experimental results on \CP violation in the \B system, primarily  from the \babar, \belle\ and \lhcb\ experiments, have transformed the understanding in this sector.
The Kobayashi-Maskawa mechanism, that provides a unique source of \CP violation within the SM, has been confirmed.
Moreover, the consistency of measurements with SM predictions leads to increasingly strong constraints on effects beyond the SM.

This remarkable progress has been achieved due to the development of new techniques in both experimental and theoretical methods.
Since the currently achieved precision is far from the level at which these approaches are expected to reach fundamental limitations, significant further progress can be anticipated.
A next generation of experiments is planned, and there are exciting prospects for significant advances in this sector in the next ten years. 
Some of these, such as the observation of \CP violation in $b$ baryon decays, and of \CP violation in mixing/decay interference in the \Bs system, are anticipated within the SM.
For some others it remains to be seen whether or not the effects seen in the data can be explained by SM dynamics alone: these include the $K\pi$ puzzle and the large asymmetries seen in regions of phase space of charmless three-body \B decays.
There are also observables where any observation of \CP violation at the achievable precision would be a clear signal of beyond SM physics: examples include the parameters of \CP violation in both \Bd and \Bs mixing, or in decays to final states involving photons or leptons.
The powerful approach of testing the consistency of the SM paradigm through global fits to CKM matrix parameters requires improved measurements of several important quantities including $\gamma$, $\beta$ and $\beta_s$.

The study of \CP violation in the \B sector will therefore remain a high-priority component of the global programme in particle physics for the foreseeable future.
Complemented by progress in other areas, including kaon and charm physics, precision measurements of low energy observables, searches for new phenomena at the energy frontier, and neutrino physics, there is a real possibility of a breakthrough.
Understanding of the physics that will be probed in these studies may help to resolve the shortcomings of the SM, and to explain the matter-antimatter asymmetry in the Universe.